\documentclass[12pt]{article}

\usepackage{amsfonts}
\usepackage{amssymb}
\usepackage{amsthm}
\usepackage[mathscr]{eucal}

\input{epsf} 

\makeatletter
\renewcommand \thesection {\@arabic\c@section}
\renewcommand\thesubsection   {\thesection.\@arabic\c@subsection}
\renewcommand\thesubsubsection{\thesubsection .\@arabic\c@subsubsection}
\renewcommand\theparagraph    {\thesubsubsection.\@arabic\c@paragraph}
\renewcommand\section{\@startsection {section}{1}{\z@}%
                                   {-3.5ex \@plus -1ex \@minus -.2ex}%
                                   {1.9ex \@plus.2ex}%
                                   {\normalfont\large\bfseries\centering}}
\renewcommand\subsection{\@startsection{subsection}{2}{\z@}%
                                     {-2ex\@plus -1ex \@minus -.2ex}%
                                     {1.2ex \@plus .2ex}%
                                    {\normalfont\normalsize\bfseries\centering}
}
\renewcommand\subsubsection{\@startsection{subsubsection}{3}{\z@}%
                                     {-2ex\@plus -1ex \@minus -.2ex}%
                                     {.5ex \@plus .2ex}%
                                     {\normalfont\normalsize\em}}
\renewcommand\paragraph{\@startsection{paragraph}{4}{\z@}%
                                    {3.25ex \@plus1ex \@minus.2ex}%
                                    {-1em}%
                                    {\normalfont\normalsize\em}}
\renewcommand\subparagraph{\@startsection{subparagraph}{5}{\parindent}%
                                       {3.25ex \@plus1ex \@minus .2ex}%
                                       {-1em}%
                                      {\normalfont\normalsize\em}}
\makeatother

\newcounter{subequation}
	\newenvironment{subequation}%
	{\addtocounter{equation}{-1}%
	\stepcounter{subequation}%
	\begin{equation}}%
	{\end{equation}%
}

\newcommand{\beq}{\begin{equation}}
\newcommand{\eeq}{\end{equation}}
\newcommand{\bseq}{\begin{subequation}}
\newcommand{\eseq}{\end{subequation}}
\newcommand{\bea}{\begin{eqnarray}}
\newcommand{\eea}{\end{eqnarray}}

\newcommand{\refeq}[1]{(\ref{#1})}

\newcommand{\eps}{\epsilon}

\newcommand{\cA}{{\cal A}}

\newcommand{\cE}{{\cal E}}

\newcommand{\cF}{{\cal F}}

\newcommand{\cN}{{\cal N}}
\newcommand{\cP}{{\cal P}}
\newcommand{\cS}{{\cal S}}

\newcommand{\NN}{{\mathbb N}}

\newcommand{\RR}{{\mathbb R}}
\newcommand{\TT}{{\mathbb T}}
\newcommand{\ZZ}{{\mathbb Z}}

\newcommand{\pr}{\prime}
\newcommand{\ppr}{{\prime\prime}}
\newcommand{\pppr}{{\prime\prime\prime}}

\newcommand{\QED}{\hfil\hbox to 0pt{}\ \hbox to 2em{\hss}\ 
         \hbox to 0pt{}\hskip-2em plus 1fill
         \vrule height6pt depth1pt width7pt\par\medskip}
\newcommand{\eQED}{\hfil\hbox to 0pt{}\ \hbox to 2em{\hss}\ 
          \hbox to 0pt{}\hskip-2em plus 1fill
          \vbox{\hrule height .25pt depth 0pt width 7pt
            \hbox{\vrule height 6.5pt depth 0pt width .25pt
              \hskip 6.5pt\vrule height 6.5pt depth 0pt width .25pt}
            \hrule height .25pt depth 0pt width 7pt}\par\medskip}

\newcommand{\tst}{\textstyle}
\newcommand{\sst}{\scriptstyle}
\newcommand{\ssst}{\scriptscriptstyle}

\newcommand{\kB}{k_{\rm B}}

\newcommand{\N}{{\ssst\rm N}}
\newcommand{\Y}{{\ssst\rm Y}}
\newcommand{\W}{{\ssst\rm W}}
\newcommand{\vdW}{{\ssst\rm vdW}}
\newcommand{\CS}{{\ssst\rm CS}}
\newcommand{\dB}{{\ssst\rm dB}}
\newcommand{\GC}{{\ssst\rm GC}}
\newcommand{\PC}{{\ssst\rm PC}}

\newcommand{\Sp}{{\ssst\rm Sp}}

\newcommand{\alg}{{\ssst\rm alg}}
\newcommand{\tr}{{\sst\rm tr}}

\newcommand{\ul}{\underline}
\newcommand{\ol}{\overline}

\newcommand{\ave}[1]{ [\! [{#1}]\! ]}
\newcommand{\Ave}[1]{ \{\! [{#1}]\! \}}

\newcommand{\norm}[2]{ \left\Vert{#1}\right\Vert_{_{#2}} }

\newcommand{\llangle}{\left\langle}
\newcommand{\rrangle}{\right\rangle}

\newcommand{\lra}{{\leftrightarrow}}

\newcommand{\bfr}{{\bf r}}

\newcommand{\ag}{\alpha,\gamma}	
\newcommand{\bg}{\beta,\gamma}	
\newcommand{\aN}{\alpha,N}

\newcommand{\aV}{{\alpha V}}

\newcommand{\wtilde}{\widetilde}

\newcommand{\diam}{{\oslash}} 

\newcommand{\dist}{{\rm dist}}

\newcommand{\esssup}{{{\rm ess\ sup}}}

\newcommand{\Lbd}{\Lambda}

\newcommand{\bullf}{{\bullet\rm\ssst f}}
\newcommand{\bulls}{{\bullet\rm\ssst s}}

\newcommand{\fs}{{\rm\ssst fs}}
\newcommand{\gl}{{\rm\ssst g\ell}}
\newcommand{\vd}{{\rm\ssst vd}}

\begin{document}

\title{Hard-sphere fluids with chemical self potentials\thanks{\copyright{2010} 
		The copyright for this preprint resides with the authors. 
		Its reproduction, in its entirety, for non-commercial purposes is permitted.
		The copyright for the version published in Journal of Mathematical Physics 
		will reside with the AIP.}}
\author{M.K.-H. Kiessling$^a$ and J.K. Percus$^b$\\
\textit{$^a$ Department of Mathematics, Rutgers, The} \\
\textit{State University of New Jersey, Piscataway, NJ 08854}\\
\textit{$^b$ Courant Institute of Mathematical Sciences and Physics} \\
\textit{Department, New York University, New York, NY 10012}}
\bigskip\bigskip\bigskip
\date{\small Version of Jan. 22, 2010\\ In press at: Journal of Mathematical Physics}
\maketitle
\newpage
\begin{abstract}
\noindent
	Existence, uniqueness and stability of solutions is studied for a set of 
nonlinear fixed point equations which define self-consistent hydrostatic equilibria 
of a classical continuum fluid that is confined inside a container $\ol\Lbd\subset \RR^3$ 
and in contact with either a heat and a matter reservoir, or just a heat reservoir.
	The local thermodynamics is furnished by the statistical mechanics of a system 
of hard balls, in the approximation of Carnahan-Starling.
	The fluid's local chemical potential per particle at $\bfr\in\Lbd$ is the sum of the matter 
reservoir's contribution and a self contribution $-(V*\rho)({\bf r})$, where $\rho$ is the 
fluid density function and $V$ a non-negative linear combination of
the Newton kernel $V_{\N}(|\bfr|) = -|\bfr|^{-1}$,
the Yukawa kernel $V_{\Y}(|\bfr|) = -|\bfr|^{-1}e^{-\kappa |\bfr|}$, and
a van der Waals kernel $V_{\W}(|\bfr|) = -  (1 + \varkappa^2 |\bfr|^2)^{-3}$.
	The fixed point equations involving  the Yukawa and Newton kernels are equivalent 
to semilinear elliptic PDEs of second order with a nonlinear, nonlocal boundary condition.
	We prove the existence of a grand canonical phase transition, and of a petit 
canonical phase transition which is embedded in the former.
	The proofs suggest that, except for  boundary layers, the grand canonical transition is of the type 
``all gas $\lra$ all liquid'' while the petit canonical one is of the type ``all vapor $\lra$ liquid drop with
vapor atmosphere.'' 
	The latter proof in particular suggests the existence of solutions with interface structure 
which compromise between the all-liquid and all-gas density solutions.
\end{abstract}
\bigskip
\bigskip
\noindent
{\bf KEYWORDS:} 

\noindent
{\sl Nonlinear analysis}: fixed point problems, integral and partial differential equations;

\noindent
{\sl Classical fluids}: liquid vs. gas phase transition, liquid drops, liquid-vapor interface; 

\noindent
{\sl Particle systems}: hard sphere-, Yukawa-, Newton-, and van der Waals-interactions;

\noindent
{\sl Statistical mechanics}: petit- and grand-canonical ensembles, van-der-Waals limit.

\newpage
\noindent
\section{INTRODUCTION} 

\noindent
	The interface between physically co\"{e}xisting thermodynamic (locally) pure phases poses a 	
challenging array of problems in statistical mechanics which fall somewhere inbetween the micro- and 
macroscopic realms.
	The large scale (macroscopic) geometry of the interface can 
be successfully modeled as a sharp {\it Gibbs interface}, computed from some constrained principle of 
minimum surface area (Wulf shape); this generalizes to the dynamical domain in form of motion by mean 
curvature and related principles.
	The transversal structure of the interface, which obviously is not resolved when the interface is 
modeled as a Gibbs interface, is the hard problem that lives at the fringe of the macro-world and for which 
there is no definitive answer yet.

	To get a hand on the transversal structure it is customary to invoke a van der Waals (for fluids)
or Weiss (for magnets) mean-field approximation which allows one to study both the large scale geometry and the 
transversal interface structure.
	While this approximation overly simplifies the problem, it is far from being understood completely
and continues to attract the attention of mathematical 
physicists.$^{\cite{albertibellettiniA,albertibellettiniB,asselah,carlenETalA,carlenETalB,cop}}$
	For the liquid-vapor interface at equilibrium, which is the motivation for this paper, the
van der Waals type models emerge in Kac$^{\cite{kacetal}}$ type scaling limits from the statistical 
mechanics of systems of interacting microscopic particles, with particle densities resolved on the  long 
distance scale of the attractive part, $V_A$,  of the particle interaction $V_R +V_A$, while the short distance 
repulsive part $V_R$ is absorbed into the local thermodynamics.$^{\cite{vankampen,percusA,mkjkp}}$ 

	The local thermodynamics of an $N$-body system with repulsive pair interaction $V_R$ 
is given by a pure phase, defined in the thermodynamic limit of a macroscopically spatially 
uniform system in thermal equilibrium with a heat reservoir at reciprocal temperature $\beta \in \RR^+$ 
and a matter reservoir at logarithmic fugacity (i.e., chemical potential per particle~:~temperature ratio) 
$\gamma \in \RR$, characterized by a position-independent pressure~:~temperature ratio $p=\wp(\bg)$ 
and  particle density $\ol\eta =\partial_\gamma \wp(\bg)$ at 
all points of differentiability of $\gamma \mapsto \wp(\bg)$.
	On general thermodynamic grounds, $(\bg)\mapsto\wp(\bg)$ is strictly positive, 
increasing in $\gamma\in \RR$, and convex in $\beta\in\RR^+$ and $\gamma$. 
	By convexity, $\gamma \mapsto \wp(\bg)$ is differentiable a.e., but the 
models from physics are expected to be better behaved and feature only finitely many points of 
non-differentiability, at $\gamma_1(\beta),\, \gamma_2(\beta), ... $, say.
	At such a $\gamma_k(\beta)$ typically two different pure phases are equally likely, one of them 
denser than the other, and one needs to select the one which furnishes the local thermodynamics. 

	In this paper the local thermodynamics is chosen to represent a continuum formed by 
many identical hard microscopic balls, known (in a fluid state) as a hard-sphere fluid and more
generally as a hard-sphere system.
	A hard-sphere system is characterized by a $\beta$-independent 
pressure~:~temperature ratio, i.e. $\wp(\bg) =\wp_{\bullet}^{\phantom{b}}(\gamma)$.
	We will write $\wp_{\bullet}^{\prime}(\gamma)$ for 
$\partial_\gamma\wp_{\bullet}^{\phantom{b}}(\gamma)$.
	The function $\wp_{\bullet}^{\phantom{b}}(\gamma)$ has a point of non-differentiability at
$\gamma_\fs$ associated with a fluid-versus-solid transition.
	Here we are interested in studying the fluid phases, but for our investigations we do need
to have control over this singularity.

	Physically, a hard-sphere fluid may model the short distance repulsion between the spherical atoms 
in a noble gas or between neutrons in a neutron fluid.
	Over somewhat larger distances $r$ any two such atoms or neutrons also feel attractive 
forces $-\nabla V_A$, the van der Waals (Jr.) force in the case of atoms, which is due to self-induced 
dipole-dipole interactions associated with their first excited configurations, and  the Yukawa force
in the case of neutrons, which is explained in terms of the pion exchange of the strong nuclear forces.
	When the number of atoms or neutrons becomes too large,
as in (helium) brown dwarf stars or in neutron stars, Newtonian gravity has to be added.
	We choose the $V_A$ interaction to mimic any of these  physical systems.
	More precisely, writing $\aV$ for $V_A$, we let $\aV$ stand for any non-negative-linear 
combination of the form
\beq
\aV(r) =  A_{\W}V_{\W}(r) + A_{\Y}V_{\Y}(r) + A_{\N}V_{\N}(r),
\label{lincomV}
\eeq 
where
\bea
V_{\W} (r) \!\!\!&=&  -  (1 + \varkappa^2 r^2)^{-3}\, ,
\label{vdWaalsV}\\
V_{\Y} (r) \!\!\!&=& - e^{- \kappa r}/r \, ,
\label{YukawaV}\\
V_{\N} (r) \!\!\!&=& - 1/r\, ,
\label{NewtonV}
\eea
are integral kernels of strictly negative definite compact operators on $L^2(\Lbd)$ for any 
bounded $\Lbd\subset\RR^3$, and where $A_{\W}\in\{0,\alpha_{\W}\}$,
$A_{\Y}\in\{0,\alpha_{\Y}\}$, and
$A_{\N}\in\{0,\alpha_{\N}\}$, 
while $\alpha_{\W}$, $\alpha_{\Y}$, and $\alpha_{\N}$
are strictly positive coupling constant~:~temperature ratios.
	In the van der Waals approximation the effect of $\aV$ on the system is accounted for 
by adding to the externally generated chemical potential per particle~:~temperature ratio $\gamma$ 
the {\it chemical self potential per particle~:~temperature ratio at $\bfr$}, given by
$-(\aV*\eta)_{_{\!\Lbd}}(\bfr)$, where
\beq
\left(V*\eta\right)_{_{\!\Lbd}}\!\!(\bfr) =\int_{\!\Lbd} V(|\bfr-\tilde\bfr|)\eta(\tilde\bfr) d^3\tilde{r}.
\label{VconvRHO}
\eeq
	We refer to $(V_{\N}*\eta)_{_{\!\Lbd}}(\bfr)$ as {\it the Newton --}, 
to $(V_{\Y}*\eta)_{_{\!\Lbd}}(\bfr)$ as {\it the Yukawa --}, and
to $(V_{\W}*\eta)_{_{\!\Lbd}}(\bfr)$ as {\it the van der Waals potential of $\eta$ at $\bfr$}.

	In his original study, van der Waals$^{\cite{vdWaals}}$ assumed boundary effects to be
negligible and the density function $\eta(\bfr)$ to be spatially uniform, i.e. $\eta(\bfr) \equiv \ol\eta$.
	These assumptions can be rigorously correct only in the infinite volume limit$^{\cite{JOELuOLI}}$
when the fluid fills all space $\RR^3$ uniformly, with gravity ``switched off;'' 
note that $V_{\W}(|\,\cdot\,|)\in L^1(\RR^3)$ and $V_{\Y}(|\,\cdot\,|)\in L^1(\RR^3)$, 
while $V_{\N}(|\,\cdot\,|)\in L_{\rm loc}^1(\RR^{\rm 3})$ merely. 
	When $A_{\N}=0$ and the constant function $\eta({\bfr})\equiv\ol\eta$ is substituted 
in \refeq{VconvRHO} with $\Lbd=\RR^3$, then 
$-(V*\ol\eta)_{_{\RR^3}} = \ol\eta \norm{V(|\,\cdot\,|)}{^{L^1(\RR^3)}}$ is a constant function, too.
	Setting $\norm{V(|\,\cdot\,|)}{L^1(\RR^3)}=:\norm{V}{1}$ for short, 
the van der Waals densities $\ol\eta_{\vdW}$ are then computed from the van der Waals fixed 
point equation\footnote{In textbooks (e.g. \cite{huang}, \cite{mattisswendsen}) 
		 one usually finds discussions of \refeq{fixptEQvdW}
		with $\wp^\pr_\bullet(\gamma)$ replaced by van der Waals' $\wp^\pr_{\vdW}\!(\gamma)$ 
		which corresponds to a system of many hard rods on a line. While for
		systems of hard balls it gives quantitatively wrong answers, qualitatively they reproduce 
		those obtained with the correct $\wp^\pr_\bullet(\gamma)$.
		Also, usually a value for $\|V\|_{L^1}$ is given without specifying $V$.}
\beq
\ol\eta  = \wp_{\bullet}^{\prime}\big(\gamma + \alpha \norm{V}{1}\!\ol\eta \big).
\label{fixptEQvdW}
\eeq 
	In $(\ag)$-parameter space there are disjoint, open two-dimensional domains
where the algebraic equation \refeq{fixptEQvdW} has one or three solutions in the fluid density regime, 
respectively (see also sections IV \&\ V); these regions are separated by a closed one-dimensional subset
featuring two solutions of \refeq{fixptEQvdW}, except for one point (the critical point) at which only one 
solution exists.
	Constant (large enough) $\alpha$ sections and constant (intermediate size) $\gamma$ sections
through the fluid solution manifold over the $(\ag)$ half plane each produce an $S$-shaped curve 
associated with the famous ``van der Waals loop.'' 
	In the region with three fluid density solutions, the largest solution is interpreted as the 
liquid density phase, the smallest as the gas (a.k.a. vapor) density phase of the fluid, and the 
intermediate density solution as a thermodynamically unstable artifact of the model.
	The liquid and the gas density solutions are each stable fixed points of \refeq{fixptEQvdW} under 
iteration, the intermediate density solution is not.
	However, thermodynamically the liquid and gas density solutions are simultaneously stable only 
along the {\it gas} $\&$ {\it liquid co\"{e}xistence curve} $\alpha\mapsto\gamma =\gamma_\gl^{\vdW}(\alpha)$
of the model, determined by Maxwell's equal-areas construction,$^{\cite{maxwell}}$ rigorously vindicated in Ref.\cite{JOELuOLI},
while away from this curve (still in the three-solutions region) thermodynamically only one of these two solutions 
is stable, the other one at most metastable.
	Here, thermodynamic stability and metastability are understood with $\alpha$ and $\gamma$ fixed, 
and explained below.

	More interesting than \refeq{fixptEQvdW} is the nonlinear fixed point problem
\beq
\eta(\bfr)  = \wp_{\bullet}^{\prime}\big(\gamma - (\aV*\eta)_{_{\!\RR^3}}(\bfr)\big)
\label{fixptEQ}
\eeq
in the positive	cone of the non-separable Banach space $C^0_b(\RR^3)$ of bounded continuous functions 
$\eta(\bfr)$, $\bfr\in \RR^3$.
	If $V\in L^1(\RR^3)$, then \refeq{fixptEQ} can be solved with the Ansatz 
$\eta(\bfr) \equiv \ol\eta$, which leads back to the algebraic fixed point equation \refeq{fixptEQvdW}.
	Yet, for a hard-sphere fluid with $V(|\bfr|) = V_{\Y}(|\bfr|)$, whenever $(\ag)$ is 
a point on the gas $\&$ liquid co\"{e}xistence curve of locally uniform phases computed from \refeq{fixptEQvdW}, 
then (modulo translations and rotations) a unique monotonic planar interface solution 
$\eta(\bfr)\equiv \wtilde\eta(x)$ exists, where $x\in\RR$ is a cartesian coordinate of $\RR^3$.
	This can be shown by adapting the ODE arguments on p.~40-41 of Ref.\cite{percusB}, which are
available because $(-\Delta +\kappa^2)V_{\Y}(|\bfr|)= -4\pi \delta(\bfr)$.
	A monotonic planar interface solution illustrates the physical phenomenon of
co\"existence of the liquid and the gas density phases; they have been extensively
studied in one dimensional models.$^{\cite{uhlenbecketal,percusB,asselah,cop}}$
	Furthermore, using the equivalent radial ODE problem obtained with the help of
$(-\Delta +\kappa^2)V_{\Y}(|\bfr|)= -4\pi \delta(\bfr)$, Mironescu$^{\cite{mironescu}}$ has shown 
that solutions in $\RR^3$ with spherical droplet / bubble geometry exist; these solutions do {\it not}
exist exactly on the gas $\&$ liquid co\"{e}xistence curve for the uniform phases, yet are nearby.
	Such ODE arguments are not available for a hard-sphere fluid with 
$V(|\bfr|) = V_{\W}(|\bfr|)$, and the existence and classification of the non-constant
solutions in $\RR^3$ of \refeq{fixptEQ} in this case is largely unexplored territory.
	We also note that since $V_{\N}\!(|\,\cdot\,|) \not\in L^1(\RR^3)$, 
the fixed point problem \refeq{fixptEQ} is not well defined in $\RR^3$ as it stands with $V_{\N}*\eta$ 
given by \refeq{VconvRHO}; however, replacing $V_{\N}*\eta$ by $\phi_{\N}$ and stipulating the familiar
Poisson equation $\Delta\phi_{\N} = 4\pi \eta$, solutions in $\RR^3$ for the related PDE problem 
$\Delta\phi_{\N} = 4\pi \wp_{\bullet}^{\prime}(\gamma - \alpha \phi)$ do exist; it is easy to numerically
compute radial solutions, which have applications in planetary science.$^{\cite{bsmkks,mkRMP}}$
	To summarize, non-uniform van der Waals fluid theory furnishes an accessible model 
to study the structure of non-uniform density functions $\eta(\bfr)$ of a hard-sphere fluid in $\RR^3$. 
	Evidently, boundary effects are absent in $\RR^3$.
	Moreover, for $V=V_{\Y}$ and $V=V_{\N}$ simple ODE techniques greatly facilitate the computation 
of solutions in $\RR^3$.

	Beside the structure of interfaces, their fluctuations  are of interest.
	Unfortunately, in unbounded space $\RR^3$ all interface solutions are thermodynamically neither 
stable nor metastable and interface fluctuations diverge,$^{\cite{percusB,percusC,percusD}}$ 
even though droplets may be quite long-lived in dynamical calculations with the 
Alan--Cahn and related evolution equations.
	To obtain finite fluctuations one needs to stabilize the interfaces.

	An intuitively obvious way to obtain a stable fluid interface is to enclose the fluid inside 
a container $\ol\Lbd$ (with either wetting, non-wetting, or neutral mechanical boundary conditions) and 
to replace the thermodynamic contact condition of prescribed logarithmic fugacity $\gamma$ by the stricter
one of prescribed amount of matter $\int_{\!\Lbd} \eta(\bfr)d^3r = N$. 
	When $\Lbd$ is macroscopic and $N$ halfway inbetween the values of 
$|\Lbd|\ol\eta$ for the large and small fluid density values $\ol\eta$ solving \refeq{fixptEQvdW}, 
then there is too much matter in the container to be all vapor, and too little to be all liquid.
	In this case the system must find a compromise structure: either a drop of liquid surrounded
by vapor or a bubble of vapor inside liquid, depending on the mechanical boundary conditions. 
	It is reassuring to find this scenario confirmed numerically for $V=V_{\W}$ and neutral 
mechanical boundary conditions,$^{\cite{mkjkp,mkRMP}}$ and in particle simulations of many hard balls
with attractive $-r^{-6}$ interactions.$^{\cite{kalosetal,koplik}}$ 
	Interestingly enough, in Ref.\cite{mkjkp} it was found numerically that the thermodynamic transition
from the vapor state to the liquid-drop state is not gradual but occurs at a petit-canonical first-order 
phase transition which is embedded in the grand-canonical first order phase transition between vapor and 
liquid.
	To rigorously prove this empirical picture correct is an interesting mathematical problem
which is still largely open.

	To make a modest contribution toward its solution we here continue our previous study$^{\cite{mkjkp}}$ 
where we numerically evaluated the fixed point problem 
\beq
\eta(\bfr)  = \wp_{\bullet}^{\prime}\big(\gamma - (\aV*\eta)_{_{\!\Lbd}}\!(\bfr) \big)
\label{fixptEQinLambda}
\eeq
for a cohesive ``hard-sphere continuum'' inside a container $\ol\Lbd\subset\RR^3$ with neutral boundary.
	In this paper we study the existence, uniqueness, structure, and stability of fluid solutions to 
\refeq{fixptEQinLambda} with rigorous analysis.
	Stability is defined as follows.

	The stability of solutions of \refeq{fixptEQinLambda} for the thermodynamic contact conditions 
``heat and matter reservoirs'' is determined by the functional
\beq
\cP^{_\Lbd}_{\ag}[\eta] =  
\int_{\!\Lbd} \wp_\bullet^{\phantom{b}}\bigl(\gamma - \left(\aV* \eta\right)_{_{\!\Lbd}}\!\!(\bfr)\bigr) d^3r 
+ \frac{1}{2}
\int_{\!\Lbd}\int_{\!\Lbd} \aV(|\bfr-\tilde\bfr|) \eta(\bfr) \eta(\tilde\bfr)\, d^3r\, d^3\tilde{r},
\label{Pfctnl}
\eeq
which we rigorously derived from the grand-canonical ensemble in Ref.\cite{mkjkp}. 
	Solutions of \refeq{fixptEQinLambda} are critical points of \refeq{Pfctnl} in the positive cone of the 
separable Banach space $C^0_b(\ol\Lbd)$. 
	A solution $\eta_{_\Lbd}$ of \refeq{fixptEQinLambda} is {\it globally $\cP$ stable} if 
$\cP^{_\Lbd}_{\ag} [\eta_{_\Lbd}] = P_{_\Lbd} (\ag)$, where 
\beq
P_{_\Lbd} (\ag) 
:=  \max_\eta \{\cP^{_\Lbd}_{\ag} [\eta]\};
\label{vpPinLambda}
\eeq
global maximizers are denoted\footnote{``GC'' stands for grand canonical because 
	those densities comprise the support of the grand canonical measure in the van der Waals limit.}
$\eta^{\GC}_{\ssst\Lbd}(\bfr)$, their dependence on $\ag$ implied.
	A solution $\eta_{_\Lbd}$ of \refeq{fixptEQinLambda} is {\it locally $\cP$ stable} if 
\beq
{\cP^{_\Lbd}_{\ag}}^{\!\!\!\!\ppr}(\sigma,\sigma)\bigl\vert_{\eta_{_\Lbd}}\bigr. < 0 
\label{linstab}
\eeq
for all $\sigma \not\equiv 0$ such that  $0\leq \eta_{_\Lbd} + \sigma \in C^0_b(\ol\Lbd)$. 
	Here,
\bea
{\cP^{_\Lbd}_{\ag}}^{\!\!\!\!\ppr}(\sigma,\sigma)\Bigl\vert_{\eta_{_\Lbd}}\Bigr. 
\!\!\!&=& 
\frac{1}{2} \int_{\!\Lbd}  
\wp_\bullet^\ppr\bigl(\gamma -
\left(\aV*\eta_{_\Lbd}\right)_{_{\!\Lbd}}\!\!(\bfr)\bigr)\left(\aV* \sigma\right)_{_{\!\Lbd}}^2(\bfr)\, d^3r 
\nonumber\\ 
\!\!\!&& \qquad\quad
+ \frac{1}{2}
\int_{\!\Lbd}\int_{\!\Lbd} \aV(|\bfr-\tilde\bfr|) \sigma(\bfr) \sigma(\tilde\bfr)\,d^3r\,d^3\tilde{r}
\label{linPfctnl}
\eea
is the diagonal part of the second G\^{a}teaux derivative of $\cP^{_\Lbd}$ at $\eta_{_\Lbd}$. 
	A solution $\eta_{_\Lbd}$ of \refeq{fixptEQinLambda} satisfying \refeq{linstab} but not \refeq{vpPinLambda}
is called {\it $\cP$ metastable}. 
	If ${\cP^{_\Lbd}_{\ag}}^{\!\!\!\!\ppr}(\sigma,\sigma)\bigl\vert_{\eta_{_\Lbd}}\bigr. > 0$ ($\,=0\,$)
for at least one $\sigma$, then $\eta_{_\Lbd}$ is called {\it $\cP$ unstable} ({\it locally $\cP$ indifferent}).

	In Ref.\cite{mkjkp} we also explained that stability of solutions of \refeq{fixptEQinLambda} for a given amount 
of matter in thermodynamic contact with a ``heat reservoir'' at inverse temperature $\beta\, (\propto \alpha)$ 
is defined in terms of the following thermodynamic free energy functional.
	For each density function $\eta\in C^0_b(\ol\Lbd)$ we define:

\noindent
(i) its amount of matter in $\Lbd$,

\beq
\cN^{_\Lbd}[\eta] =  \int_{\!\Lbd} \eta(\bfr) d^3r;
\label{Nfunc}
\eeq
(ii) its energy~:~temperature ratio,
\beq
\cE^{_\Lbd}_{\alpha}[\eta] = {\tst\frac{3}{2}} \cN^{_\Lbd}[\eta]
+ 
{\tst\frac{1}{2}}\int_{\!\Lbd}\int_{\!\Lbd}\aV(|\bfr-\tilde\bfr|) \eta(\bfr)\eta(\tilde\bfr)\, d^3r\, d^3\tilde{r};
\label{Efunc}
\eeq
(iii) its (strictly) classical entropy,
\beq
\cS^{_\Lbd}_\bullet[\eta] = {\tst\frac{5}{2}}\cN^{_\Lbd}[\eta]
 -
\int_{\!\Lbd} \eta(\bfr) \ln \eta(\bfr) d^3r
-
\int_{\!\Lbd} \eta(\bfr)\int_0^{\eta(\bfr)}
\!\! \big(p_\bullet^{\phantom{b}}(\ol\eta)-\ol\eta\big)/{\ol\eta}^{2}\, d\ol\eta\, d^3r,
\label{Sfunc}
\eeq
where $p_\bullet^{\phantom{b}}(\ol\eta)$ is the hard-sphere  pressure~:~temperature ratio as function of $\ol\eta$;

\noindent
(iv) its free energy~:~temperature ratio,
\beq
\cF^{_\Lbd}_\alpha[\eta] = \cE^{_\Lbd}_{\alpha}[\eta] - \cS^{_\Lbd}_\bullet[\eta].
\label{Ffunc}
\eeq
	The thermodynamic free energy~:~temperature ratio $F_{_\Lbd}(\aN)$ is then given by 
\beq
F_{_\Lbd}(\aN) 
= \min_\eta \{ \cF^{_\Lbd}_{\alpha}[\eta]\; |\; \cN^{_\Lbd}[\eta] = N\}.
\label{PCstab}
\eeq
	Solutions of \refeq{fixptEQinLambda} which saturate \refeq{PCstab} are called {\it globally $\cF$ stable} 
and denoted\footnote{``PC'' stands for petit canonical because those densities comprise 
	the support of the petit canonical measure in the van der Waals limit.}
$\eta^{\PC}_{\ssst\Lbd}(\bfr)$, their dependence on $\alpha, N$ implied.
	{\it Local $\cF$ stability} etc. is defined in terms of the diagonal
part of the second G\^{a}teaux derivative of $\cF^{_\Lbd}_{\alpha}$,
\beq
{\cF^{_\Lbd}_{\alpha}}^{\!\ppr}(\sigma,\sigma)\Bigl\vert_{\eta_{_\Lbd}}\Bigr. 
= 
-\frac{1}{2} \int_{\!\Lbd}  
s_\bullet^\ppr\bigl(\eta_{_\Lbd}(\bfr)\bigr) \sigma^2(\bfr)\, d^3r 
+ \frac{1}{2}
\int_{\!\Lbd}\int_{\!\Lbd} \aV(|\bfr-\tilde\bfr|) \sigma(\bfr) \sigma(\tilde\bfr)\,d^3r\,d^3\tilde{r}
\label{linFfctnl}
\eeq
where $s_\bullet(\eta)$ is defined by $\cS^{_\Lbd}_\bullet[\eta]=\int_{\!\Lbd} s_\bullet(\eta(\bfr))d^3r$.
	Variation is to be carried out under the constraint $\int_{\!\Lbd}\sigma d^3 r = 0$ so that the 
$\eta_{_\Lbd}$-disturbances preserve the number of particles.
\smallskip

	We close this introduction by re-emphasizing that our study of the finite volume fixed point 
problem \refeq{fixptEQinLambda} is not motivated by trying to understand so-called finite-size
{\sl{corrections}} to dominant infinite volume results. 
	Rather it is motivated by the physical phenomenon of stable interface solutions in finite
containers holding a fixed amount of fluid.
	What makes such a study difficult are the following two points: (i) boundary layer 
effects are as big as interface effects, and one has to separate these in the analysis; 
(ii) one has to rule out competing
solutions which take values in the crystal regime, and this leaves little wiggle room in parameter space.
	To level the ground we first study the simpler problem when the amount of fluid is controlled
by a matter reservoir before turning to the problem with a fixed amount of fluid.
	
	The rest of this article is structured as follows:

$\bullet$ 
In section II we define the Carnahan--Starling approximation to the fluid part and the
Speedy approximation to the solid part of the function $\gamma\mapsto\wp_{\bullet}(\gamma)$.

$\bullet$ 
	In section III we identify a parameter region in the $(\ag)$ (half) space in which 
all solutions of \refeq{fixptEQinLambda} take values exclusively in the fluid density range.

$\bullet$ 
	In section IV we identify a region in which fluid solutions are unique.

$\bullet$ 
	In section V we identify a region where various fluid solutions exists.

$\bullet$ 
	In section VI we study the thermodynamic stability of the fluid solutions in contact with
heat-plus-matter reservoirs.
	We prove the existence of a vapor $\lra$ liquid phase transition in the finite-volume 
grand-canonical ensemble.

$\bullet$ 
	In section VII we address the thermodynamic stability of the fluid solutions in contact only 
with a heat reservoir and
also explain the relationship with the infinite-volume Lebowitz-Penrose results.$^{\cite{JOELuOLI}}$
	We prove the existence of a petit-canonical finite-volume phase transition between the vapor 
state and a state for which we present evidence that it is of liquid-drop type.

$\bullet$ 
	Appendix A supplies some explicit evaluations valid for spherical geometry. 

$\bullet$ 
	Appendix B lists the nonlinear partial differential equations associated with our fixed point integral equations.

$\bullet$ 
	Appendix C provides a ``dictionary'' to translate our dimensionless notation into conventional 
dimensional notation used in the physics literature.

$\bullet$ 
	Appendix D contains a brief list of (minor) errata for our previous papers on the subject, 
Ref.\cite{mkJSPa} and Ref.\cite{mkjkp}.

\noindent
\section{THERMODYNAMICS OF HARD SPHERE SYSTEMS}
\medskip

\noindent
	Numerical simulations$^{\cite{hansenmacdonald,hooverree}}$ of the dynamics of many identical hard balls
indicate a thermodynamically stable fluid phase only when the dimensionless density (the volume fraction)
$\ol\eta$ stays below $\ol\eta_\fs^{\ssst\, <}\approx 0.49$, with numerical errors reportedly less than $1\%$. 
	Numerical simulations$^{\cite{hooverree,speedy}}$ also indicate a thermodynamically stable solid phase 
above $\ol\eta_\fs^{\ssst\, >}\approx 0.54$ all the way up to 
${\ol\eta_{\ssst fcc}^{\ssst\, cp}}  = \pi \sqrt{2\,}/6 \approx 0.7405$, 
the fcc crystal close packing fraction,
although the system may ``jam'' into a glassy structure.$^{\cite{rintoultorquato}}$
	The interval $(\ol\eta_\fs^{\ssst\, <},\ol\eta_\fs^{\ssst\, >})$ is interpreted as furnishing
the co\"{e}xistence of both fluid and solid phases.$^{\cite{hansenmacdonald,hooverree}}$  
	In the absence of empirical evidence for any further phase transition of the hard-sphere system,$^{\cite{speedy}}$
one may thus assume that the map $\gamma \mapsto p=\wp_\bullet^{\phantom{b}}(\gamma)$ for a hard-sphere 
system is a positive, increasing, convex function over $\RR$, which is asymptotic to a 
straight line with slope equal to ${\ol\eta_{\ssst fcc}^{\ssst\, cp}}$ when $\gamma\uparrow\infty$.
	Moreover, the map $\gamma \mapsto p=\wp_\bullet^{\phantom{b}}(\gamma)$ has a kink at 
$\gamma =\gamma_\fs$ but otherwise is locally real analytic, such that away from the kink we
have $\ol\eta = \wp_\bullet^\pr(\gamma)$. 
	At $\gamma=\gamma_\fs$ the left-derivative
$\lim_{\gamma\uparrow \gamma_\fs}\wp_\bullet^\pr(\gamma)= \ol\eta_\fs^{\ssst\, <}$, and
the right-derivative
$\lim_{\gamma\downarrow \gamma_\fs}\wp_\bullet^\pr(\gamma)= \ol\eta_\fs^{\ssst\, >}$.
	Unfortunately, no manageable formula is known for the exact $\wp_{\bullet}^{\phantom{b}}(\gamma)$, 
but convenient formulas for very accurate approximations to $\wp_{\bullet}^{\phantom{b}}(\gamma)$ are known.

	For the fluid regime $\gamma\in (-\infty,\gamma_\fs]$ we resort to a formula by 
Carnahan and Starling,$^{\cite{carnahanstarling}}$ whose numerological
manipulations have led to a simple approximation $\wp_{\CS}(\gamma)$ to\footnote{Carnahan 
	and Starling$^{\ssst\cite{carnahanstarling}}$ proposed $p = g_{_1}(\ol\eta)$, with 
	$ g_{_1}$ given in Definition 2.1, as the explicit sum of an approximate virial 
	series for the equation of state $ p=g_\bullf^{\phantom{b}} (\ol\eta)$ for a hard-sphere fluid, 
	inspired by the few available terms in the actual virial expansion. Quantitatively their 
	equation of state slightly improves over the Percus--Yevick$^{\ssst\cite{percusyevick}}$ 
	equation of state under compressibility closure$^{\ssst\cite{wertheim}}$ (identical to the equation 
	of state obtained from scaled particle theory$^{\ssst\cite{reissfrischjll}}$), 
	from which it differs by the extra $-\ol\eta^4$ term in the numerator.}
$\wp_\bullet^{\phantom{b}}(\gamma)\big|_{\gamma\leq\gamma_\fs}\big.=:\wp_\bullf^{\phantom{b}}(\gamma)$
which remarkably accurately fits the empirical data obtained in computer simulations.
	Graphs of the function $\wp_{\CS}(\gamma)$ and its derivative $\wp_{\CS}^\pr(\gamma)$ are displayed 
in Figs.1 $\&$ 2 of Ref.\cite{mkjkp}.

\smallskip
\noindent
{\bf Definition 2.1:} {\it The Carnahan-Starling approximation to 
$\wp_\bullf^{\phantom{b}}(\gamma)$ 
is defined by the map $\gamma\mapsto p=\wp_{\CS}(\gamma)$, given by the parameter representation 
$p= g_{_1}(\ol\eta)$ and $\gamma =g_{_2}(\ol\eta)$, with$^{\cite{carnahanstarling,balescu,hansenmacdonald}}$ 
\bea
g_{_1} (\ol\eta) \!\!\!&=& \frac{{\ol\eta + \ol\eta^2 +\ol\eta^3 - \ol\eta^4}}{(1-\ol\eta)^3}
\label{pOFeta}\\
g_{_2}(\ol\eta) \!\!\!&=& \ln \ol\eta + \frac{8\ol\eta -9\ol\eta^2 +3 \ol\eta^3}{(1-\ol\eta)^3},
\label{gammaOFeta}
\eea
where $\ol\eta$ is a real parameter in the  interval $0<\ol\eta \leq  \ol\eta_\fs^{\ssst\, <} \approx 0.49$.
	This gives $\gamma_\fs = g_{_2}(\ol\eta_\fs^{\ssst\, <}) \approx 15.208$ as 
right limit for the domain of definition $(-\infty,\gamma_\fs]$ of $\wp_{\CS}(\gamma)$.}

\noindent
{\bf Remark:} 
	Note that \refeq{pOFeta} and \refeq{gammaOFeta} are related by a thermodynamic identity for a 
system of many identical hard balls,
\beq
\ol\eta g_{_2}^\prime(\ol\eta) = g_{_1}^\prime(\ol\eta).
\label{diffID}
\eeq
	Indeed, \refeq{gammaOFeta} is obtained from \refeq{pOFeta} by integrating \refeq{diffID} and conveniently choosing
the integration constant.  
	As a consequence we have that $\wp_{\CS}^\pr(\gamma)=g_{_2}^{-1}(\gamma)=\ol\eta$ is
a dimensionless particle density --- as already implied by the stipulated notation.~\eQED

\noindent
{\bf Remark:} 
	Formally \refeq{pOFeta} and \refeq{gammaOFeta} are well defined for all $\ol\eta\in (0,1)$, and one
may want to study this mathematical model in its own right.
	Whenever we use $\wp_{\CS}(\gamma)$ as defined for all $\ol\eta\in (0,1)$
we will refer to it as the {\it Carnahan-Starling model}, to distinguish 
this mathematical model from the actual hard-sphere physics.~\eQED

	For the solid regime $\gamma\in [\gamma_\fs,\infty)$ we resort to Speedy's 
effective approximation $\wp_{\Sp}(\gamma)$ to 
$\wp_\bullet^{\phantom{b}}(\gamma)\big|_{\gamma\geq\gamma_\fs}\big.=:\wp_\bulls^{\phantom{b}}(\gamma)$,
whose leading term is determined theoretically while the other terms invoke a Pad\'e approximation to fit
the numerical simulation data.\footnote{Speedy reports that his formula agrees to within less than 
	half the reported error with the results of Alder et al.$^{\ssst\cite{alderetal}}$, which 
	are given by their formula (1), an asymptotic expansion 
	in powers of ${\ol\eta_{\ssst fcc}^{\ssst\, cp}}-\ol\eta$, viz.
\beq
{g_3 (\ol\eta) = {\ol\eta_{\ssst fcc}^{\ssst\, cp}}\Big[
 3\frac{1}{{1-\ol\eta /\ol\eta_{\ssst fcc}^{\ssst\, cp}}} +
 K_0\,  +
 K_1(1-\ol\eta/{\ol\eta_{\ssst fcc}^{\ssst\, cp}}) + 
 O\bigl(({\ol\eta_{\ssst fcc}^{\ssst\, cp}}-\ol\eta)^2 \bigr)}\Big]\, , 
\label{SPEEDYformula}
\eeq
with $ K_0\approx -3.44$ and $ K_1\approx 1$ taken from table III in Ref.\cite{alderetal}.}

\smallskip
\noindent
{\bf Definition 2.2:} {\it The Speedy approximation to $\wp_\bulls^{\phantom{b}}(\gamma)$ 
is defined by the map $\gamma\mapsto p=\wp_{\Sp}(\gamma)$, given by the parameter representation 
$p= g_{_3}(\ol\eta)$ and $\gamma =g_{_4}(\ol\eta)$, with$^{\cite{speedy}}$ 
\beq
g_{_3} (\ol\eta) = 
	3\frac{{\ol\eta_{\ssst fcc}^{\ssst\, cp}}}{{1-\ol\eta/\ol\eta_{\ssst fcc}^{\ssst\, cp}}} 
+ a \frac{{b-\ol\eta/\ol\eta_{\ssst fcc}^{\ssst\, cp}}}{{c-\ol\eta/\ol\eta_{\ssst fcc}^{\ssst\, cp}}},
\label{pOFetaSOLIDfcc}
\eeq
with $a=0.5921$, $b=0.7072$, $c=0.601$, and
\beq
g_{_4} (\ol\eta) = \gamma_\fs + \int_{\ol\eta_\fs^{\ssst\, >}}^{\ol\eta}\frac{g_3^\prime(x)}{x}dx
\label{gammaOFetaSOLIDfcc}
\eeq
where $\ol\eta$ ranges in the  interval 
$0.54 \approx  \ol\eta_\fs^{\ssst\, >} \leq \ol\eta <  {\ol\eta_{\ssst fcc}^{\ssst\, cp}}\approx 0.7402$.
	Note that both $g_{_3}(\ol\eta)$ and $g_{_4}(\ol\eta)$ are monotonic increasing on 
$(\ol\eta_\fs^{\ssst\, >},\ol\eta_{\ssst fcc}^{\ssst\, cp})$, diverging $\uparrow\infty$ for 
$\ol\eta\uparrow\ol{\eta}_{\ssst fcc}^{\ssst\, cp}$. }

	Speedy's paper$^{\cite{speedy}}$ features formula \refeq{pOFetaSOLIDfcc}, while \refeq{gammaOFetaSOLIDfcc} 
follows from postulating the thermodynamic relation $\ol\eta g_{_4}^\pr(\ol\eta) = g_{_3}^\pr(\ol\eta)$ 
for $\ol\eta_\fs^{\ssst\, >} < \ol\eta < \ol\eta_{\ssst fcc}^{\ssst\, cp}$; 
as a consequence, $\wp_\bulls^\pr(\gamma)= g_{_4}^{-1}(\gamma)$ for $\gamma >\gamma_\fs$.
	The integration constant is chosen such that $\gamma_\fs = g_{_4}(\ol\eta_\fs^{\ssst\, >}) \approx 15.208$ 
is the left limit for the domain of definition $[\gamma_\fs,\infty)$ of $\wp_{\Sp}(\gamma)$.

	To summarize, we stipulate the following:

\medskip
\noindent
{\bf Convention 2.3:} {\it In the remainder of the paper, for $\gamma\leq \gamma_\fs$, i.e. in the fluid
phase, we take 
$\wp_\bullf^{\phantom{b}}(\gamma) := \wp_{\CS}(\gamma) \equiv (g_{_1}\circ g_{_2}^{-1})(\gamma)$, 
with $g_{_1}$ and $g_{_2}$ given by \refeq{pOFeta} and \refeq{gammaOFeta} in Definition 2.1. 
	For $\gamma\geq \gamma_\fs$, i.e. in the solid phase, we take
$\wp_\bulls^{\phantom{b}}(\gamma) := \wp_{\Sp}(\gamma) \equiv (g_{_3}\circ g_{_4}^{-1})(\gamma)$, with 
$g_{_3}$ and $g_{_4}$ given by \refeq{pOFetaSOLIDfcc} and \refeq{gammaOFetaSOLIDfcc} in Definition 2.2.} 
\medskip

	In the next two figures we display the hard-sphere pressure~:~temperature ratio (Fig.1)
and the hard-sphere chemical potential per particle~:~temperature ratio  (Fig.2), 
both as functions of $\ol\eta$.
	The second figure in particular will be very helpful to consult when reading our proofs 
in the ensuing sections. 

\epsfxsize=7cm
\epsfysize=7cm
\centerline{\epsffile{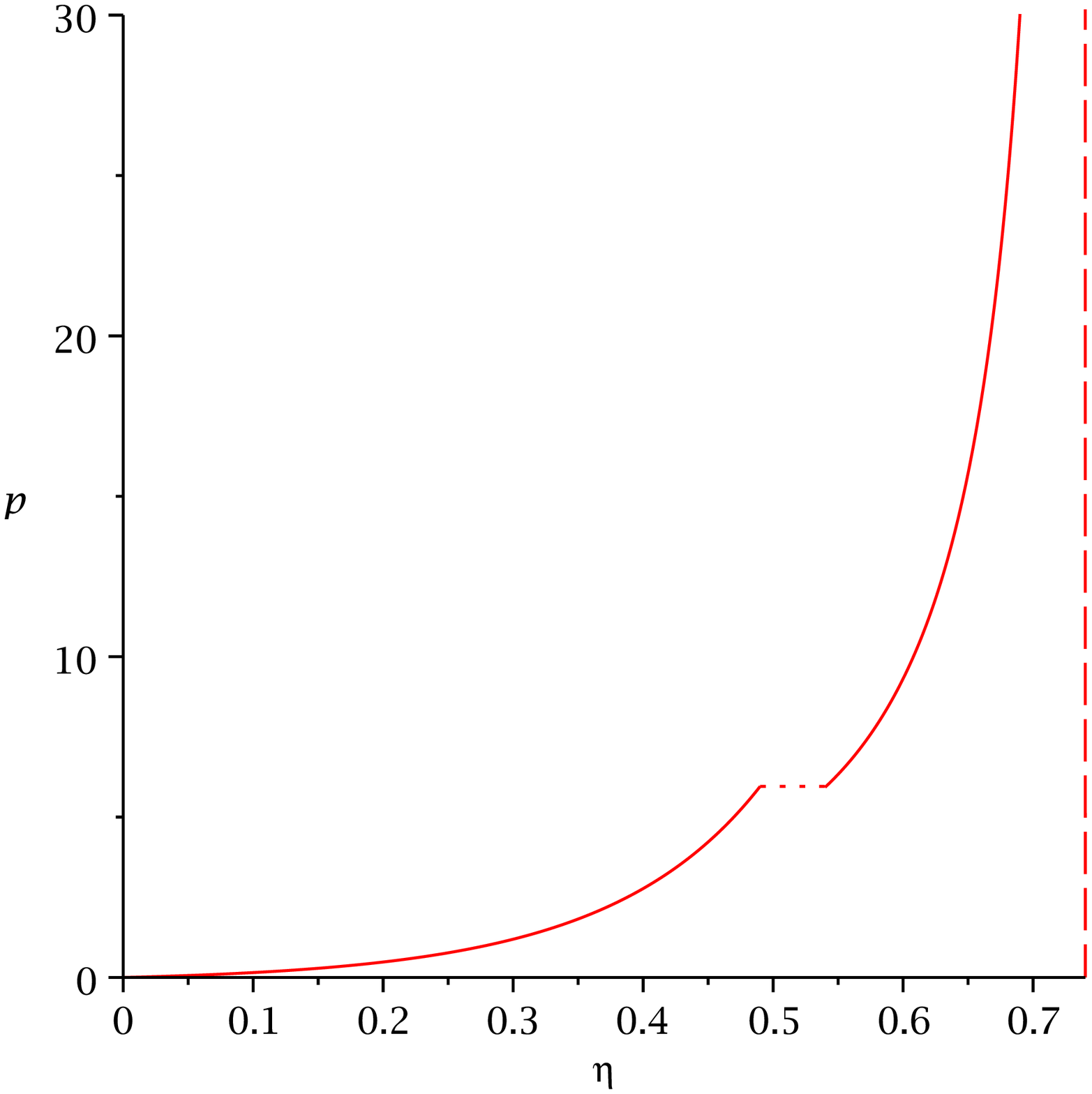}}
{\noindent\footnotesize{Fig.1. Equation of state $p$ vs. $\eta$ for a classical hard-balls continuum. Displayed 
		are fluid branch ($0<\eta<0.49$) and solid branch ($0.54<\eta<0.74$) together with the 
		coexistence line ($0.49\leq\eta\leq0.54$; dotted) and the fcc crystal close packing limit
		(broken vertical line at $\eta=0.74019$).}}

\hrule
\noindent
\epsfxsize=7cm
\epsfysize=7cm
\centerline{\epsffile{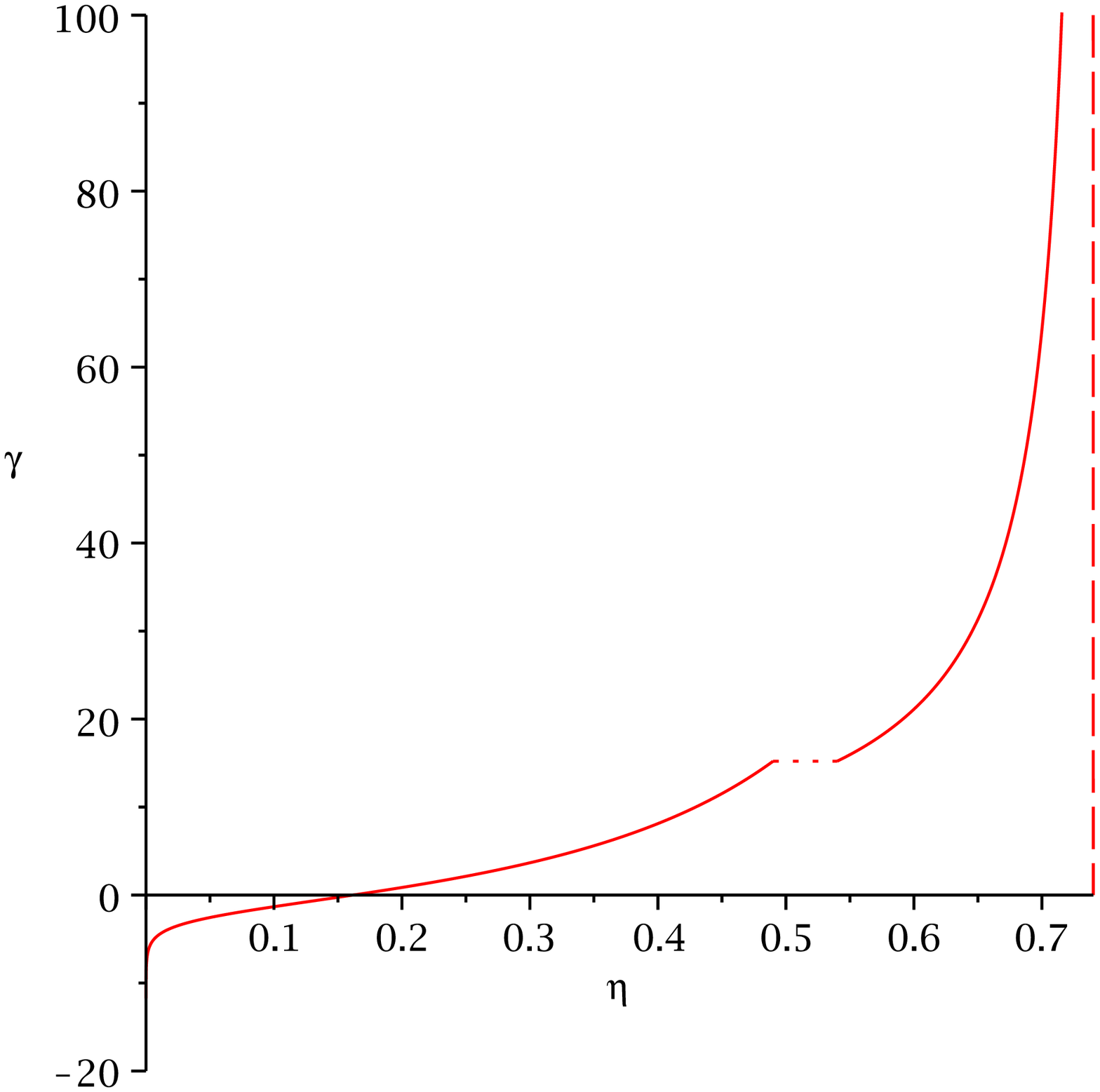}}
{\noindent\footnotesize{Fig.2. Chemical-potential-per-particle~:~temperature ratio $\gamma$ vs. volume
		fraction $\eta$ for a classical hard-balls continuum. Displayed are fluid branch 
		($0<\eta<0.49$) and solid branch ($0.54<\eta<0.74$) together with the coexistence line 
		($0.49\leq\eta\leq0.54$; dotted) and the fcc crystal close packing limit (broken vertical 
		line at $\eta=0.74019$).}}
\newpage

	We end this subsection by emphasizing that we do need to know what we stipulated about 
$\wp_\bullet^{\phantom{b}}(\gamma)$ for the solid phase even though in this paper we are studying only fluid solutions.
	In particular, ``all the hard work'' in our paper is caused by the following dilemma: 
to prove a first-order phase transition between two different stable fluid solutions one must find suitable $(\ag)$ 
pairs for which \refeq{fixptEQinLambda} has at least three all-fluid solutions, but one also must rule out 
any other globally stable solution which takes solid values somewhere in the container.
	Clearly, a sufficient though not necessary condition for the absence of a globally stable solution 
which takes solid values somewhere is the complete absence of any solution taking non-fluid values somewhere.
	This sufficient condition is simpler to implement, but is of course also more restrictive.

	The space-uniform van der Waals theory gives a good indication of the difficulties ahead.
	Recall that the space-uniform solutions to the van der Waals problem \refeq{fixptEQvdW} for given $(\ag)$ and 
$\norm{V}{1}<\infty$ can be graphically determined as the abscissa values of the intersection points of the 
graph displayed in the second figure with the straight line $\ol\eta\mapsto \gamma + \alpha \norm{V}{1}\!\ol\eta$. 
	The $(\ag)$ pairs for which a phase transition in the fluid regime occurs while no solid solution exists at all
lie in the $(\ag)$ domain which corresponds graphically to the family of straight lines 
$\ol\eta\mapsto \gamma + \alpha \norm{V}{1}\!\ol\eta$ 
which have three intersections with the fluid branch but no intersection with the solid branch in the second figure
	Inspection of the second figure makes it obvious that this leaves us with only very little ``room to wiggle'' 
in $(\ag)$ space, so that we need to develop delicate analytical estimates to accomplish our feat of proving 
the grand canonical gas vs. liquid transition and the petit canonical vapor vs. drop transition within the 
non-uniform van der Waals model for a hard-sphere fluid with chemical self-potential confined to
a container.

\medskip
\noindent
\section{LOCATING THE FLUID SOLUTIONS IN $(\ag)$ HALF SPACE} 

	In this section we give some sufficient and some necessary a priori conditions concerning 
the existence of solutions $\eta$ of \refeq{fixptEQinLambda} which do not take values outside the 
fluid regime, i.e. for which $\gamma - (\aV*\eta)_{_{\!\Lbd}} < \gamma_\fs$. 
	We shall write $V*\eta$ for either $(V*\eta)_{_{\!\Lbd}}$ or $(V*\eta)_{_{\!\RR^3}}$
whenever it is clear from the context what we mean. 
	We set $\|V*1\|_{C^0_b(\Lbd)}= \Phi_{_\Lbd}$, where $\|\,\cdot\,\|_{C^0_b(\Lbd)}$ 
denotes the uniform (supremum) norm for $C^0_b(\Lbd)$; 
notice that $\Phi_{\!{\RR^3}}\!=\norm{V}{1}$.
	We also introduce the notation 
$\ol{\cal B}_{\xi}=\{\eta :\,\norm{\eta}{C^0_b(\Lbd)}\!\! \leq \xi\}$ for the
closed ball of radius $\xi$ in $C^0_b(\ol\Lbd)$.

	We begin with some sufficient conditions for existence.

\medskip\noindent
{\bf Proposition 3.1:} {\it Assume that the inequality 
\beq
\gamma +  \alpha \Phi_{_\Lbd} \ol\eta - g_{_2}(\ol\eta) \leq 0 \, 
\label{Aestim}
\eeq
is satisfied for at least one $\ol\eta\in (0,\ol\eta_\fs^{\ssst\, <}]$, so that 
the algebraic fixed point equation
\beq
\ol\eta = g_{_2}^{-1}\bigl(\gamma +\alpha\Phi_{_\Lbd} \ol\eta \bigr)\, 
\label{FIXetaPHIeq}
\eeq
has at least one solution $\in (0,\ol\eta_\fs^{\ssst\, <}]$.
	Let $\ol\eta^{\ssst m}_{\ssst\Lbd}$ be the minimal and
$\ol\eta^{\ssst M}_{\ssst\Lbd}$ the maximal solution in 
$[0,\ol\eta_\fs^{\ssst\,<}]$ of \refeq{FIXetaPHIeq}.
	Then in the truncated positive cone 
$C^0_{b,+}(\ol\Lbd) \cap \ol{\cal B}_{\ol\eta^{\ssst M}_{\ssst\Lbd}}$
there exists a pointwise minimal and a pointwise maximal fluid solution of \refeq{fixptEQinLambda}, 
denoted ${\eta}^{\ssst m}_{\ssst\Lbd}(\bfr)$ 
and ${\eta}^{\ssst M}_{\ol\eta^{\ssst M}_{\ssst\Lbd}}(\bfr)$,
respectively. 
	In particular, the iteration sequences $\{\eta^{(n)}\}_{n=0}^\infty$ defined by
\beq
\eta^{(n+1)} = g_{_2}^{-1}\bigl(\gamma - \alpha V*\eta^{(n)}\bigr)
\label{iteraRHO}
\eeq
with starting densities $\eta^{(0)} = \ol\eta^{\ssst m}_{\ssst\Lbd}$ and 
$\eta^{(0)}= g_{_2}^{-1}(\gamma)$, respectively, both converge pointwise to the minimal solution 
${\eta}^{\ssst m}_{\ssst\Lbd}(\bfr)$, the former monotone downward and the latter monotone upward.
	Starting the iteration map \refeq{iteraRHO} with $\eta^{(0)} = \ol\eta^{\ssst M}_{\ssst\Lbd}$ 
yields a sequence which converges pointwise monotone downward to the maximal solution
${\eta}^{\ssst M}_{\ol\eta^{\ssst M}_{\ssst\Lbd}}(\bfr)$.}

\medskip\noindent
{\bf Remarks:} 
	(a) Since, by hypothesis, \refeq{Aestim} is satisfied, and since $g_{_2}$ is
continuous with $\lim_{\ol\eta\downarrow 0}g_{_2}(\ol\eta) = -\infty$, 
the straight line $\ol\eta \mapsto \gamma  + \alpha \Phi_{_\Lbd} \ol\eta$ 
intersects the curve $\ol\eta \mapsto g_{_2}(\ol\eta)$ at least once 
(and at most three times) in $(0, \ol\eta_\fs^{\ssst\,<}]$. 
	Therefore a maximal point of intersection 
$\ol\eta^{\ssst M}_{\ssst\Lbd} \leq \ol\eta_\fs^{\ssst\,<}$ 
does exist.
	(b) Proposition 3.1 does not state that ${\eta}^{\ssst M}_{\ol\eta^{\ssst M}_{\ssst\Lbd}}$ is 
the maximal solution in $C^0_{b,+} \cap \ol{\cal B}_{\ol\eta_\fs^{\ssst\,<}}$; however,
${\eta}^{\ssst m}_{\ssst\Lbd}$ is automatically the minimal solution in 
$C^0_{b,+} \cap \ol{\cal B}_{\ol\eta_\fs^{\ssst\,<}}$.
	(c) Maximal and minimal solution, 
${\eta}^{\ssst M}_{\ol\eta^{\ssst M}_{\ssst\Lbd}}$ and ${\eta}^{\ssst m}_{\ssst\Lbd}$, may coincide.
\eQED

To prove Proposition 3.1, all we need to know about $V$ is $V\in L^1(\Lbd)$ and $V<0$. 

\smallskip
{\it Proof of Proposition 3.1:} 
	Consider first the case $\eta^{(0)}=\ol\eta^{\ssst M}_{\ssst\Lbd}$.
	Since $\gamma \mapsto g_{_2}^{-1}(\gamma)$ is strictly monotonic increasing, and since
$-(V*1)(\bfr) \leq \norm{V*1}{C^0_b(\Lbd)}=\Phi_{_\Lbd}$, but with 
$-(V*1)(\bfr) \not\equiv\Phi_{_\Lbd}$, the iteration \refeq{iteraRHO} yields
$\eta^{(n)}(\bfr) \leq\eta^{(n-1)}(\bfr)\,\forall\,n\in\NN$, and even
$\eta^{(n+1)}(\bfr) <\eta^{(n)}(\bfr)\,\forall\,n\in\NN$ and all $\bfr\in\Lbd$.
	Since
\beq
g_{_2}^{-1}(\gamma - \alpha V*\eta) \geq g_{_2}^{-1}(\gamma) >0\, ,
\label{uSCHRANKE}
\eeq 
the iterates are bounded below by a positive number. 
	Hence, the iterates converge pointwise down to a strictly 
positive function $\eta^{\ssst M}_{\ol\eta^{\ssst M}_{\ssst\Lbd}}$ which clearly is entirely fluid. 
	By the $C^0_b(\ol\Lbd)$ continuity of the operator $g_{_2}^{-1}(\gamma -\alpha V*\, \cdot\,)$,
the function $\eta^{\ssst M}_{\ol\eta^{\ssst M}_{\ssst\Lbd}}$ solves \refeq{fixptEQinLambda}.

	As in Ref.\cite{amannC} it can be shown that 
$\eta^{\ssst M}_{\ol\eta^{\ssst M}_{\ssst\Lbd}}$ is the pointwise maximal solution in 
$C^0_{b,+} \cap \ol{\cal B}_{\ol\eta^{\ssst M}_{\ssst\Lbd}}$.
	For suppose that $\eta < \ol\eta^{\ssst M}_{\ssst\Lbd}$ is any solution of \refeq{fixptEQinLambda},
then by the monotonic increase of $g_{_2}^{-1}(\gamma -\alpha V*\, \cdot\,)$ and by the fact that 
$\ol\eta^{\ssst M}_{\ssst\Lbd}$ is a strict supersolution for \refeq{fixptEQinLambda}, we can conclude that 
$g_{_2}^{-1}(\gamma -\alpha V*\, \cdot\,)$ maps $[\eta,\ol\eta^{\ssst M}_{\ssst\Lbd}]$ into itself. 
	Therefore, $\eta\leq \eta^{\ssst M}_{\ol\eta^{\ssst M}_{\ssst\Lbd}}$, and this proves that 
$\eta^{\ssst M}_{\ol\eta^{\ssst M}_{\ssst\Lbd}}$ is the pointwise maximal solution in
$C^0_{b,+} \cap \ol{\cal B}_{\ol\eta^{\ssst M}_{\ssst\Lbd}}$.

	By essentially the same arguments, starting the iteration 
with $\eta^{(0)}=\ol\eta^{\ssst m}_{\ssst\Lbd}$ yields a monotone downward converging sequence of
iterates with limit $\eta^{\ssst m}_{\ssst\Lbd}$, and $\eta^{\ssst m}_{\ssst\Lbd}$ 
is the pointwise maximal solution in
$C^0_{b,+} \cap \ol{\cal B}_{\ol\eta^{\ssst m}_{\ssst\Lbd}}$.

	Next consider the case $\eta^{(0)} =g_{_2}^{-1}(\gamma)$. 
	Using again the strict monotonic increase of $\gamma \mapsto g_{_2}^{-1}(\gamma)$, 
this time combined with the positivity of $-(V*1)(\bfr)$, we conclude that the sequence 
\refeq{iteraRHO} iterates pointwise monotone upward. 
	By \refeq{uSCHRANKE} all iterates are strictly positive. 
	Moreover, by induction it follows that, if $\eta^{(n)}<\ol\eta^{\ssst m}_{\ssst\Lbd}$, then 
\beq
	\eta^{(n+1)} 
= 
	g_{_2}^{-1}\big(\gamma-\alpha V*\eta^{(n)}\big) 
< 
	g_{_2}^{-1}\big(\gamma + \alpha\Phi_{_\Lbd}\ol\eta^{\ssst m}_{\ssst\Lbd}\big) 
= 
	\ol\eta^{\ssst m}_{\ssst\Lbd} \, .
\label{oSCHRANKE}
\eeq
	Clearly, $\eta^{(0)}<\ol\eta^{\ssst m}_{\ssst\Lbd}$, so the sequence is bounded above by 
$\ol\eta^{\ssst m}_{\ssst\Lbd}$. 
	It now follows that it converges pointwise to a strictly positive solution
$\dot\eta^{\ssst m}_{\ssst\Lbd}\leq \eta^{\ssst m}_{\ssst\Lbd}$ of \refeq{fixptEQinLambda}, 
and also that this solution is entirely fluid. 
	Moreover, similarly as for the maximal solution it now follows that 
$\dot\eta^{\ssst m}_{\ssst\Lbd}$ is the pointwise minimal solution in 
$C^0_{b,+} \cap \ol{\cal B}_{\ol\eta^{\ssst m}_{\ssst\Lbd}}$,
hence in $C^0_{b,+} \cap \ol{\cal B}_{\ol\eta_\fs^{\ssst\,<}}$.

	Lastly, the proof that $\dot\eta^{\ssst m}_{\ssst\Lbd}=\eta^{\ssst m}_{\ssst\Lbd}$ is
a minor variation on the proof of Corollary 4.5 in section IV.\QED

	By a slight sharpening of \refeq{Aestim} we can improve Proposition 3.1 to the following.

\smallskip\noindent
{\bf Proposition 3.2:} {\it Assume that 
\beq
\gamma - \gamma_\fs +  \alpha \Phi_{_\Lbd}\ol\eta_\fs^{\ssst\,<} \leq 0 \, .
\label{AAestim}
\eeq
	Then \refeq{Aestim} is satisfied for $\ol\eta =\ol\eta_\fs^{\ssst\,<}$, so
Proposition 3.1 applies.
	Now the pointwise maximal fluid solution $\eta^{\ssst M}_{\ol\eta^{\ssst M}_{\ssst\Lbd}}$
of \refeq{fixptEQinLambda} in $C^0_{b,+} \cap \ol{\cal B}_{\ol\eta^{\ssst M}_{\ssst\Lbd}}$ is in 
fact the pointwise maximal fluid solution in $C^0_{b,+} \cap \ol{\cal B}_{\ol\eta_\fs^{\ssst\,<}}$.
}

{\it Proof of Proposition 3.2:} 
	Since $\gamma_\fs = g_{_2}(\ol\eta_\fs^{\ssst\,<})$, \refeq{AAestim} implies 
that \refeq{Aestim} is satisfied by $\ol\eta = \ol\eta_\fs^{\ssst\,<}$, so
all conclusions in Proposition 3.1 apply. 

	To show that the pointwise maximal solution in 
$C^0_{b,+} \cap \ol{\cal B}_{\ol\eta^{\ssst M}_{\ssst\Lbd}}$ 
is in fact the pointwise maximal solution in $C^0_{b,+} \cap \ol{\cal B}_{\ol\eta_\fs^{\ssst\,<}}$,
we notice that $\ol\eta_\fs^{\ssst\,<}$ is a strict supersolution a.e. for \refeq{fixptEQinLambda}.
	This implies that the sequence $\{\eta^{(n)}\}_{n=0}^\infty$ defined 
by \refeq{iteraRHO} with initial value $\eta^{(0)} = \ol\eta_\fs^{\ssst\,<}$ iterates pointwise 
monotonically downward, strictly monotonically a.e., to the  pointwise maximal
solution in $C^0_{b,+} \cap \ol{\cal B}_{\ol\eta_\fs^{\ssst\,<}}$ of \refeq{fixptEQinLambda}. 
	We show that this solution is in $C^0_{b,+} \cap \ol{\cal B}_{\ol\eta^{\ssst M}_{\ssst\Lbd}}$,
and so, {\it a forteriori}, it is also the  pointwise maximal solution in 
$C^0_{b,+} \cap \ol{\cal B}_{\ol\eta^{\ssst M}_{\ssst\Lbd}}$.
	
	By \refeq{AAestim}, $\ol\eta_\fs^{\ssst\,<}$ is a supersolution for \refeq{FIXetaPHIeq}.
	Therefore, either $\ol\eta_\fs^{\ssst\,<}$ is itself the largest fixed point in 
$[0,\ol\eta_\fs^{\ssst\,<}]$ of \refeq{FIXetaPHIeq}, or else the sequence $\{\ol\eta^{(n)}\}_{n=0}^\infty$ 
defined by
\beq
\ol\eta^{(n+1)} = g_{_2}^{-1}\bigl(\gamma +\alpha\Phi_{_\Lbd} \ol\eta^{(n)} \bigr)\, 
\label{ITERAeta }
\eeq
with initial value $\ol\eta^{(0)} = \ol\eta_\fs^{\ssst\,<}$ iterates strictly monotonically downward 
to the largest fixed point in $[0,\ol\eta_\fs^{\ssst\,<})$ of \refeq{FIXetaPHIeq}, which in either case
is  $\ol\eta^{\ssst M}_{\ol\eta_\fs^{\ssst\,<}}$.
	Moreover, with $\ol\eta^{(0)} = \ol\eta_\fs^{\ssst\,<} = \eta^{(0)}$, for each $n>0$ we have 
\beq
\eta^{(n)} \leq \ol\eta^{(n)} \, ,
\label{dominance}
\eeq 
because $\eta^{(n_0)} \leq \ol\eta^{(n_0)}$ for some $n_0\geq 0$ implies that
\beq
	\eta^{(n_0+1)} 
 = 
	g_{_2}^{-1}\big(\gamma - \alpha V* \eta^{(n_0)}\big) 
\leq 
	g_{_2}^{-1}\big(\gamma + \alpha \Phi_{_\Lbd} \ol\eta^{(n_0)}\big) 
= 
	\ol\eta^{(n_0+1)}\, . 
\label{dominancePROOF}
\eeq
	We conclude that 
\beq
	\eta^{\ssst M}_{ \ol\eta_\fs^{\ssst\,<}}:=
\lim_{n\to \infty} \eta^{(n)} \leq  \lim_{n\to \infty} \ol\eta^{(n)} = 
	\ol\eta^{\ssst M}_{\ssst\Lbd} \, .
\label{maxESTIM}
\eeq
	Hence,
$\eta^{\ssst M}_{ \ol\eta_\fs^{\ssst\,<}}=\eta^{\ssst M}_{\ol\eta^{\ssst M}_{\ssst\Lbd}}$, so
$\eta^{\ssst M}_{\ol\eta^{\ssst M}_{\ssst\Lbd}}$ is the pointwise maximal
solution in $C^0_{b,+} \cap \ol{\cal B}_{ \ol\eta_\fs^{\ssst\,<}}$.\QED

\smallskip\noindent
{\bf Remark:} 
	For our $V$, the dominance can be sharpened from ``$\leq$'' to
``$<\ a.e.$'' by noting that obviously $\eta^{(1)}<\ol\eta^{(1)}$ a.e.
\eQED

	If we consider the extension of \refeq{fixptEQinLambda} to all $\gamma\in\RR$, with 
$\wp_\bullet^\prime(\,\cdot\,)=\wp_{\CS}^\prime(\,\cdot\,)= g_{\ssst 2}^{-1}(\,\cdot\,)$ 
for $\,\cdot\, \leq \gamma_\fs$ with $\wp_\bullet^\prime$ meaning left derivative, 
and with $\wp_\bullet^\prime(\,\cdot\,)= g_{\ssst 4}^{-1}(\,\cdot\,)$ 
when $\,\cdot\, >\gamma_\fs$, with  $\wp_\bullet^\prime$ now meaning right derivative, 
covering fluid and solid branch as explained in Convention 2.3, then the existence results of 
Propositions 3.1 and  3.2 can be complemented by a result about the non-existence 
of solutions to the so extended \refeq{fixptEQinLambda} which are not fluid somewhere in $\Lbd$.
 
\medskip\noindent
{\bf Proposition 3.3:} {\it If the inequality 
\beq
\gamma - \gamma_\fs + \alpha \Phi_{_\Lbd} {\ol\eta_{\ssst fcc}^{\ssst\, cp}} \leq  0\, ,
\label{NOnonFLUIDcondition}
\eeq
holds, then the extended fixed point problem \refeq{fixptEQinLambda} does not have any solution 
that takes values outside the hard-sphere fluid regime somewhere in $\Lbd$.}

\medskip
{\it Proof of Proposition 3.3:}
	Since $\eta\leq {\ol\eta_{\ssst fcc}^{\ssst\, cp}}$, and since $g_4(\ol\eta) > \gamma_\fs$ 
for all $\ol\eta\in(\ol\eta_\fs^{\ssst\,>},{\ol\eta_{\ssst fcc}^{\ssst\, cp}}]$, we conclude that 
\refeq{NOnonFLUIDcondition} implies for all 
$\ol\eta\in(\ol\eta_\fs^{\ssst\,>},{\ol\eta_{\ssst fcc}^{\ssst\, cp}}]$ that
\beq
\gamma  + \alpha \Phi_{_\Lbd}\ol\eta < g_4(\ol\eta).
\label{gbfour}
\eeq

	Now suppose a solution $\eta$ of the extended \refeq{fixptEQinLambda} would exist which in some open 
subdomain $\Lbd_s$ of $\Lbd$ is solid. 
	Then, clearly, 
$\ol\eta_\fs^{\ssst\,>}\leq\|{\eta}\|_{C^0_b(\Lbd)}\leq {\ol\eta_{\ssst fcc}^{\ssst\, cp}}$, 
and since $\gamma\mapsto g_4^{-1}(\gamma)$ 
is increasing, we conclude that in the solid region (i.e., in $\Lbd_s$) we have
\beq
\norm{\eta}{C^0_b(\Lbd)}
\leq 
g_4^{-1}\big(\gamma + \alpha \Phi_{_\Lbd} \norm{\eta}{C^0_b(\Lbd)}\big)
\label{normfourest}
\eeq
as a consequence of the extended \refeq{fixptEQinLambda}. 
	But by applying $g_4$ to both sides of \refeq{normfourest}, this leads to a contradition with \refeq{gbfour}.
	Hence, no solution of the extended \refeq{fixptEQinLambda} can exist which somewhere in $\Lbd$ is not 
a hard-sphere fluid.
\QED
\medskip

	The next result requires $V\in L^1(\RR^3)$.
	It relates the algebraic fixed point problem \refeq{fixptEQvdW} for constant solutions in $\RR^3$ of 
\refeq{fixptEQ} to the problem \refeq{fixptEQinLambda} in bounded $\Lbd \subset \RR^3$.

\medskip\noindent
{\bf Proposition 3.4: } {\it 
	Let $\alpha\norm{V}{1} =  A_{\W}(\pi^2/ 4\varkappa^3) +  A_{\Y}(4\pi /\kappa^2)$. 
	Suppose the algebraic fixed point problem \refeq{fixptEQvdW} has a solution 
$\ol\eta_{\vdW} \leq \ol\eta_\fs^{\ssst\,<}$, so that $\ol\eta_{\vdW}$ satisfies
\beq
\ol\eta = g_{_2}^{-1}(\gamma + \alpha \norm{V}{1} \ol\eta). 
\label{vdWaalsFIXPTEQ}
\eeq
	Then for all domains $\Lbd\subset \RR^3$ the fixed point problem  \refeq{fixptEQinLambda} with 
$\aV=A_{\W}V_{\W}+ A_{\Y}V_{\Y}$
and $\wp_\bullet^{\phantom{b}}(\gamma)$ given in Definition 2.1 has a hard-sphere fluid solution.}
\medskip

{\it Proof of Proposition 3.4:} 
	By subadditivity of the norm, we have
\beq
\alpha\norm{V*1}{C^0_b(\Lbd)} 
\leq    
A_{\W}\norm{V_{\W}\!\!*1}{C^0_b(\Lbd)} 
+  
A_{\Y}\norm{V_{\Y}\!\!*1}{C^0_b(\Lbd)} .
\label{TRIANGLEineq}
\eeq
	Since $V_{\W}(|\,\cdot\,|)\in L^1(\RR^3)$ and $V_{\Y}(|\,\cdot\,|)\in L^1(\RR^3)$, we have 
\bea
\norm{V_{\W}\!\!*1}{C^0_b(\Lbd)} 
\!\!\!&\leq& 
 \norm{V_{\W}*1}{C^0_b(\RR^3)}
= 
\frac{\pi^2 }{ 4\varkappa^3} \, ,
\label{vdWaNORMsup}\\
\norm{V_{\Y}\!\!*1}{C^0_b(\Lbd)} 
\!\!\!&\leq& 
\norm{V_{\Y}*1}{C^0_b(\RR^3)}
= \frac{4\pi}{\kappa^2}\, .
\label{YukaNORMsup}
\eea
	With \refeq{TRIANGLEineq}, \refeq{vdWaNORMsup}, and \refeq{YukaNORMsup}, we thus have
\beq
\alpha\norm{V*1}{C^0_b(\Lbd)} 
\leq    A_{\W}\frac{\pi^2}{ 4\varkappa^3} +  
	A_{\Y}\frac{4\pi}{\kappa^2}
= \norm{V}{1},
\label{Phiineq}
\eeq
valid for all $\Lbd \subset \RR^3$.
	Hence, if \refeq{vdWaalsFIXPTEQ} has a solution $\ol\eta_{\vdW} \leq \ol\eta_\fs^{\ssst\,<}$, 
then by \refeq{Phiineq} this $\ol\eta_{\vdW}$ is a supersolution for \refeq{FIXetaPHIeq}, and 
Proposition 3.1 now concludes the proof. 
\QED  

	We turn to the necessary conditions for the existence of fluid solutions. 

\medskip\noindent
{\bf Proposition 3.5: } {\it If the inequality 
\beq
\gamma - \gamma_\fs + \alpha \Phi_{_\Lbd} \wp^\pr_\bullet(\gamma) \geq  0\, ,
\label{Cestim}
\eeq
holds, then the extended \refeq{fixptEQinLambda} does not have a hard-sphere fluid solution.}
\medskip

{\it Proof of Proposition 3.5:}
	Since $V< 0$ and $\alpha >0$, and since $\wp^\pr_\bullet(\gamma)>0$, it follows 
directly from \refeq{fixptEQinLambda} that any solution $\eta$ of the extended \refeq{fixptEQinLambda} 
satisfies the lower estimate
\beq
	\eta(\bfr) > \wp^\pr_\bullet\big(\gamma\big)
\label{etaLOWbound}
\eeq
for all $\bfr\in \Lbd$.
	Convoluting \refeq{etaLOWbound} with $-V$ ($>0$) and multiplying by $\alpha$ yields
\beq
- (\aV*\eta)(\bfr) 
> 
- (\aV*1)(\bfr)  \wp^\pr_\bullet(\gamma) 
\label{gbLOWbound}
\eeq
for all $\bfr\in \Lbd$, from which it follows that
\beq
\gamma +  \|\aV*\eta\|_{C^0_b(\Lbd)} 
> 
\gamma + \alpha \Phi_{_\Lbd} \wp^\pr_\bullet(\gamma) .
\label{gbLOWboundsup}
\eeq
	If \refeq{Cestim} holds, then from  \refeq{gbLOWboundsup} it follows that 
$\gamma +  \|\aV*\eta\|_{C^0_b(\Lbd)} >\gamma_\fs$, and so
$\|\eta\|_{C^0_b(\Lbd)} >\ol\eta_\fs^{\ssst\,<}$.
	Therefore, violation of \refeq{Cestim} is a necessary condition for the existence of an all 
fluid solution of \refeq{fixptEQinLambda}.
\QED

	We conclude this section with an obvious non-existence result.

\smallskip\noindent
{\bf Proposition 2.6: } {\it If the inequality 
\beq
\gamma - \gamma_\fs  >  0\, ,
\label{Destim}
\eeq
holds, then the extended \refeq{fixptEQinLambda} does not have a solution which is fluid somewhere in $\Lbd$.}
\smallskip

{\it Proof of Proposition 2.6:}
Trivial.
\QED
\newpage

\noindent
\section{A $(\ag)$ REGION WITH UNIQUE FLUID SOLUTIONS}
\smallskip

	We now locate  a connected region in $(\ag)$ space in which there exists 
a unique fluid solution for each pair of $(\ag)$ parameter values.
	The pertinent unique fluid solution need not be the unique solution per se,
yet any other solution of \refeq{fixptEQinLambda} would necessarily take nonfluid values somewhere in $\Lbd$. 

	Our existence and uniqueness results are based on the following theorem, for which 
much less is assumed about $\wp_\bullet^{\phantom{b}}(\gamma)$ than stipulated in Convention 2.3. 

\medskip\noindent
{\bf Lemma 4.1} {\it Consider \refeq{fixptEQinLambda} for a map $\gamma\mapsto \wp_\bullet(\gamma)$ of class 
$C^2(-\infty, {\tilde\gamma})$ which is strictly positive, increasing, and convex, and for which
\beq
	K(\tilde\gamma) 
:=
	\sup_{\gamma\in (-\infty,{\tilde\gamma})}\wp^\ppr_\bullet(\gamma) 
< \infty\, .
\label{Mdefine}
\eeq 
	Assume $\gamma\ (< {\tilde\gamma})$ and $\alpha (>0)$ are  such that the operator 
$\wp^\pr_\bullet(\gamma -\alpha V*\, \cdot\, )$ maps $C^0_{b,+} \cap \ol{\cal B}_{{\tilde\eta}}$ 
into itself, where 
$\ol{\cal B}_{{\tilde\eta}} = \{\eta :\, \norm{\eta}{C^0_b(\Lbd)} \leq {\tilde\eta}\}$ 
is a ball of radius ${\tilde\eta} = \wp^\pr_\bullet({\tilde\gamma})$.  
	Assume furthermore that 
\beq
K(\tilde\gamma) \alpha \Phi_{_\Lbd} < 1 \, ,	
\label{Mbound}
\eeq 
with $\Phi_{_\Lbd} = \norm{V*1}{C^0_b(\Lbd)}$, as defined above Proposition 3.1.
	Then there exists a unique solution $\in C^0_{b,+}\cap\ol{\cal B}_{{\tilde\eta}}$ 
of \refeq{fixptEQinLambda}.
	In particular, the iteration sequence \refeq{iteraRHO}, starting with any 
$\eta^{(0)} \in C^0_{b,+}\cap \ol{\cal B}_{{\tilde\eta}}$, converges strongly in $C^0_b(\ol\Lbd)$ 	
to the unique solution.
}

\medskip
\noindent
{\bf Remark:} Lemma 4.1 improves over Theorem 6.4 of Ref.\cite{mkjkp}, where uniqueness and strong ${L^1}$ 
convergence are established under the same condition \refeq{Mbound}.
\eQED

\medskip
{\it Proof of Lemma 4.1:} By hypothesis, the operator 
$\wp^\pr_\bullet(\gamma - \alpha V*\, \cdot\, )$ maps the $\|\, .\, \|_{C^0_b(\Lbd)}$ closed set 
$C^0_{b,+}\cap \ol{\cal B}_{{\tilde\eta}}$ into itself. 
	This implies that $\gamma - \alpha V*\eta \leq \tilde\gamma$ for any
$\eta\in C^0_{b,+}\cap \ol{\cal B}_{{\tilde\eta}}$. 
	This together with \refeq{Mdefine} in turn implies that 
$\wp^\ppr_\bullet(\gamma - \alpha V*\eta) \leq K(\tilde\gamma)$ for any 
$\eta\in C^0_{b,+}\cap \ol{\cal B}_{{\tilde\eta}}$. 

	Consider now two sequences $\{\eta_i^{(n)}\in C^0_{b,+}\cap\ol{\cal B}_{\tilde\eta}\}_{n =0}^\infty$,
$i = 1,2$, defined by \refeq{iteraRHO}, with $\eta_1^{(0)} \neq \eta_2^{(0)}$ on a fat set. 
	Set $-(V*\eta_i^{(n)})(\bfr) = \phi_i^{(n)}(\bfr)$. 
	Pick any $1 < q <\infty$. 
	Then, by the fact that $\wp^\ppr_\bullet(\gamma - \alpha V*\eta) \leq K(\tilde\gamma)$
for any $\eta\in C^0_{b,+}\cap \ol{\cal B}_{{\tilde\eta}}$,  we estimate
\bea 
	\norm{\eta_2^{(n+1)} -\eta_1^{(n+1)}}{L^q(\Lbd)}^{q}
\!\!\!&=& 
	\int_{\!\Lbd} \left|\wp^\pr_\bullet\left(\gamma + \alpha\phi_2^{(n)}(\bfr)\right)
	- \wp^\pr_\bullet\left(\gamma + \alpha \phi_1^{(n)}(\bfr)\right)\right|^q d^3r 
\nonumber\\
\!\!\!&=&
	\int_{\!\Lbd} \Bigl| \int_{\phi_1^{(n)}(\bfr)}^{\phi_2^{(n)}(\bfr)} 
	\alpha \wp^\ppr_\bullet(\gamma + \alpha \varphi)\, d\varphi\Bigr|^q d^3r 
\nonumber\\
\!\!\!&\leq& 
	K^q \alpha^q \int_{\!\Lbd}  \Bigl|
	\int_{\phi_1^{(n)}(\bfr)}^{\phi_2^{(n)}(\bfr)}\!\! d\varphi\Bigr|^q d^3r 
\nonumber\\
\!\!\!&=& 
	K^q\alpha^q \norm{\phi_2^{(n)} - \phi_1^{(n)}}{L^q(\Lbd)}^{q} 
\label{chainA}
\eea
	By the definition of $\phi_i^{(n)}$, followed by an obvious estimate 
and then by an application of H\"older's inequality with conjugate exponents $q, q^\pr$, we estimate
\bea
	\int_{\!\Lbd} \left|\phi_2^{(n)}(\bfr) - \phi_1^{(n)}(\bfr) \right|^q d^3r 
\!\!\!&=& 
	\int_{\!\Lbd} \left|\int_{\!\Lbd} 
   -V(|\bfr-\tilde\bfr|)\left(\eta_2^{(n)} -\eta_1^{(n)} \right)(\tilde\bfr)\, d^3\tilde{r} \right|^q d^3r 
\nonumber\\
\!\!\!&\leq& 
	\int_{\!\Lbd} \left(\int_{\!\Lbd} 
     -V(|\bfr-\tilde\bfr|)\Bigl|\eta_2^{(n)} -\eta_1^{(n)}\Bigr|(\tilde\bfr)\, d^3\tilde{r}\right)^q d^3r 
\nonumber\\
\!\!\!&\leq&  
	\norm{(-V)^{q^\pr} * 1}{L^{q/q^\pr}(\Lbd)}^{q/{q^\pr}} 
	\norm{\eta_2^{(n)} -\eta_1^{(n)}}{L^q(\Lbd)}^{q}
\label{chainB}
\eea
	Combining \refeq{chainA} and \refeq{chainB} gives, after taking the $q$th root,
\beq
	\norm{\eta_2^{(n+1)} -\eta_1^{(n+1)}}{L^q(\Lbd)} 
\leq  
	K(\tilde\gamma) \alpha \norm{(-V)^{q^\pr} * 1}{L^{q/q^\pr}(\Lbd)}^{1/{q^\pr}}
	\norm{\eta_2^{(n)} -\eta_1^{(n)}}{L^{q}(\Lbd)}
\label{chainC}
\eeq
for all $q\in(1,\infty)$. 
	By taking $q\to\infty$, and noting that here $\esssup=\sup$, we get 
\beq
	\norm{\eta_1^{(n+1)} - \eta_2^{(n+1)}}{C^0_b(\Lbd)} 
\leq 
K(\tilde\gamma)
\norm{\aV*1}{C^0_b(\Lbd)}\norm{\eta_1^{(n)}-\eta_2^{(n)}}{C^0_b(\Lbd)}
\label{chainD}
\eeq

	By hypothesis \refeq{Mbound}, we have $K(\tilde\gamma)\alpha \norm{V* 1}{C^0_b(\Lbd)} < 1$, 
whence from \refeq{chainD} we conclude that the map $\eta \mapsto \wp^\pr_{\bullet}(\gamma - \alpha V*\eta)$ is a
$C^0_b$ contraction map in the closed truncated cone $C^0_{b,+}\cap \ol{\cal B}_{{\tilde\eta}}$. 
	We now  apply the contraction mapping principle$^{\cite{cbdwdb,nussbaum}}$
and conclude  that a unique fixed point of $\eta \mapsto \wp^\pr_{\bullet}(\gamma - \alpha V*\eta)$ exists 
in $C^0_{b,+}\cap \ol{\cal B}_{{\tilde\eta}}$.
	In addition, the proof of the contraction mapping principle implies the $C^0_b$ convergence of the 
iteration sequence \refeq{iteraRHO} for any initial density $\eta^{(0)}\in C^0_{b,+}\cap \ol{\cal B}_{{\tilde\eta}}$.~\QED
\medskip

	We now return to our $\wp_\bullet(\gamma)$ given by Convention 2.3.
	In our first application of Lemma 4.1 we set $\tilde\gamma =\gamma_\fs\ (\approx 15.208)$. 
	The following input from Ref.\cite{mkjkp} 
capitalizes on the fact that the graph of $\ol\eta\mapsto g_{_2}(\ol\eta)$ has a 
unique inflection point at $\ol\eta=\ol\eta_{\wr}\approx 0.130$.
\medskip
\noindent
{\bf Lemma 4.2:} {\it 
	The regular global maximum $K(\gamma_\fs)$ of $\wp^\ppr_{\CS}(\gamma)$ over the
set $(-\infty, \gamma_\fs)$  occurs at $\gamma_{\wr} \approx - 0.67$ at which 
$\ol\eta_{\wr} \equiv g_{_2}^{-1}\left(\gamma_{\wr} \right) \approx 0.130$
and $\wp^\ppr_{\CS}(\gamma_{\wr})=K(\gamma_\fs) \approx 0.047$.}
\medskip

	We are now in the position to state the following Corollary of Lemma 4.1. 

\medskip
\noindent
{\bf Corollary 4.3:} {\it 
	Let the parameters $(\ag)$ satisfy the bound \refeq{Aestim}, and let 
$\alpha$ satisfy the inequality $\norm{\aV*1}{C^0_b(\Lbd)} <21.20$.
	Then there exists a unique fluid solution of \refeq{fixptEQinLambda}.}
\medskip

{\it Proof:} By hypothesis, the parameters $(\ag)$ satisfy \refeq{Aestim}. 
	This implies that the operator $\wp^\pr_{\CS}(\gamma - \alpha V*\, \cdot\, )$ maps 
$C^0_{b,+} \cap \ol{\cal B}_{\ol\eta_\fs^{\ssst\,<}}$ into itself. 
	Next, using Lemma 4.2 and $1/0.047\approx 21.20$, we conclude that 
$\norm{\aV*1}{C^0_b(\Lbd)} <21.20$ implies \refeq{Mbound}. 
	Lemma 4.1 now guarantees us a unique solution
$\in C^0_{b,+} \cap \ol{\cal B}_{\ol\eta_\fs^{\ssst\,<}}$ of \refeq{fixptEQinLambda}. 
\QED
\medskip
\noindent

	It is interesting to compare \refeq{Mbound} to the {\it sharp} criterion for uniqueness, 
irrespective of $\gamma$,  of a solution $\eta <\ol\eta_\fs^{\ssst\,<}$ to the associated 
algebraic fixed point problem \refeq{FIXetaPHIeq}.
	Geometrically, this criterion for uniqueness is that the slope of the straight line 
$\ol\eta \mapsto \gamma  + \alpha \Phi_{_\Lbd} \ol\eta$ may not surpass the smallest derivative of 
the curve $\ol\eta \mapsto g_{_2}(\ol\eta)$, or $\alpha\Phi_{_\Lbd} \leq g_{_2}^\pr(\ol\eta_{\wr})$, 
with $\ol\eta_{\wr} \approx 0.130$ defined in Lemma 4.2. 
	From the definition of $\wp_{\CS}(\gamma)$ we then see that this criterion is precisely 
$K(\gamma_\fs)\alpha \Phi_{_\Lbd} \leq 1$, with 
$K(\gamma_\fs) =\wp^\ppr_{\CS}(\gamma_{\wr})\ (\approx 0.047)$ 
given in Lemma 4.2.
	Thus, \refeq{Mbound} is the direct analog of the geometric criterion that governs the associated 
algebraic fixed point problem \refeq{FIXetaPHIeq}, except for the case of equality 
$K(\gamma_\fs)\alpha \Phi_{_\Lbd} = 1$, about which the contraction mapping principle is silent.

	If $K(\gamma_\fs)\alpha \Phi_{_\Lbd} > 1$, then there exist values of $\gamma$ for
which  \refeq{FIXetaPHIeq} has either two or three solutions. 
	In that case we can still arrive at a uniqueness theorem for \refeq{FIXetaPHIeq} under the condition on 
$\gamma$ that it be not too large. 
	Similarly, if \refeq{Mbound} is violated, Lemma 4.1 still gives a uniqueness result 
for \refeq{fixptEQinLambda} by appropriately restricting $\gamma$ from above. 
	For this second application of our Lemma 4.1 we introduce the following.

\smallskip\noindent
{\bf Definition 4.4:} {\it Given $\Lbd$, for each $\alpha$ we define $\gamma_{\!\ssst\Lbd}(\alpha)$ to be the
largest upper bound on $\gamma$ such that for each $\gamma<\gamma_{\!\ssst\Lbd}(\alpha)$ 
there exists a unique positive solution $\ol\eta(\ag)$ of \refeq{FIXetaPHIeq}.}
\smallskip

\noindent
{\bf Remarks:}
(a) 
Since $g_{_2}^\prime(\eta)>0$ and $g_{_2}((0,\ol\eta_\fs^{\ssst\,<}])=(-\infty,\gamma_\fs]$, 
clearly $\gamma_{\!\ssst\Lbd} > - \infty$; 
(b) $\gamma_{\!\ssst\Lbd}(\alpha)$ has a discontinuity when
$ \alpha\Phi_{_\Lbd}K(\gamma_\fs)=1$. 
\eQED

\noindent
{\bf Corollary 4.5:} {\it 
	Let $\alpha$ satisfy $K(\gamma_\fs)\alpha \Phi_{_\Lbd} > 1$, and 
let $\gamma < \gamma_{\!\ssst\Lbd}(\alpha)$. 
	Then $\ol\eta(\ag) < \ol\eta_{\wr}$, and \refeq{fixptEQinLambda} has a unique fluid solution 
$\eta_{\ssst\Lbd}\in C^0_{b,+} \cap \ol{\cal B}_{\ol\eta_\fs^{\ssst\,<}}$; in fact, 
$\eta_{\ssst\Lbd}\in C^0_{b,+} \cap \ol{\cal B}_{\ol\eta(\ag)}$.
	Moreover, the iteration sequence defined by 
$\eta^{(n+1)} = \wp^\pr_{\CS}\big(\gamma - \alpha V*\eta^{(n)}\big)$, starting with any 
$\eta^{(0)}\in C^0_{b,+} \cap\ol{\cal B}_{\ol\eta_\fs^{\ssst\,<}}$, converges in supnorm 
to this unique fixed point.
}

\medskip
{\it Proof:} 
	Since $K(\gamma_\fs)\alpha \Phi_{_\Lbd} > 1$, by definition of $\gamma_{\!\ssst\Lbd}$ we see that 
$\ol\eta(\ag) < \ol\eta_{\wr}$.
	Therefore all $\ol\eta\in [\ol\eta_\alpha^\gamma,\ol\eta_\fs^{\ssst\,<}]$ are supersolutions
for \refeq{FIXetaPHIeq}, and thus strict supersolutions for \refeq{fixptEQinLambda}.
	By the type of argument presented in the proof of Proposition 3.2 we conclude that no fluid 
solution of \refeq{fixptEQinLambda} exists which is somewhere larger than $\ol\eta(\ag)$.

	Now pick any $\eta \in C^0_{b,+} \cap \ol{\cal B}_{\ol\eta(\ag) }$. 
	Since $\gamma < \gamma_{\!\ssst\Lbd}(\alpha)$, $(g_{_2}^{-1})^\pr(\gamma)>0$, $V<0$, we have
\bea
	\norm{ g_{_2}^{-1}\big(\gamma - \alpha V*\eta\big) }{C^0_b(\Lbd)} 
\!\!\!&\leq&
  g_{_2}^{-1}\Big(\gamma+\alpha\Phi_{_\Lbd}\norm{\eta }{C^0_b(\Lbd)}\Big)
\nonumber\\
\!\!\!&\leq&  
	g_{_2}^{-1}\big(\gamma + \alpha \Phi_{_\Lbd} \ol\eta(\ag)\big) 
 = 
	\ol\eta(\ag)\, .
\label{zweiINEQ}
\eea
	Therefore, the operator $g_{_2}^{-1}\big(\gamma - \alpha V*\, \cdot\, \big)$ maps 
$C^0_{b,+} \cap \ol{\cal B}_{\ol\eta(\ag) }$ into itself. 

	We observe that ${g_{_2}}^\pr(\ol\eta(\ag)) > \alpha \Phi_{_\Lbd}$ so that 
$\ol\eta(\ag)$ is a stable fixed point of \refeq{FIXetaPHIeq}.	 
	The stability of $\ol\eta(\ag)$ and the convexity of $g_{_2}^{-1}(\nu)$ for 
$\nu< \gamma + \alpha \Phi_{_\Lbd}\ol\eta(\ag)$ implies that 
\beq
K(\gamma_{\!\ssst\Lbd}) \alpha \Phi_{_\Lbd} < 1\, ,
\label{Mestimate }
\eeq
where
\beq
	K(\gamma_{\!\ssst\Lbd})
:= 
	\sup_{\gamma, \nu} \Bigl\{(g_{_2}^{-1})^\prime(\gamma +\nu ):\
	\gamma\in \big(-\infty,\gamma_{\!\ssst\Lbd}(\alpha)\big) \wedge\  \nu 
	\leq \alpha \Phi_{_\Lbd} \ol\eta(\ag)\Bigr\}
\label{Mdefinition}
\eeq
	We now can apply Lemma 4.1 to $\eta\in C^0_{b,+} \cap \ol{\cal B}_{\ol\eta(\ag) }$.
	The proof is complete.\QED

\noindent
\section{A $(\ag)$ REGION WITH SEVERAL FLUID SOLUTIONS}
\smallskip
	When  $V\in L^1(\RR^3)$, it is readily shown that there is a connected region in $(\ag)$ parameter 
space in which the van der Waals' algebraic fixed point equation \refeq{fixptEQvdW} for constant density functions 
in $\RR^3$ has three solutions inside the fluid regime, 
$\ol\eta_{\vdW}^{\ssst m} <\ol\eta_{\vdW}^{u} <\ol\eta_{\vdW}^{\ssst M} \leq \ol\eta_\fs^{\ssst\,<}$,
so these solutions satisfy \refeq{vdWaalsFIXPTEQ}.
	The smallest and the largest ones are stable under iterations while the intermediate one is 
unstable.
	Intuitively one expects that when $\ol\Lbd\subset\RR^3$ is a container of macroscopic 
proportions, and $\kappa^{-1}$ and $\varkappa^{-1}$ are molecular distances, then for most $(\ag)$ 
in the three fluid solutions region for the algebraic \refeq{vdWaalsFIXPTEQ} our nonlinear integral equation 
\refeq{fixptEQinLambda} should also have a small and a large fluid solution which are stable under iterations, 
while the unstable solution $\ol\eta_{\vdW}^{u}$ of \refeq{vdWaalsFIXPTEQ} should be replaced by an interface 
type solution of \refeq{fixptEQinLambda} which is unstable under iterations.
	Numerical integrations of \refeq{fixptEQinLambda} with $V=V_{\W}$ for a ball domain $\Lbd=B_R$ with 
moderately large $R= 50/\varkappa$ support this expectation.$^{\cite{mkjkp}}$
	A rigorous proof is desirable.

	In this section we use monotone iteration techniques with sub- and supersolutions to
show that at least part of this multiplicity region for the algebraic equation \refeq{vdWaalsFIXPTEQ} 
corresponds to a multiplicity region of the integral equation \refeq{fixptEQinLambda} ---
for certain sufficiently {\it small} $\Lbd$.
	We will prove that at least three hard-sphere fluid solutions exist in some region
of $(\ag)$ parameter space, two of them stable under iteration and one unstable.
	We will not show that exactly three fluid solutions exist; in fact, it might not be true 
that exactly three fluid solutions of \refeq{fixptEQinLambda} exist whenever it has at least three 
fluid solutions.

	Recall that the starting function $\eta^{(0)}=g_{_2}^{-1}(\gamma)$ is a subsolution for 
\refeq{fixptEQinLambda} in any $\Lbd$, and it launches an iteration sequence which converges upward
toward the pointwise minimal solution; see Proposition 3.1.
	We also know from Proposition 3.4 that when 
$\ol\eta_{\vdW}^{\ssst M}\leq\ol\eta_\fs^{\ssst\,<}$, then
any starting function $\ol\eta^{(0)}\in [\ol\eta_{\vdW}^{\ssst M},\ol\eta_\fs^{\ssst\,<}]$ is a 
supersolution for \refeq{fixptEQinLambda} in any $\Lbd$, and it launches an iteration sequence 
which converges downward toward the pointwise maximal fluid solution. 
	One can rule out that the pointwise maximal solution coincides with the pointwise minimal solution 
if a {\it sufficiently large subsolution} of \refeq{fixptEQinLambda} in $\Lbd$ is available
from which the iteration $\eta^{(n+1)} = \wp^\pr_{\bullet}\big(\gamma - \alpha V*\eta^{(n)}\big)$
converges upward toward a fluid solution which is larger than the pointwise minimal 
solution to \refeq{fixptEQinLambda}.

	Constructing suitable subsolutions that imply a $(\ag)$ region of multiple 
hard-sphere fluid solutions is a very difficult business, yet much easier for the Carnahan--Starling model.
	We will take advantage of this fact and, until further notice, first discuss \refeq{fixptEQinLambda} 
with $\wp_\bullet^{\phantom{b}}(\gamma)$ replaced by $\wp_{\CS}(\gamma)$ for all $\gamma\in\RR$, viz.
\beq
\eta(\bfr)  =  \wp_{\CS}^{\prime}\big(\gamma - (\aV*\eta)_{_{\!\Lbd}}\!(\bfr) \big).
\label{fixptEQinLambdaCS}
\eeq
	Subsequently we seek those solutions which nowhere in $\Lbd$ surpass 
$\ol\eta_\fs^{\ssst\,<}$.
	We emphasize that our multiplicity results for the Carnahan--Starling model  in general
have no bearing on the hard-sphere fluid; however, there will be a small sliver in $(\ag)$ space
for which our Carnahan--Starling multiplicity results yield multiple hard-sphere fluid solutions.

	So recall that $\wp_{\CS}^\pr(\,\cdot\,) = g_{_2}^{-1}(\,\cdot\,)$ and consider the algebraic 
fixed point problem
\beq
\ol\eta = g_{_2}^{-1}\bigl(\gamma +\alpha\tau \ol\eta \bigr)
\label{FIXetaTAUeq}
\eeq
for $\gamma\in\RR$ and $\alpha\in\RR_+$, where $\tau\in\RR_+$.
	Multiplicity of solutions of \refeq{FIXetaTAUeq} can only occur if $\alpha$ is large enough, namely 
(recalling Lemma 4.2 and Corollary 4.3) if
\beq
\alpha\tau > \min_{\ol\eta\in(0,1)}g_{_2}^\prime(\ol\eta)= g_{_2}^\pr(\ol\eta_{\wr})\approx 21.20.
\label{alphaCONDITIONfortriplesTAU}
\eeq
	In addition, $\gamma$ needs to satisfy
$\gamma\in(\check\gamma^{\alg}_{\CS}(\alpha\tau),\hat\gamma^{\alg}_{\CS}(\alpha\tau))$,
with upper and lower interval limits given by
\beq
\hat\gamma^{\alg}_{\CS}(\alpha\tau) = g_{_2}(\ol{\eta}_{\ssst\, <})-\alpha\tau\ol{\eta}_{\ssst\, <},
\label{gammaRIGHT}
\eeq
\beq
\check\gamma^{\alg}_{\CS}(\alpha\tau) = g_{_2}(\ol{\eta}_{\ssst\, >})-\alpha\tau\ol{\eta}_{\ssst\, >},
\label{gammaLEFT}
\eeq
where $\ol{\eta}_{\ssst\, <}<\ol{\eta}_{\ssst\, >}$ are the two {\it distinct} solutions to the equation
\beq
\alpha\tau = g_{_2}^\prime(\ol\eta),
\label{etaGROSSkleinEQ}
\eeq
which exist only when \refeq{alphaCONDITIONfortriplesTAU} is satisfied, in which case 
$\ol{\eta}_{\ssst\, <}<\ol\eta_{\wr}$
is a decreasing, and $\ol{\eta}_{\ssst\, >}>\ol\eta_{\wr}$ an increasing function of $\alpha\tau$.
	While it does not seem feasible to write down  closed form expressions of
the functions $\alpha\tau\mapsto \ol{\eta}_{\ssst\, <}$ and $\alpha\tau\mapsto \ol{\eta}_{\ssst\, >}$, 
their asymptotics for $\alpha\tau\approx g_{_2}^\pr(\ol\eta_{\wr})$ (recall \refeq{alphaCONDITIONfortriplesTAU}) 
and $\alpha\tau\gg g_{_2}^\pr(\ol\eta_{\wr})$ can easily be worked out, which gives us
\bea
\!\!\!\!\!\!&&
\hat\gamma^{\alg}_{\CS}(\alpha\tau) \asymp \left\{\begin{array}{rc}
\phantom{niixxxxxxxxxxxxxxxxxxxx}
g_{_2}(\ol\eta_{\wr}) - \ol\eta_{\wr}\alpha\tau; &\alpha\tau\approx  g_{_2}^\pr(\ol\eta_{\wr}) \\
           -\ln(\alpha\tau) -1 +O[1/\alpha\tau]; &\alpha\tau\gg  g_{_2}^\pr(\ol\eta_{\wr}) 
\end{array}\right.
\label{hatGAMMA}\\
\!\!\!\!\!\!&&
\check\gamma^{\alg}_{\CS}(\alpha\tau) \asymp \left\{\begin{array}{rc}
   g_{_2}(\ol\eta_{\wr}) - g_{_1}^\pr(\ol\eta_{\wr}) - 
	2^{\sst\frac{3}{2}}{\tst\frac{g_{_2}^\pr(\ol\eta_{\wr})}{ g_{_2}^\pppr(\ol\eta_{\wr})^{1/2}}}
	[\alpha\tau- g_{_2}^\pr(\ol\eta_{\wr})]^{1/2};
&\alpha\tau\approx g_{_2}^\pr(\ol\eta_{\wr}) \\
 -{\tst\frac{2}{3}}\alpha\tau + O([\alpha\tau]^{3/4}) ;
&\alpha\tau\gg  g_{_2}^\pr(\ol\eta_{\wr}) 
\end{array}\right.
\label{checkGAMMA}
\eea
where we used the identity $g_{_1}^\pr(\ol\eta_{\wr})= \ol\eta_{\wr} g_{_2}^\pr(\ol\eta_{\wr})$
to simplify.
	Numerically, $g_{_2}^\pppr(\ol\eta_{\wr})\approx 1235.22$.

	So the algebraic fixed point equation \refeq{FIXetaTAUeq} has three solutions for all 
$(\ag)\in \Theta^{\alg}_{\CS}(\tau)$, where
$\Theta^{\alg}_{\CS}(\tau)\equiv \{(\ag): \alpha\tau > g_{_2}^\pr(\ol\eta_{\wr})\wedge
		\check\gamma^{\alg}_{\CS}(\alpha\tau)<\gamma<\hat\gamma^{\alg}_{\CS}(\alpha\tau)\}$.
	Note that the boundary $\partial\Theta^{\alg}_{\CS}(\tau)$ is given by two functions of $\alpha$ 
which depend on $\alpha$ only through the product $\alpha\tau$. 
	Hence, for \refeq{FIXetaTAUeq}, triple solution regions in the $(\ag)$ half plane for any 
two different $\tau=\tau_1$ and $\tau=\tau_2$ differ from each other only by some scaling along the
$\alpha$ axis, viz. they are affine similar.
	Since for fixed $\tau$ the upper boundary curve $\hat\gamma^{\alg}_{\CS}(\alpha\tau)$ diverges
to $-\infty$ logarithmically while the lower boundary curve $\check\gamma^{\alg}_{\CS}(\alpha\tau)$ does
so linearly when $\alpha$ becomes large, it follows that any pair of triple domains 
$\Theta^{\alg}_{\CS}(\tau_1)$ and $\Theta^{\alg}_{\CS}(\tau_2)$ has a non-empty intersection.

	We next identify functionals of $V$ which can be substituted for $\tau$ to construct 
sub- and supersolutions for \refeq{fixptEQinLambdaCS}.
	Since both our van der Waals kernel $V_{\W}(|\bfr|)$ and the Yukawa kernel $V_{\Y}(|\bfr|)$ 
are monotonic increasing negative functions of $|\bfr|=r$, for $\aV= A_{\W}V_{\W}+ A_{\Y}V_{\Y}$ and 
any container $\ol\Lbd$ with diameter $\diam(\Lbd)$, we have
\beq
V(|\bfr-\bfr^\pr|) \leq V(\diam(\Lbd))\quad\forall\ \bfr,\bfr^\pr \in \Lbd.
\label{VupperEST}
\eeq
	We define the abbreviation 
\beq
\Psi_{_\Lbd} 
:= - V(\diam(\Lbd))|\Lbd|.
\label{VupperESTmodLAMBDA}
\eeq
	Subsolutions for \refeq{fixptEQinLambdaCS} can be constructed by setting $\tau=\Psi_{_\Lbd}$,
supersolutions by setting $\tau=\Phi_{_\Lbd}$ or $\tau=\norm{V}{1}$.
	Note that for bounded $\Lbd\subset\RR^3$ we have the chain of inequalities
\beq
\Psi_{_\Lbd}< \Phi_{_\Lbd}<\Phi_{_{\RR^3}}=\norm{V}{1}.
\label{PsiPhiCHAINa}
\eeq 
	Since our findings about the triple algebraic solutions domain for \refeq{FIXetaTAUeq} imply 
in particular that for any bounded domain $\Lbd\subset\RR^3$ we have
$\Theta^{\alg}_{\CS}(\Phi_{_\Lbd})\cap\Theta^{\alg}_{\CS}(\Psi_{_\Lbd})\neq\emptyset$ 
and also 
$\Theta^{\alg}_{\CS}(\norm{V}{1})\cap\Theta^{\alg}_{\CS}(\Psi_{_\Lbd})\neq\emptyset$,
one can now show, with the help of monotone iterations and the mountain pass lemma, that for each 
$(\ag)\in \Theta^{\alg}_{\CS}(\Phi_{_\Lbd})\cap\Theta^{\alg}_{\CS}(\Psi_{_\Lbd})$ 
and each 
$(\ag)\in \Theta^{\alg}_{\CS}(\norm{V}{1})\cap\Theta^{\alg}_{\CS}(\Psi_{_\Lbd})$ the 
fixed point equation \refeq{fixptEQinLambdaCS} 
has at least three solutions in $C^0_b(\Lbd)$, which are ordered. 
	However, for physically interesting domains $\Lbd$ the sets
$\Theta^{\alg}_{\CS}(\Phi_{_\Lbd})\cap\Theta^{\alg}_{\CS}(\Psi_{_\Lbd})$ 
and
$\Theta^{\alg}_{\CS}(\norm{V}{1})\cap\Theta^{\alg}_{\CS}(\Psi_{_\Lbd})$ are generally 
very bad approximations to the full set of such $(\ag)$ points.
	The reason is that for physically interesting, i.e. macroscopic domains $\Lbd$, the ratio
$\Psi_{_\Lbd}/\Phi_{_\Lbd}$ is tiny, converging to zero as $\Lbd\uparrow\RR^3$.
	Worse, $\Theta^{\alg}_{\CS}(\Phi_{_\Lbd})\cap\Theta^{\alg}_{\CS}(\Psi_{_\Lbd})$ may even 
be a totally useless estimate of the three hard-sphere fluid solutions regime of \refeq{fixptEQinLambda},
in the sense that the largest solution of \refeq{fixptEQinLambdaCS} obtained by this method may {\it always}
take values outside the physical range of hard-sphere fluid densities. 
	
	The following variation on our strategy yields more desirable multiplicity results.
	For bounded $\Lbd\subset\RR^3$, let $\varsigma\Lbd\subset\Lambda$ denote a rescaling of 
$\Lbd$ into $\Lbd$ by a factor $\varsigma\leq 1$, so that 
$\diam(\varsigma\Lbd)=\varsigma\,\diam(\Lbd)$ and $|\varsigma\Lbd|=\varsigma^3|\Lbd|$.
	Then for $\aV= A_{\W}V_{\W}+ A_{\Y}V_{\Y}$ the map 
$\varsigma\mapsto \Psi_{_{\varsigma\Lbd}} = - \varsigma^3V(\varsigma\,\diam(\Lbd))|\Lbd|$ takes 
a global maximum at $\varsigma=\grave{\varsigma}$ (which might not be unique; it is unique when 
$V=V_{\W}$ or $V=V_{\Y}$).
	We always mean the largest $\grave{\varsigma}$.
	Suppose now that $\Lbd$ is a container domain of macroscopic proportions, 
and that $\kappa^{-1}$ and $\varkappa^{-1}$ are molecular distances.
	Then $\grave{\varsigma}\ll 1$, and $\forall\,\varsigma>\grave{\varsigma}$  we have the ordering
\beq
\Psi_{_\Lbd}\ll \Psi_{_{\grave{\varsigma}\Lbd}}<
\Phi_{_{\grave{\varsigma}\Lbd}}< \Phi_{_{\varsigma\Lbd}}
<\Phi_{_{\RR^3}}=\norm{V}{1}.
\label{PsiPhiCHAINb}
\eeq 
	For spherical macroscopic $\Lbd$ (see Appendix A), and with 
$\kappa=\varkappa=1/2$, we have $\Phi_{_\Lbd} \approx 50 \Psi_{_{\grave{\varsigma}\Lbd}}$, while 
the first inequality separates two quantities ``a universe apart.''

\smallskip
\noindent
{\bf Proposition 5.1: } {\it 
	Let $\aV= A_{\W}V_{\W}+ A_{\Y}V_{\Y}$.
	Let $\ol\Lbd\subset\RR^3$ be a container for which $\grave{\varsigma}\ol\Lbd\subset\ol\Lbd$.
	Then for each $\varsigma\in[\grave{\varsigma},1]$ and
$(\ag)\in
		\Theta^{\alg}_{\CS}(\Phi_{_{\varsigma\Lbd}})\cap
		\Theta^{\alg}_{\CS}(\Psi_{_{\grave{\varsigma}\Lbd}})$ 
the equation 
\beq
\eta(\bfr)  = \wp_{\CS}^{\prime}\big(\gamma - (\aV*\eta)_{_{\!\varsigma\Lbd}}(\bfr)\big)
\label{fixptEQinLambdaCSvarsigma}
\eeq
has at least three distinct solutions in $C^0_b(\varsigma\ol\Lbd)$.
	In particular, \refeq{fixptEQinLambdaCSvarsigma} has a pointwise minimal and a pointwise maximal 
solution, both of which are stable under iteration, and a third, unstable solution which 
is sandwiched inbetween.}
\smallskip

{\it Proof:}
	For each 
$(\ag)
	\in\Theta^{\alg}_{\CS}(\Phi_{_{\varsigma\Lbd}})\cap
	   \Theta^{\alg}_{\CS}(\Psi_{_{\grave{\varsigma}\Lbd}})$ 
the algebraic fixed point equation \refeq{FIXetaTAUeq} has three solutions for $\tau=\Phi_{_{\varsigma\Lbd}}$ 
and for $\tau=\Psi_{_{\grave{\varsigma}\Lbd}}$, denoted 
$\ol\eta^{\ssst m}_{\ssst\varsigma\Lbd} 
<\ol\eta^u_{\ssst\varsigma\Lbd}
<\ol\eta^{\ssst M}_{\ssst\varsigma\Lbd}$ 
and 
$\ol\eta^{\ssst m}_{\ssst\grave{\varsigma}\Lbd}
<\ol\eta^u_{\ssst\grave{\varsigma}\Lbd}
<\ol\eta^{\ssst M}_{\ssst\grave{\varsigma}\Lbd}$, 
respectively (suppressing their dependence on $(\ag)$ from being displayed).
	Moreover, since $\Psi_{_{\grave{\varsigma}\Lbd}}<\Phi_{_{\varsigma\Lbd}}$, we have 
$\ol\eta^{\ssst m}_{\ssst\grave{\varsigma}\Lbd}<\ol\eta^{\ssst m}_{\ssst\varsigma\Lbd}$ and 
$\ol\eta^{\ssst M}_{\ssst\grave{\varsigma}\Lbd} <\ol\eta^{\ssst M}_{\ssst\varsigma\Lbd}$; 
the ordering of the unstable solutions 
is $\ol\eta^u_{\ssst\grave{\varsigma}\Lbd}>\ol\eta^u_{\ssst\varsigma\Lbd}$, but this is irrelevant for our arguments.

	Now consider the iteration 
$\eta^{(n+1)} = \wp^\pr_{\CS}\big(\gamma - (\aV*\eta^{(n)})_{_{\!\varsigma\Lbd}}\big)$
in $C^0_{b,+}(\varsigma\ol\Lbd)\cap\ol{\cal B}_1$, with 
\beq
\eta^{(0)}_\mu(\bfr) 
= 
\wp^\pr_{\CS}
\Big(\gamma -\ol\eta^\mu_{\ssst\grave{\varsigma}\Lbd}\int_{\grave{\varsigma}\Lbd} \aV(|\bfr-\tilde\bfr|) 
d^3\tilde{r}\Big)\ \forall\ \bfr\in \varsigma\ol\Lbd
\label{etaNULLmu}
\eeq
and either $\mu=m$ or $M$.
	It is easily verified that $\eta^{(0)}_\mu(\bfr)$ is a subsolution of \refeq{fixptEQinLambdaCSvarsigma}.
	Since $\wp^\ppr_{\CS}(\,\cdot\,)>0$ and $\wp^\pr_{\CS}(\,\cdot\,)<1$, each
$\eta^{(0)}_\mu(\bfr)$ launches a monotonic increasing sequence 
$\{\eta^{(n)}_\mu\}_{n=0}^\infty\in C^0_{b,+}(\ol\Lbd)\cap\ol{\cal B}_1$ 
which converges pointwise to some solution $\eta_\mu^\varsigma(\bfr)$ of \refeq{fixptEQinLambdaCSvarsigma}.

	Moreover, we have $\eta_{\ssst m}^\varsigma(\bfr)<\eta_{\ssst M}^\varsigma(\bfr)$.
	To verify this claim, we note on the one hand that in Proposition 3.4 we already 
showed that the constant function $\bfr\mapsto \ol\eta^{\ssst m}_{\ssst\varsigma\Lbd}$ is a 
supersolution of \refeq{fixptEQinLambdaCSvarsigma} for any $\varsigma\Lbd$ (recall, this follows from 
$-(V*1)_{_{\!\Lbd}}\leq\Phi_{_\Lbd}$ for any $\Lbd$), so that with 
$\ol\eta^{\ssst m}_{\ssst\grave{\varsigma}\Lbd}<\ol\eta^{\ssst m}_{\ssst\varsigma\Lbd}$ 
we find $\eta^{(0)}_m(\bfr)<\ol\eta^{\ssst m}_{\ssst\varsigma\Lbd}$, and now we conclude as in the 
proof of Proposition 3.1 that $\eta_{_m}^\varsigma(\bfr)<\ol\eta^{\ssst m}_{\ssst\varsigma\Lbd}$; 
incidentally, $\ol\eta^{\ssst m}_{\ssst\varsigma\Lbd}<\ol\eta_{\wr}$.
	On the other hand, 
$\eta^{(0)}_{\ssst M}(\bfr) > \ol\eta^{\ssst M}_{\ssst\grave{\varsigma}\Lbd}>
\ol\eta_{\wr}\,\forall\,\bfr\in \grave{\varsigma}\Lbd$,
and since the iteration $\{\eta^{(n)}_{\ssst M}\}_{n=0}^\infty$ is monotone upwards, it follows that
$\eta_{_M}^\varsigma(\bfr)>\eta_{_m}^\varsigma(\bfr)\,\forall\,\bfr\in \grave{\varsigma}\Lbd$. 
	In addition, $\eta^{(0)}_{\ssst M}(\bfr) > \eta^{(0)}_m(\bfr)\,\forall\,\bfr\in \varsigma\Lbd$, 
so the strict monotonic increase of the iterations now guarantees that 
$\eta_{_M}^\varsigma(\bfr)\geq\eta_{_m}^\varsigma(\bfr)\,\forall\,\bfr\in \varsigma\Lbd$, and since  
$\eta_{_M}^\varsigma(\bfr)>\eta_{_m}^\varsigma(\bfr)\,\forall\,\bfr\in \grave{\varsigma}\Lbd$, it even 
follows that $\eta_{_M}^\varsigma(\bfr)>\eta_{_m}^\varsigma(\bfr)\,\forall\,\bfr\in \varsigma\Lbd$.

	Standard results about monotone iterations in ordered Banach spaces show
that $\eta_{_m}^\varsigma(\bfr)$ and $\eta_{_M}^\varsigma(\bfr)$ are stable under iterations, and also 
locally $\cP$ stable; see Proposition 3.1 in Ref.\cite{amannC}.
	The existence of a third, unstable (under iterations and in $\cP$ sense)
solution sandwiched between $\eta_{_m}^\varsigma(\bfr)$ and $\eta_{_M}^\varsigma(\bfr)$ now follows,
via the mountain pass lemma, from the local $\cP$ stability of $\eta_{_m}^\varsigma(\bfr)$ and 
$\eta_{_M}^\varsigma(\bfr)$ and the strong $C^0_b(\ol\Lbd)$ differentiability of the functional 
$\cP^{_\Lbd}_{\ag}[\eta]$.

	Lastly, {\it a forteriori} the unstable solution sandwiched between $\eta_{_m}^\varsigma(\bfr)$ 
and $\eta_{_M}^\varsigma(\bfr)$ is also sandwiched between the pointwise smallest and the pointwise largest 
solutions, $\eta_{\ssst\varsigma\Lbd}^{\ssst m}(\bfr)$ and $\eta_{\ssst\varsigma\Lbd}^{\ssst M}(\bfr)$, 
of \refeq{fixptEQinLambdaCSvarsigma}, obtained by the iteration 
$\eta^{(n+1)} = \wp^\pr_{\CS}\big(\gamma - (\aV*\eta^{(n)})_{_{\!\varsigma\Lbd}}\big)$
from, respectively, $\eta^{(0)}\equiv \wp^\pr_{\CS}\big(\gamma\big)$ and 
any $\eta^{(0)}\equiv \ol\eta^{(0)}> \ol\eta^{\ssst M}_{\ssst\varsigma\Lbd}$; cf., Proposition 3.1 with 
$(0,\ol\eta_\fs^{\ssst\,<}]$ replaced by $(0,\ol\eta^{\ssst M}_{\ssst\varsigma\Lbd}]$
or by $(0,1)$), and which are stable under iterations.$^{\cite{amannB,amannC}}$
\QED

	Our proof of Proposition 5.1 reveals the ordering 
\beq
 \wp^\pr_{\CS}\big(\gamma\big)   < 
\eta_{\ssst\varsigma\Lbd}^{\ssst m}(\bfr) \leq 
	\eta_{_m}^\varsigma(\bfr) < 
\eta_{_M}^\varsigma(\bfr)\leq \eta_{\ssst\varsigma\Lbd}^{\ssst M}(\bfr)<\ol\eta^{\ssst M}_{\ssst\varsigma\Lbd}.
\label{etaMmMmORDER}
\eeq
	Our next proposition shows that the first ``$\leq$'' actually is an identity.

\smallskip
\noindent
{\bf Proposition 5.2:} {\it Under the hypotheses of Proposition 5.1, we have}
\beq
\eta_{\ssst\varsigma\Lbd}^{\ssst m}(\bfr)\equiv \eta_{_m}^\varsigma(\bfr)
\label{etaMMmm}
\eeq

{\it Proof of Proposition 5.2:}
An obvious variation on the proof of Corollary 4.5.
\QED

\noindent
{\bf Remark:}
We surmise that also 
$\eta_{_M}^\varsigma\!(\bfr)\equiv \eta_{\ssst\varsigma\Lbd}^{\ssst M}\!(\bfr)$ but have not been able to
prove it.\eQED

	For a macroscopic container $\ol\Lbd$, Proposition 5.1 tells us in particular 
that the Carnahan--Starling model \refeq{fixptEQinLambdaCSvarsigma} with $\varsigma=1$ has at 
least three ordered solutions when
$(\ag)\in 
\Theta^{\alg}_{\CS}(\Phi_{_\Lbd})\cap\Theta^{\alg}_{\CS}(\Psi_{_{\grave{\varsigma}\Lbd}})$.
	One of these solutions is bounded above by $\ol\eta_{\wr}$, while another one takes (some) 
values larger than $\ol\eta_{\wr}$. 
	For large enough $\alpha$ and negatively large enough $\gamma$ (recall that 
$\Theta^{\alg}_{\CS}(\Phi_{_\Lbd})\cap\Theta^{\alg}_{\CS}(\Psi_{_{\grave{\varsigma}\Lbd}})$ 
is unbounded) this large solution will take values larger than $\ol\eta_\fs^{\ssst\,<}$, 
possibly even larger than $\ol\eta_{\ssst fcc}^{\ssst\,cp}\approx 0.7402$.
	Those solutions do not seem to have an interpretation in terms of hard-sphere systems.

	We now return to our task of finding multiple solutions of
\refeq{fixptEQinLambda} which all take only hard-sphere fluid density values.
	Unfortunately our analytical control is not good enough to find a subset of 
$\Theta^{\alg}_{\CS}(\Phi_{_\Lbd})\cap\Theta^{\alg}_{\CS}(\Psi_{_{\grave{\varsigma}\Lbd}})$ 
which satisfies our wishes, and it's even more hopeless to na\"{\i}vely seek an admissible
subset of $\Theta^{\alg}_{\CS}(\Phi_{_\Lbd})\cap\Theta^{\alg}_{\CS}(\Psi_{_\Lbd})$.
	However, if we shrink the size of $\Lbd$ by choosing a
suitable $\varsigma\in(\grave{\varsigma},1)$, then we can find a subset of 
$\Theta^{\alg}_{\CS}(\Phi_{_{\varsigma\Lbd}})\cap
 \Theta^{\alg}_{\CS}(\Psi_{_{\grave{\varsigma}\Lbd}})$ 
for which at least three solutions of \refeq{fixptEQinLambdaCSvarsigma} 
take values only in the hard-sphere fluid regime, 
i.e. for which $\eta_{\sst \varsigma}^{\ssst M}\leq \ol\eta_\fs^{\ssst\,<}\approx 0.49$.
	So we impose the restriction 
$\ol\eta^{\ssst M}_{\sst \varsigma}\leq \ol\eta_\fs^{\ssst\,<}$ 
on $\Theta^{\alg}_{\CS}(\Phi_{_{\varsigma\Lbd}})\cap
	\Theta^{\alg}_{\CS}(\Psi_{_{\grave{\varsigma}\Lbd}})$ 
and seek admissible $\varsigma$.

	To analyze the effect of this restriction we impose it on $\Theta^{\alg}_{\CS}(\tau)$.
	Let $\Theta^{\alg}_{\bullet\rm \ssst f}(\tau)$ denote the $(\ag)$
domain featuring three solutions of \refeq{FIXetaTAUeq} in the hard-sphere fluid regime.
	Recalling the proof of Proposition 3.2,	it is readily verified that 
$\Theta^{\alg}_{\bullet\rm \ssst f}(\tau)$ is given by 
$\Theta^{\alg}_{\bullet\rm \ssst f}(\tau)
\equiv \{(\ag): g_{_2}^\pr(\ol\eta_{\wr})< \alpha\tau < g_{_2}^\pr(\ol\eta_{\fs}^{\ssst\,<})
\wedge
 \check\gamma^{\alg}_{\bullet\rm \ssst f}(\alpha\tau)<\gamma<\hat\gamma^{\alg}_{\bullet\rm \ssst f}(\alpha\tau)\}$,
where
\bea
\check\gamma^{\alg}_{\bullet\rm \ssst f}(\alpha\tau)
\!\!\!&=& \check\gamma^{\alg}_{\CS}(\alpha\tau)\,,
\label{gammaLEFTbull}\\
\hat\gamma^{\alg}_{\bullet\rm \ssst f}(\alpha\tau)
\!\!\!&=&
\min\{
\hat\gamma^{\alg}_{\CS}(\alpha\tau)\,,\, 
\gamma^{\alg}_\fs(\alpha\tau)\}\,,
\label{gammaRIGHTbull}
\eea
with
\beq
\gamma^{\alg}_\fs(\alpha\tau)
=
\gamma_\fs - \ol\eta_\fs^{\ssst\,<}\alpha \tau\, .
\label{HATgammaFofBETAtau}
\eeq
	We note that the two boundary curves $\hat\gamma^{\alg}_{\bullet\rm \ssst f}(\alpha\tau)$ and
$\check\gamma^{\alg}_{\bullet\rm \ssst f}(\alpha\tau)$ intersect at the endpoints of the allowed $\alpha\tau$
interval, i.e. at $\alpha\tau = g_{_2}^\pr(\ol\eta_{\wr})$ and
$\alpha\tau = g_{_2}^\pr(\ol\eta_\fs^{\ssst\,<})$.
	So also the boundary $\partial\Theta^{\alg}_{\bullet\rm \ssst f}(\tau)$ 
is given by two functions of $\alpha$ which depend on $\alpha$ only through the product $\alpha\tau$, 
and this implies for \refeq{FIXetaTAUeq} that also triple hard-sphere fluid solution regions in the $(\ag)$ 
half plane for any two different $\tau=\tau_1$ and $\tau=\tau_2$ differ from each other only by some 
scaling along the $\alpha$ axis, i.e. once again these triple regions are affine similar.
	{\it However,} distinct from the set $\Theta^{\alg}_{\CS}(\tau)$, the set 
$\Theta^{\alg}_{\bullet\rm \ssst f}(\tau)$ is bounded, and since it is also bounded away from $\alpha\tau=0$,
if $\tau_1$ and $\tau_2$ differ by too much then 
$\Theta^{\alg}_{\bullet\rm \ssst f}(\tau_1)\cap\Theta^{\alg}_{\bullet\rm \ssst f}(\tau_2)=\emptyset$.

	Thus, to carry out our construction of subsolutions presented in the proof of Proposition 5.1 
we need to limit the size of $\varsigma\Lbd$ to make sure that 
$\Theta^{\alg}_{\bullet\rm \ssst f}(\Psi_{_{\grave{\varsigma}\Lbd}})/
 \Theta^{\alg}_{\bullet\rm \ssst f}(\Phi_{_{\varsigma\Lbd}})$ 
is not too small.
	Since $\Lbd$ is supposed to be a macroscopic container domain, this means that 
$\varsigma>\grave{\varsigma}$ has to be chosen sufficiently small.
	Recall that the maximum of $\Psi_{_{\varsigma\Lbd}}$ then occurs for one or more 
$\grave{\varsigma}\ll 1$, and we stipulated that we mean the largest $\grave{\varsigma}$ in 
case $\grave{\varsigma}$ is not unique.
	We can precisely, though only implicitly characterize the range of scaled domains 
$\varsigma\Lbd$ for which our construction of subsolutions presented in the proof of 
Proposition 5.1 can be carried out.
	Namely, the intersection 
$\Theta^{\alg}_{\bullet\rm \ssst f}(\Psi_{_{\grave{\varsigma}\Lbd}})\cap
 \Theta^{\alg}_{\bullet\rm \ssst f}(\Phi_{_{\varsigma\Lbd}})\neq\emptyset$
for all $\varsigma\in[\grave{\varsigma},\acute{\varsigma})$, where $\acute{\varsigma}>\grave{\varsigma}$ 
is the unique $\varsigma$ value for which the lower boundary
of $\Theta^{\alg}_{\bullet\rm \ssst f}(\Psi_{_{\grave{\varsigma}\Lbd}})$ only touches the upper boundary of
$\Theta^{\alg}_{\bullet\rm \ssst f}(\Phi_{_{\varsigma\Lbd}})$ (possibly more than once), determined by
\bea
\check\gamma^{\alg}_{\bullet\rm \ssst f}(\alpha\Psi_{_{\grave{\varsigma}\Lbd}})
\!\!\!&=&\quad \hat\gamma^{\alg}_{\bullet\rm \ssst f}(\alpha\Phi_{_{\varsigma\Lbd}})\,,
\label{gammaBULLs}\\
\partial_\alpha\check\gamma^{\alg}_{\bullet\rm \ssst f}(\alpha\Psi_{_{\grave{\varsigma}\Lbd}})
\!\!\!&=& \partial_\alpha \hat\gamma^{\alg}_{\bullet\rm \ssst f}(\alpha\Phi_{_{\varsigma\Lbd}})\,.
\label{gammaBULLSPRIME}
\eea
	The upshot is:

\smallskip
\noindent
{\bf Proposition 5.3: } {\it Let $\dot{\varsigma}\in[\grave{\varsigma},\acute{\varsigma})$ and
$(\ag)\in
		\Theta^{\alg}_{\bullet\rm \ssst f}(\Psi_{_{\grave{\varsigma}\Lbd}})\cap
		\Theta^{\alg}_{\bullet\rm \ssst f}(\Phi_{_{\dot{\varsigma}\Lbd}})$.
	Then
\beq
\eta(\bfr)  = \wp_{\bullet}^{\prime}\big(\gamma - (\aV*\eta)_{_{\!\dot{\varsigma}\Lbd}}(\bfr)\big)
\label{fixptEQinsLambdaCS}
\eeq
has at least three ordered solutions in 
$C^0_b(\dot{\varsigma}\Lbd)\cap \ol{B}_{\ol\eta_\fs^{\ssst\,<}}$,
two of which can be computed by iterating with r.h.s.\refeq{fixptEQinsLambdaCS}, starting from \refeq{etaNULLmu}
with $\mu=m$ and $\mu=M$, respectively.\hfill}

\noindent
{\bf Remark:}  It is helpful to have a geometric illustration of the situation. 
	Recall that $\Theta^{\alg}_{\bullet\rm \ssst f}(\tau)$ is the bounded domain in $(\ag)$
half space determined by \refeq{gammaLEFTbull}
, \refeq{gammaRIGHTbull}, \refeq{HATgammaFofBETAtau} for which the 
algebraic fixed point equation \refeq{FIXetaTAUeq} has exactly three solutions in the hard-sphere fluid regime.
	For the various $\tau>0$ values associated with $\Lbd$ which we have encountered in this section, 
all the domains $\Theta^{\alg}_{\bullet\rm \ssst f}(\tau)$ are located in the negative $\gamma$ half of 
$(\ag)$ half space.
	They have roughly the shape of a receding moon crescent, being affine similar to each
other by horizontal scaling (along the $\alpha$ axis).
	The domains we have encountered are arranged as follows: 
$\Theta^{\alg}_{\bullet\rm \ssst f}(\norm{V}{1})$ is the leftmost domain, followed by
$\Theta^{\alg}_{\bullet\rm \ssst f}(\Phi_{_\Lbd})$, then 
$\Theta^{\alg}_{\bullet\rm \ssst f}(\Phi_{_{\acute{\varsigma}\Lbd}})$, 
then 
$\Theta^{\alg}_{\bullet\rm \ssst f}(\Phi_{_{\dot{\varsigma}\Lbd}})$, 
and
finally $\Theta^{\alg}_{\bullet\rm \ssst f}(\Psi_{_{\grave{\varsigma}\Lbd}})$.
	For macroscopic $\Lbd$ we have
$\Theta^{\alg}_{\bullet\rm \ssst f}(\Phi_{_\Lbd})\cap\Theta^{\alg}_{\bullet\rm \ssst f}(\norm{V}{1})
	\neq\emptyset$,
in fact 
$\Theta^{\alg}_{\bullet\rm \ssst f}(\Phi_{_\Lbd})\approx \Theta^{\alg}_{\bullet\rm \ssst f}(\norm{V}{1})$, 
and we have
$\Theta^{\alg}_{\bullet\rm \ssst f}(\Phi_{_{\varsigma\Lbd}})\cap
  \Theta^{\alg}_{\bullet\rm \ssst f}(\Psi_{_{\grave{\varsigma}\Lbd}})\neq\emptyset$
for all $\varsigma\in[\grave{\varsigma},\acute{\varsigma})$; 
however, 
$\Theta^{\alg}_{\bullet\rm \ssst f}(\Phi_{_\Lbd})\cap
 \Theta^{\alg}_{\bullet\rm \ssst f}(\Phi_{_{\acute{\varsigma}\Lbd}})=\emptyset$,
and there is much space inbetween.~\eQED

	For general macroscopic domains $\Lbd$ it is not easy to come up with good explicit
estimates on $\acute{\varsigma}$, but in our section on spherical domains we will see that 
$\acute{\varsigma}\Lbd$ is not exactly what one would call a macroscopic domain. 
	So Proposition 5.3 falls far short of our ideal goal, which is to construct suitable 
subsolutions in macroscopic $\Lbd$ which imply that for most if not all 
$(\ag)\in \Theta^{\alg}_{\bullet\rm \ssst f}(\Phi_{_\Lbd})$ equation \refeq{fixptEQinLambda} 
has (at least) three solutions whose range is in $(0,\ol\eta_\fs^{\ssst\,<})$.
	On the other hand, with the help of variational arguments we will be able to show that
for a significant fraction of pairs $(\ag)\in \Theta^{\alg}_{\bullet\rm \ssst f}(\Phi_{_\Lbd})$ 
the fixed point equation \refeq{fixptEQinLambda} has at least three solutions whose range is in 
$(0,\ol\eta_\fs^{\ssst\,<})$, indeed. 
	These arguments invoke our functional $\cP^{_\Lbd}_{\ag}[\eta]$ given in \refeq{Pfctnl}.

\noindent
\section{$\cP$ STABILITY AND THE GAS $\lra$ LIQUID PHASE TRANSITION}
\smallskip
	Consider first $V\in L^1(\RR^3)$ and recall that 
$\Theta^{\alg}_{\bullet\rm \ssst f}(\norm{V}{1})$ is the bounded domain in 
$(\ag)$ half space determined by \refeq{gammaLEFTbull}, \refeq{gammaRIGHTbull}, \refeq{HATgammaFofBETAtau} 
with $\tau=\norm{V}{1}$ for which the algebraic fixed point equation \refeq{fixptEQvdW} has exactly 
three solutions in the hard-sphere fluid regime which are spatially uniform solutions of \refeq{fixptEQ}. 
	This triplicity region of uniform hard-sphere fluid solutions contains a phase 
transition curve $\gamma=\gamma_\gl^{\vdW}(\alpha)$ along which the {\it mean pressure functional} 
$\Pi_{\ag}(\eta):= \lim_{\Lbd\to\RR^3} |\Lbd|^{-1}\cP^{_\Lbd}_{\ag}[\eta]$ has an uncountable 
family of global maximizers {\it for each} $(\ag)=(\ag_\gl^{\vdW}(\alpha))$ 
--- the variational problem for $\Pi_{\ag}(\eta)$ is degenerate!
	Amongst its global maximizers are a small ($\ol\eta_{\vdW}^{\ssst,m}$)
and a large ($\ol\eta_{\vdW}^{\ssst M}$) spatially uniform solution of \refeq{fixptEQ}.
	For spatially uniform density functions $\ol\eta$, the functional $\Pi_{\ag}$ takes 
the simple form
\beq
\Pi_{\ag}(\ol\eta) = 
\wp_\bullet^{\phantom{b}}
\bigl(\gamma +\alpha \norm{V}{1}\ol\eta\bigr)
- {\tst\frac{1}{2}}
\alpha\norm{V}{1}\ol\eta^2,
\label{vdWfunc}
\eeq
and it is an elementary exercise to show that \refeq{fixptEQvdW} is the Euler--Lagrange equation for 
\beq
\pi_\bullet(\ag):= {\sup}_{_{\ol\eta}}\left\{\Pi_{\ag}^{\phantom{b}}(\ol\eta) \right\}.
\label{vdWvp}
\eeq
	Van der Waals$^{\cite{vdWaals}}$ interpreted the existence of two global maximizers of \refeq{vdWfunc} 
as a phase transition between a uniform gas and a uniform liquid phase of the hard-sphere fluid; 
however, since \refeq{fixptEQ} also has uncountably many interface type solutions which maximize 
$\Pi_{\ag}(\eta)$, eventually the uniform solutions were interpreted as {\it pure--}, the 
interface type solutions as {\it mixed phases} describing the physical co\"existence of 
{\it locally pure} phases.

	Our goal in this section is to prove the finite volume analog of this gas $\lra$ liquid phase transition 
when the fluid is confined in a macroscopic container $\ol\Lbd$ and in contact with both heat and 
matter reservoirs.
	Of course, the analogy can go only so far:
with our neutral mechanical boundary conditions there are no spatially uniform solutions to 
\refeq{fixptEQinLambda}, so that the thermodynamic notion of a ``pure phase'' cannot apply in the strict 
sense of its original definition.
	Yet, empirically$^{\cite{mkjkp}}$ (and intuitively) finite size distortions of the spatially uniform 
pure phases are limited to boundary layer effects near the container walls, so that in a macroscopic 
container which is connected to a matter reservoir the pure phases of the infinite volume thermodynamic 
formalism are approximately achieved in most of the container's interior by quasi-uniform density functions. 
	On the other hand, interface type solutions will not maximize  $\cP^{_\Lbd}_{\ag}[\eta]$, for the 
formation of an interface comes at the price of an ``interface penalty'' which becomes negligible only
in the thermodynamic (infinite volume) limit.
	To be sure, we have not been able to verify all those details.
	What we have been able to prove is stated in our

\smallskip
\noindent
{\bf Theorem 6.1:} {\it Let $\ol\Lbd$ be a convex container of macroscopic proportions, i.e.
$\diam(\Lbd)\gg 1$ and $\diam(\Lbd)/|\Lbd|^{1/3}= O(1)$.
 	Let $V\in L^1(\RR^3)$.
	Then for a subset of $\Theta^{\alg}_{\bullet\rm \ssst f}(\norm{V}{1})$ at least three ordered 
hard-sphere fluid solutions of \refeq{fixptEQinLambda} exist, (at least) two of which are locally $\cP$ stable. 
	The extension of this subset of $\Theta^{\alg}_{\bullet\rm \ssst f}(\norm{V}{1})$ to the open set
$\Theta_\bullf^{\ssst\Lbd}$ of  (at least) triplicity of hard-sphere fluid solutions of 
\refeq{fixptEQinLambda} contains a phase transition curve along which (at least) two distinct hard-sphere fluid
solutions maximize $\cP^{_\Lbd}_{\ag}[\eta]$ globally.
	The transition is of first order in the sense of Ehrenfest, i.e. the partial derivatives of 
$(\ag)\mapsto P_{_\Lbd} (\ag)$ are discontinuous across this grand canonical phase transition curve.}

\smallskip
{\it Proof of Theorem 6.1:} 
	For the proof we adapt the line of reasoning of Ref.\cite{mkJSPa} where a canonical phase transition 
is proved for $V$ given by a class of regularizations of $V_{\N}$ and $\wp$ given by the perfect gas law.
	Yet many more technical estimates are needed for the current proof, which makes it somewhat long, 
and so we begin with its outline.

	In the first part of the proof we establish the multiplicity of solutions claimed in Theorem 6.1.
	We use Propositions 3.1 and 3.4 according to which a pointwise smallest hard-sphere fluid solution 
$\eta^{\ssst m}_{\ssst\Lbd}(\bfr)$ of \refeq{fixptEQinLambda} exists when 
$(\ag)\in\Theta^{\alg}_{\bullet\rm \ssst f}(\norm{V}{1})$, and that
$\eta^{\ssst m}_{\ssst\Lbd}(\bfr)$ is locally $\cP$ stable (see Prop.3.1 in Ref.\cite{amannC}, and also below).
	Also by Propositions 3.1 and 3.4, a locally $\cP$ stable pointwise largest hard-sphere fluid 
solution $\eta^{\ssst M}_{\ssst\Lbd}(\bfr)$ of \refeq{fixptEQinLambda} exists, but Proposition 3.1 left open 
the possibility that $\eta^{\ssst m}_{\ssst\Lbd}(\bfr)$ and  $\eta^{\ssst M}_{\ssst\Lbd}(\bfr)$ are 
identical.
	We will show that when $\Lbd$ is a convex container domain of macroscopic proportions, 
then $\eta^{\ssst m}_{\ssst\Lbd}(\bfr) < \eta^{\ssst M}_{\ssst\Lbd}(\bfr)$
for a subset of pairs $(\ag)\in\Theta^{\alg}_{\bullet\rm \ssst f}(\norm{V}{1})$.
	This will be achieved by showing that for the favorable subset of 
$(\ag)\in\Theta^{\alg}_{\bullet\rm \ssst f}(\norm{V}{1})$ the
pressure functional $\cP^{_\Lbd}_{\ag}[\eta]$ evaluated with $\eta^{\ssst m}_{\ssst\Lbd}(\bfr)$ 
is bounded above by a bound which is surpassed by the evaluation of $\cP^{_\Lbd}_{\ag}[\eta]$ with 
$\ol\eta^{\ssst M}_{\vdW}$.
	This implies that the locally $\cP$ stable pointwise minimal solution is not a global maximizer, 
so another solution of \refeq{fixptEQinLambda} exists which is, yet it does not establish that this solution 
is a hard-sphere fluid solution.
	This in turn is guaranteed by imposing the ``no non-fluid solutions condition'' 
\refeq{NOnonFLUIDcondition} of Proposition 3.3 on $(\ag)\in\Theta^{\alg}_{\bullet\rm \ssst f}(\norm{V}{1})$, 
which leaves us with a bounded but non-empty set of favorable $(\ag)$ values for sufficiently large $\Lbd$.
	This set is then extended by continuity to the multiplicity set $\Theta_\bullf^{\ssst\Lbd}$ 
introduced in Theorem 6.1.

	In the second part of the proof we establish the existence of the phase transition 
in $\Theta_\bullf^{\ssst\Lbd}$.
	Having already established, in part one, that the locally $\cP$ stable pointwise minimal solution is not 
a global maximizer of $\cP^{_\Lbd}_{\ag}(\eta)$ when ``$\alpha$ and $\gamma$ are big enough,'' we recall our
uniqueness results to establish that the pointwise minimal solution is in fact the unique
global maximizer of $\cP^{_\Lbd}_{\ag}(\eta)$ when ``$\alpha$ and $\gamma$ are small enough.'' 
	The rest of the proof consists in using continuity arguments to show that for favorable 
$(\ag)$ the pointwise minimal solution is a global maximizer of $\cP^{_\Lbd}_{\ag}(\eta)$ but not the only one.

	This ends the outline of our strategy of proof.

	So our first task is to estimate $\cP^{_\Lbd}_{\ag}[\eta^{\ssst m}_{\ssst\Lbd}]$ from above.
	Since each solution $\ol\eta_{\vdW}$ of \refeq{fixptEQvdW} is a constant solution 
$\bfr\mapsto\ol\eta_{\vdW}$ of \refeq{fixptEQ}, when restricted to $\Lbd$, this constant solution
is a strict supersolution for \refeq{fixptEQinLambda} with the same $(\ag)$, and so the small 
solution of \refeq{fixptEQinLambda} is necessarily bounded above by the small solution of \refeq{fixptEQvdW}, 
i.e. $\eta^{\ssst m}_{\ssst\Lbd}(\bfr)\leq  \ol\eta^{\ssst m}_{\vdW}$; see Proposition 3.4.
	Incidentally, we also know that 
$\ol{\eta}^{\ssst m}_{\vdW}\leq  \ol\eta_{\wr}$ uniformly for all small solutions of \refeq{fixptEQvdW} 
when $(\ag)\in\Theta^{\alg}_{\bullet\rm \ssst f}(\norm{V}{1})$.
	Also, $-(V*1)_{_{\!\Lbd}}\!(\bfr)\leq \norm{V}{1}\forall\bfr\in\Lbd$.
	These pieces of information allow us to find the following upper estimate to 
$\cP^{_\Lbd}_{\ag}[\eta^{\ssst m}_{\ssst\Lbd}]$,
\bea
\cP^{_\Lbd}_{\ag} [\eta^{\ssst m}_{\ssst\Lbd}] =
\!\!\!&&
\int_{\!\Lbd} \wp_\bullet^{\phantom{b}}
\bigl(\gamma -\left(\aV*\eta^{\ssst m}_{\ssst\Lbd}\right)_{_{\!\Lbd}}\!\!(\bfr)\bigr)
d^3r 
+ \frac{1}{2}\int_{\!\Lbd}\int_{\!\Lbd} \aV(|\bfr-\tilde\bfr|) 
\eta^{\ssst m}_{\ssst\Lbd}(\bfr) \eta^{\ssst m}_{\ssst\Lbd}(\tilde\bfr)\, d^3r\, d^3\tilde{r}
\nonumber\\
\leq
\!\!\!&& 
\, \wp_\bullet^{\phantom{b}}\bigl(\gamma +\alpha \norm{V}{1}\ol{\eta}_{\vdW}^{\ssst\, m}\bigr)|\Lbd|
+ \frac{1}{2}\int_{\!\Lbd}\int_{\!\Lbd} \aV(|\bfr-\tilde\bfr|) 
\eta^{\ssst m}_{\ssst\Lbd}(\bfr) \eta^{\ssst m}_{\ssst\Lbd}(\tilde\bfr)\, d^3r\, d^3\tilde{r}.
\nonumber\\
=
\!\!\!&& 
\Big(\Pi_{\ag}\bigl(\ol{\eta}_{\vdW}^{\ssst\, m}\bigr) 
+ {\tst\frac{1}{2}} \alpha\norm{V}{1}{\ol\eta^{\ssst\, m}_{\vdW}}^{\!\!\!\!2}\Big)|\Lbd|
+ {\tst\frac{1}{2}} \alpha
\int_{\!\Lbd}\eta^{\ssst m}_{\ssst\Lbd}(\bfr)(V*\eta^{\ssst m}_{\ssst\Lbd})_{_{\!\Lbd}}(\bfr)\,d^3r,
\label{PfctnlATetaUPmESTabove}
\eea
and so
\beq
|\Lbd|^{-1}\cP^{_\Lbd}_{\ag} [\eta^{\ssst m}_{\ssst\Lbd}] 
\leq
\Pi_{\ag}\bigl(\ol{\eta}_{\vdW}^{\ssst\, m}\bigr) + {\tst\frac{1}{2}} \alpha
\Big(
\norm{V}{1}{\ol\eta^{\ssst\, m}_{\vdW}}^{\!\!\!\!2}
+ \llangle\eta^{\ssst m}_{\ssst\Lbd}(V*\eta^{\ssst m}_{\ssst\Lbd})_{_{\!\Lbd}}\rrangle_{_{\!\Lbd}}
\Big),
\label{PfctnlATetaUPmESTaboveREwrite}
\eeq
where $\llangle \,\cdot\,\rrangle_{_{\!\Lbd}}$ denotes average over $\Lbd$ w.r.t. 
normalized Lebesgue measure. 
	We will next show that
\beq
\norm{V}{1}{\ol\eta^{\ssst\, m}_{\vdW}}^{\!\!\!\!2} +
\llangle\eta^{\ssst m}_{\ssst\Lbd}(V*\eta^{\ssst m}_{\ssst\Lbd})_{_{\!\Lbd}}\rrangle_{_{\!\Lbd}}
=
O\big[\diam(\Lbd)^{-2/3}\big].
\label{INTetaUPmVetaUPmEST}
\eeq
	We abbreviate $\dist(\bfr,\partial\Lbd)\equiv s(\bfr)$, and 
$\norm{V(|\,\cdot\,|)}{L^1(B_{s(\bfr)})}\equiv\norm{V}{s(\bfr)}$.
	We note the obvious pointwise estimate
\beq
-(V*1)_{_{\!\Lbd}}(\bfr)
\geq
\norm{V}{s(\bfr)}.
\label{VstarONEinLambdaEST}
\eeq
	Now we add zero, in the form of $\norm{V}{1} -\norm{V}{1}$,  
to r.h.s.\refeq{VstarONEinLambdaEST}, then average the so rewritten \refeq{VstarONEinLambdaEST}
over $\Lbd$ w.r.t. normalized Lebesgue measure, multiply by
the constant function ${\ol\eta^{\ssst\, m}_{\vdW}}^{\!\!\!\!2}$, and get
\beq
-\llangle
 {\ol\eta^{\ssst\, m}_{\vdW}}\big(V* {\ol\eta^{\ssst\, m}_{\vdW}}\big)_{_{\!\Lbd}}
\rrangle_{_{\!\Lbd}}
\geq
\norm{V}{1}{\ol\eta^{\ssst\, m}_{\vdW}}^{\!\!\!\!2}
-
\llangle \norm{V}{1}-\norm{V}{s(\bfr)}\rrangle_{_{\!\Lbd}}
{\ol\eta^{\ssst\, m}_{\vdW}}^{\!\!\!\!2}.
\label{aveVstarONEestim}
\eeq
	The average at r.h.s.\refeq{aveVstarONEestim} is estimated as follows.
	The integrand $\norm{V}{1}-\norm{V}{s(\bfr)}$ depends on $\bfr$ only through 
$s(\bfr)=\dist(\bfr,\partial\Lbd)$, and when  extended to all $s>0$,
it is decreasing fast to zero (at least like $Cs^{-3}$) for $s$ large (in molecular units);
just asymptotically expand (A.4) and (A.5) for large $R$.
	Hence, and since $\ol\Lbd$ is convex, 
\beq
\int_{\!\Lbd}\!\! \left(\norm{V}{1}-\norm{V}{s(\bfr)}\right)\!d^3r
\leq
C(V)|\partial\Lbd|
\label{intVstarONEouterSPACE}
\eeq
where 
\beq
C(V) = \int_0^\infty\left(\norm{V}{1}-\norm{V(|\,\cdot\,|)}{L^1(B_R)}\right)dR
\label{CVint}
\eeq
is independent of $\Lbd$.
	With \refeq{intVstarONEouterSPACE}, \refeq{CVint} inserted into \refeq{aveVstarONEestim},
we obtain the estimate
\beq
\norm{V}{1}{\ol\eta^{\ssst\, m}_{\vdW}}^{\!\!\!\!2}
\leq
-\llangle{\ol\eta^{\ssst\, m}_{\vdW}}\big(V* {\ol\eta^{\ssst\, m}_{\vdW}}\big)_{_{\!\Lbd}}\rrangle_{_{\!\Lbd}}
+
C(V){\ol\eta^{\ssst\, m}_{\vdW}}^{\!\!\!\!2}{|\partial\Lbd|/ |\Lbd|}
\label{aveVstarONEidentityEST}
\eeq
with ${|\partial\Lbd|/|\Lbd|} =O\big(\diam(\Lbd)^{-1}\big)$, by hypothesis.
	So for the l.h.s.\refeq{INTetaUPmVetaUPmEST} we arrive at
\bea
\norm{V}{1}{\ol\eta^{\ssst\, m}_{\vdW}}^{\!\!\!\!2} 
+
\llangle
\eta^{\ssst m}_{\ssst\Lbd}\big(V*\eta^{\ssst m}_{\ssst\Lbd}\big)_{_{\!\Lbd}}
\rrangle_{_{\!\Lbd}}
\!\!\!&\leq&
\nonumber\\
-\llangle
 {\ol\eta^{\ssst\, m}_{\vdW}}\big(V* {\ol\eta^{\ssst\, m}_{\vdW}}\big)_{_{\!\Lbd}}
-
\eta^{\ssst m}_{\ssst\Lbd}\big(V*\eta^{\ssst m}_{\ssst\Lbd}\big)_{_{\!\Lbd}}
\rrangle_{_{\!\Lbd}}
+
O\big[\diam(\Lbd)^{-1}\big]
\!\!\!&=&
\nonumber\\
-\llangle\big({\ol\eta^{\ssst\, m}_{\vdW}}-\eta^{\ssst m}_{\ssst\Lbd}\big)
\big(V*({\ol\eta^{\ssst\, m}_{\vdW}}+
	        \eta^{\ssst m}_{\ssst\Lbd})\big)_{_{\!\Lbd}}\rrangle_{_{\!\Lbd}}
+
O\big[\diam(\Lbd)^{-1}\big]
\!\!\!&&.
\label{INTetaUPmVetaUPmESTest}
\eea
	The last displayed integral in \refeq{INTetaUPmVetaUPmESTest} we estimate thusly,
\bea
-\llangle\big({\ol\eta^{\ssst\, m}_{\vdW}}-\eta^{\ssst m}_{\ssst\Lbd}\big)
\big(V*({\ol\eta^{\ssst\, m}_{\vdW}}+
	        \eta^{\ssst m}_{\ssst\Lbd})\big)_{_{\!\Lbd}}\rrangle_{_{\!\Lbd}}
\!\!\!&\leq&
\nonumber\\
- 2\llangle\big({\ol\eta^{\ssst\, m}_{\vdW}}-\eta^{\ssst m}_{\ssst\Lbd}\big)
\big(V*{\ol\eta^{\ssst\, m}_{\vdW}}\big)_{_{\!\Lbd}}\rrangle_{_{\!\Lbd}}
\!\!\!&\leq&
\nonumber\\
- 2\llangle\big({\ol\eta^{\ssst\, m}_{\vdW}}-\eta^{\ssst m}_{\ssst\Lbd}\big)
\big(V*{\ol\eta^{\ssst\, m}_{\vdW}}\big)_{_{\!\RR^3}}\rrangle_{_{\!\Lbd}}
\!\!\!&=&
\nonumber\\
2\norm{V}{1}{\ol\eta^{\ssst\, m}_{\vdW}}
\llangle\big({\ol\eta^{\ssst\,m}_{\vdW}}-\eta^{\ssst m}_{\ssst\Lbd}\big)
\rrangle_{_{\!\Lbd}}
\!\!\!&=&
\nonumber\\
2\norm{V}{1}{\ol\eta^{\ssst\, m}_{\vdW}}
\left(
\ol\eta^{\ssst\,m}_{\vdW} - \llangle \eta^{\ssst m}_{\ssst\Lbd}\rrangle_{_{\!\Lbd}}
\right)
\!\!\!&&.
\label{INTetaUPmVetaUPmvdWest}
\eea
	To estimate $\llangle\eta^{\ssst m}_{\ssst\Lbd}\rrangle_{\ssst\Lbd}$, we recall that 
for $(\ag)\in\Theta^{\alg}_{\bullet\rm \ssst f}(\norm{V}{1})$ the small hard-sphere fluid
solution $\eta^{\ssst m}_{\ssst\Lbd}< \ol\eta^{\ssst\,m}_{\vdW} \leq \ol\eta_{\wr}$, and 
that for such pairs $(\ag)$ the map 
$\eta\mapsto \wp_{\bullet}^{\prime}\big(\gamma - (\aV*\eta)_{_{\!\Lbd}}(\bfr)\big)$
with $\wp_{\bullet}^{\prime}(\,\cdot\,)=g_{\ssst 2}^{-1}(\,\cdot\,)$ is convex
when restricted to $C^0_b(\Lbd)\cap \ol{B}_{\ol\eta_{\wr}}$.
	Jensen's inequality then gives
\beq
\llangle \eta^{\ssst m}_{\ssst\Lbd}\rrangle_{_{\!\Lbd}}
\geq
 \wp_{\bullet}^{\prime}
\big(\gamma - \llangle(\aV*\eta^{\ssst m}_{\ssst\Lbd})_{_{\!\Lbd}}\rrangle_{_{\!\Lbd}}\big).
\label{AVEetaUPmJENSENest}
\eeq
	We now notice that
\beq
\llangle\big(V*\eta^{\ssst m}_{\ssst\Lbd}\big)_{_{\!\Lbd}}\rrangle_{_{\!\Lbd}}
=
\llangle\eta^{\ssst m}_{\ssst\Lbd}\big(V*1\big)_{_{\!\Lbd}}\rrangle_{_{\!\Lbd}}
\label{AVEetaUPmVstarONE}
\eeq
and recall our pointwise estimate \refeq{VstarONEinLambdaEST} and that 
$\norm{V}{1}-\norm{V}{s(\bfr)}$ is decreasing to zero at least like $Cs^{-3}$ for $s$ 
large in molecular units.
	So if $\delta\Lbd\subset\Lbd$ is a corridor of thickness $O[\oslash(\Lbd)^{1/3}]$ 
next to the boundary $\partial\Lbd$, then upon splitting 
$\Lbd=\delta\Lbd\cup(\Lbd\setminus\delta\Lbd)$ we get
\beq
\llangle \eta^{\ssst m}_{\ssst\Lbd}\rrangle_{_{\!\Lbd}}
\geq
\wp_{\bullet}^{\prime}
\big(
\gamma^{\ssst \delta} +\alpha\tau^{\ssst \delta}\llangle\eta^{\ssst m}_{\ssst\Lbd}\rrangle_{_{\!\Lbd}}
\big),
\label{AVEetaUPmJENSENestEST}
\eeq
where
\beq
\gamma^{\ssst \delta} = \gamma - O[\oslash(\Lbd)^{-2/3}]
\label{gammaDELTA}
\eeq
and
\beq
\tau^{\ssst \delta}= \norm{V}{1}- O[\oslash(\Lbd)^{-1}].
\label{tauDELTA}
\eeq
	So $\llangle\eta^{\ssst m}_{\ssst\Lbd}\rrangle_{_{\!\Lbd}}$ is a supersolution for 
the algebraic fixed point problem
\beq
\ol\eta
=
\wp_{\bullet}^{\prime}\big(\gamma^{\ssst \delta} +\alpha\tau^{\ssst \delta}\ol\eta\big),
\label{deltaFIXptEQ}
\eeq
and this yields the lower bound
\beq
\llangle \eta^{\ssst m}_{\ssst\Lbd}\rrangle_{_{\!\Lbd}}
\geq
\ol\eta^{\ssst \delta}
\label{olETAest}
\eeq
where $\ol\eta^{\ssst \delta}$ is the smallest solution of \refeq{deltaFIXptEQ}.
	We also know that 
$\llangle\eta^{\ssst m}_{\ssst\Lbd}\rrangle_{_{\!\Lbd}}
			< \ol\eta^{\ssst\,m}_{\vdW} \leq \ol\eta_{\wr}$, 
so by the concavity of $\ol\eta\mapsto g_{\ssst 2}(\ol\eta)$ for $\ol\eta<\ol\eta_{\wr}$
we easily find the explicit lower bound 
\beq
\ol\eta^{\ssst \delta} > \ul\eta^{\ssst \delta},
\label{ulETAboundsOLetaDELTA}
\eeq
where $\ul\eta^{\ssst \delta}$ solves the linear algebraic equation
\beq
g_{\ssst 2}(\ol\eta^{\ssst\,m}_{\vdW})
+
g_{\ssst 2}^\pr(\ol\eta^{\ssst\,m}_{\vdW})\big(\ul\eta-\ol\eta^{\ssst\,m}_{\vdW}\big)
=
\gamma - O[\oslash(\Lbd)^{-2/3}] + \alpha(\norm{V}{1}- O[\oslash(\Lbd)^{-1}])\ul\eta,
\label{ulETAequ}
\eeq
which gives
\beq
\ul\eta^{\ssst \delta}
=
\ol\eta^{\ssst\,m}_{\vdW} - O[\oslash(\Lbd)^{-2/3}].
\label{ulETAsol}
\eeq
	Hence,
\beq
\ol\eta^{\ssst\,m}_{\vdW}-\llangle\eta^{\ssst m}_{\ssst\Lbd}\rrangle_{_{\!\Lbd}}
\leq 
O[\oslash(\Lbd)^{-2/3}].
\label{AVEolETAvdWMINUEolETAest}
\eeq
	All in all, this proves \refeq{INTetaUPmVetaUPmEST}, i.e. we have shown that
\beq
|\Lbd|^{-1}\cP^{_\Lbd}_{\ag} [\eta^{\ssst m}_{\ssst\Lbd}] 
\leq
\Pi_{\ag}\bigl(\ol{\eta}_{\vdW}^{\ssst\, m}\bigr) + O\big[\diam(\Lbd)^{-2/3}\big].
\label{PfctnlATetaUPmESTfinal}
\eeq

	We next recall that the maximum of $\cP^{_\Lbd}_{\ag}[\eta]$ is estimated below 
(in particular) by
\beq
\max_\eta \cP^{_\Lbd}_{\ag} [\eta] 
\geq 
\cP^{_\Lbd}_{\ag} [\ol\eta^{\ssst M}_{\vdW}].
\label{maxPlowerESTa}
\eeq
	We estimate $\cP^{_\Lbd}_{\ag}[\eta^{\ssst M}_{\vdW}]$ from below by using 
that our $\aV= A_{\W}V_{\W}+ A_{\Y}V_{\Y}$ is of molecular 
effective range, and that $\Lbd$ is convex and of macroscopic proportions, so that 
our earlier $\delta$-corridor estimate tells us that we can find 
$\vartheta = 1 -O[\diam(\Lbd)^{-1}]\in(0,1)$ and 
$\varsigma = 1 -O[\diam(\Lbd)^{-2/3}]\in(0,1)$ so that
$-\big(V*1\big)_{_{\!\Lbd}}\!\!(\bfr)\geq \vartheta\norm{V}{1}\forall\bfr\in\varsigma\Lbd$.
	Using also that
$-(V*1)_{_{\!\Lbd}}\!(\bfr)\leq \norm{V}{1}\forall\bfr\in\Lbd$, we find
\bea
\cP^{_\Lbd}_{\ag} [\ol\eta^{\ssst M}_{\vdW}]
=\!\!\!&&
\int_{\!\Lbd}\wp_\bullet^{\phantom{b}}
\bigl(\gamma -\big(\aV*\ol\eta^{\ssst M}_{\vdW}\big)_{_{\!\Lbd}}\!\!(\bfr)\bigr)
d^3r 
+ {\tst\frac{1}{2}}\,{\ol\eta^{\ssst M}_{\vdW}}^{\!\!\!\!2}
\int_{\!\Lbd}\int_{\!\Lbd} \aV(|\bfr-\tilde\bfr|)\, d^3r\, d^3\tilde{r}
\nonumber\\
\geq
\!\!\!&& 
\,\wp_\bullet^{\phantom{b}}
	\bigl(\gamma +\alpha \norm{V}{1}\vartheta\,\ol{\eta}^{\ssst M}_{\vdW}\bigr)
	|\varsigma\Lbd|
- {\tst\frac{1}{2}}\alpha\,{\ol\eta^{\ssst M}_{\vdW}}^{\!\!\!\!2}\norm{V}{1}|\Lbd|
\nonumber\\
=
\!\!\!&& 
\Big(\varsigma\wp_\bullet^{\phantom{b}}
	\bigl(\gamma +\alpha \norm{V}{1}\vartheta\,\ol{\eta}^{\ssst M}_{\vdW}\bigr)
	- {\tst\frac{1}{2}}\alpha\,{\ol\eta^{\ssst M}_{\vdW}}^{\!\!\!\!2}\norm{V}{1}\Big)|\Lbd|.
\label{PfctnlLOWERmaxBOUND}
\eea
	Next, a simple telescoping yields the identity
\bea
\varsigma
\wp_\bullet^{\phantom{b}}
	\bigl(\gamma +\alpha \norm{V}{1}\vartheta\,\ol{\eta}^{\ssst M}_{\vdW}\bigr)
\!\!&=&\;
\wp_\bullet^{\phantom{b}}
	\bigl(\gamma +\alpha \norm{V}{1}\ol{\eta}^{\ssst M}_{\vdW}\bigr) -
(1-\varsigma) 
\wp_\bullet^{\phantom{b}}
	\bigl(\gamma +\alpha \norm{V}{1}\ol{\eta}^{\ssst M}_{\vdW}\bigr)-
\nonumber\\
\!\!\!&&\;
\varsigma\Big(
\wp_\bullet^{\phantom{b}}
	\bigl(\gamma +\alpha \norm{V}{1}\ol{\eta}^{\ssst M}_{\vdW}\bigr)
-
\wp_\bullet^{\phantom{b}}
	\bigl(\gamma +\alpha \norm{V}{1}\vartheta\,\ol{\eta}^{\ssst M}_{\vdW}\bigr)\Big),
\label{pREWRITE}
\eea
and by the mean value theorem we have the further identity
\beq
\wp_\bullet^{\phantom{b}}
	\bigl(\gamma +\alpha \norm{V}{1}\ol{\eta}^{\ssst M}_{\vdW}\bigr)
-
\wp_\bullet^{\phantom{b}}
	\bigl(\gamma +\alpha \norm{V}{1}\vartheta\,\ol{\eta}^{\ssst M}_{\vdW}\bigr)
=
(1-\vartheta)\wp_\bullet^\pr(\ol\gamma)
\alpha\norm{V}{1}\ol{\eta}^{\ssst M}_{\vdW}
\label{mvTHMappl}
\eeq
for some $\ol\gamma$ inbetween the two arguments at the l.h.s.\refeq{mvTHMappl}.
	The monotonic increase of $\wp^\pr_\bullet$ now gives
\beq
\wp_\bullet^\pr(\ol\gamma) < \wp_\bullet^\pr
	\bigl(\gamma +\alpha \norm{V}{1}\ol{\eta}_{\vdW}^{\ssst\,m}\bigr).
\label{pESTa}
\eeq
	The r.h.s.\refeq{pESTa} is independent of $\Lbd$ and depends on $(\ag)$ 
as displayed plus implicitly through $\ol{\eta}_{\vdW}^{\ssst\,m}$.
	Since $1-\vartheta = O[\diam(\Lbd)^{-1}]$ and $1-\varsigma = O[\diam(\Lbd)^{-2/3}]$, 
we conclude that
\beq
\varsigma
\wp_\bullet^{\phantom{b}}
	\bigl(\gamma +\alpha \norm{V}{1}\vartheta\,\ol{\eta}^{\ssst M}_{\vdW}\bigr)
\geq 
\wp_\bullet^{\phantom{b}}
	\bigl(\gamma +\alpha \norm{V}{1}\ol{\eta}^{\ssst M}_{\vdW}\bigr) - O[\diam(\Lbd)^{-2/3}],
\label{pESTc}
\eeq
and so
\bea
 |\Lbd|^{-1} \max_\eta \cP^{_\Lbd}_{\ag} [\eta] 
&\geq &
\wp_\bullet^{\phantom{b}}
	\bigl(\gamma +\alpha \norm{V}{1}\ol{\eta}^{\ssst M}_{\vdW}\bigr)
	- {\tst\frac{1}{2}}\alpha\,{\ol\eta^{\ssst M}_{\vdW}}^{\!\!\!\!2}\norm{V}{1}
 - O[\diam(\Lbd)^{-2/3}]
\nonumber\\
&=&\Pi_{\ag}\bigl(\ol{\eta}_{\vdW}^{\ssst M}\bigr)  - O[\diam(\Lbd)^{-2/3}].
\label{maxPlowerESTb}
\eea

	Combining \refeq{PfctnlATetaUPmESTfinal}
with \refeq{maxPlowerESTb} now yields the desired estimate
\beq
|\Lbd|^{-1}\left(\max_\eta \cP^{_\Lbd}_{\ag} [\eta] 
-
\cP^{_\Lbd}_{\ag} [\eta^{\ssst m}_{\ssst\Lbd}] \right)
\geq
\Pi_{\ag}\bigl(\ol{\eta}_{\vdW}^{\ssst M}\bigr)  - \Pi_{\ag}\bigl(\ol{\eta}_{\vdW}^{\ssst\, m}\bigr) 
- O\big[\diam(\Lbd)^{-2/3}\big].
\label{PfctnlDIFFest}
\eeq
	But for $(\ag)$ in the triplicity set $\Theta^{\alg}_{\bullet\rm \ssst f}(\norm{V}{1})$,
the $\Lbd$-independent function 
\beq
 \varpi(\ag)
:=
\Pi_{\ag}\bigl(\ol{\eta}_{\vdW}^{\ssst\, M}\bigr) - \Pi_{\ag}\bigl(\ol{\eta}_{\vdW}^{\ssst\, m}\bigr) 
\label{PIvdWdiff}
\eeq 
vanishes {\it only} on the van der Waals gas $\&$ liquid
co\"existence curve $\alpha\mapsto\gamma=\gamma_\gl^{\vdW}(\alpha)$, thereby dividing
$\Theta^{\alg}_{\bullet\rm \ssst f}(\norm{V}{1})$ into two disjoint subsets, in one of which 
$\varpi(\ag)<0$, and $\varpi(\ag)>0$ in the other. 
	Since $\varpi(\ag)$ is independent of $\Lbd$, while 
$O\big[\diam(\Lbd)^{-2/3}\big]\downarrow 0$ as $\diam(\Lbd)\to \infty$, 
it follows that for each pair $(\ag)$ for which $\varpi(\ag)>0$, we have 
\beq
|\Lbd|^{-1}\left(\max_\eta \cP^{_\Lbd}_{\ag} [\eta] 
-
\cP^{_\Lbd}_{\ag} [\eta^{\ssst m}_{\ssst\Lbd}] \right)
\geq
\varpi(\ag) - O\big[\diam(\Lbd)^{-2/3}\big] >0
\label{PfctnlDIFFestB}
\eeq
eventually, for large enough $\Lbd$. 
	So for this subset of $\Theta^{\alg}_{\bullet\rm \ssst f}(\norm{V}{1})$, a locally 
$\cP$ stable small hard-sphere fluid solution $\eta^{\ssst m}_{\ssst\Lbd}(\bfr)<\ol\eta_{\wr}$ 
exists, but it is not globally $\cP$ stable. 

	Our result \refeq{PfctnlDIFFestB} for the $\varpi(\ag)>0$ subset of 
$\Theta^{\alg}_{\bullet\rm \ssst f}(\norm{V}{1})$ in sufficiently large $\Lbd$
does not establish that the global maximizer is a hard-sphere fluid solution, or even 
that any other hard-sphere fluid solution of \refeq{fixptEQinLambda} exists for the ``parameters''
$(\ag)$ and $\Lbd$ under consideration.
	Yet, since \refeq{PfctnlDIFFestB} holds for any particular $(\ag)$ in the $\varpi(\ag)>0$ 
subset of $\Theta^{\alg}_{\bullet\rm \ssst f}(\norm{V}{1})$ {\it whenever} $\Lbd$ is sufficiently 
large, we can impose the additional condition \refeq{NOnonFLUIDcondition} (with $\norm{V}{1}$ in 
place of $\Phi_{_\Lbd}$) on $(\ag)$, so that no solution to the extended \refeq{fixptEQinLambda} 
exists which somewhere in $\Lambda$ is not a (hard-sphere) fluid; see Proposition 3.3.
	Notice that \refeq{NOnonFLUIDcondition} is a sufficient but certainly not a necessary condition, yet
to improve on it we would need to have better control over the solid branch of 
$\gamma\mapsto\wp_\bullet(\gamma)$.
	In absence of such better control we consider
\beq
(\ag)\in \Theta^{\alg}_{\bullet\rm \ssst f}(\norm{V}{1})
\cap
	\{\varpi(\ag)>0
\;\wedge\; 
	\gamma +\alpha\norm{V}{1}\ol\eta_{\ssst fcc}^{\ssst\,cp} \leq \gamma_\fs \}.
\label{agTRIPLEsetLambda}
\eeq
	This set, which is defined entirely in terms of the algebraic van der Waals theory with spatially
uniform density functions, is non-empty, as can be verified by evaluating this van der Waals model.
	We conclude that when $(\ag)$ satisfies \refeq{agTRIPLEsetLambda} and is fixed, then for 
large enough $\Lambda$ a locally $\cP$ stable pointwise minimal hard-sphere fluid solution 
$\eta^{\ssst m}_{\ssst\Lbd}(\bfr)<\ol\eta_{\wr}$ of \refeq{fixptEQinLambda} 
exists, but the global $\cP^{_\Lbd}_{\ag}$ maximizer is given by another, pointwise larger 
hard-sphere fluid solution of \refeq{fixptEQinLambda} which is locally $\cP$ stable, or locally $\cP$ indifferent
in exceptional cases.

	The existence of a third, unstable (under iterations and in $\cP$ sense)
solution sandwiched between the locally $\cP$ stable minimal solution and the globally $\cP$ stable
solution of \refeq{fixptEQinLambda} now follows via the mountain pass lemma thanks to 
the strong $C^0_b(\ol\Lbd)$ differentiability of the functional $\cP^{_\Lbd}_{\ag}[\eta]$.
	By continuity we can extend the so constructed multiplicity sub-region of hard-sphere fluid
solutions of \refeq{fixptEQinLambda} to a larger set $\Theta_\bullf^{\ssst\Lbd}$, which is the open set of 
pairs $(\ag)$ for which at least three ordered hard-sphere fluid solutions of \refeq{fixptEQinLambda} exist, 
(at least) two of which are locally $\cP$ stable (or, exceptionally, locally $\cP$ indifferent), and
no non-fluid solution.

	This completes the part of our proof of Theorem 6.1 which establishes multiplicity of hard-sphere 
fluid solutions of \refeq{fixptEQinLambda} in a certain domain in $(\ag)$ space.
	We next prove that somewhere in this multiplicity region of hard-sphere fluid solutions 
a first-order phase transition occurs between a small and a large(r) hard-sphere fluid solution. 

	We already know from part one of our current proof that for any $(\ag)$ satisfying \refeq{agTRIPLEsetLambda},
whenever $\Lambda$ is large enough, then the global $\cP^{_\Lbd}_{\ag}$ maximizer is given by a hard-sphere fluid 
solution of \refeq{fixptEQinLambda} which is  pointwise larger than the locally $\cP$  stable pointwise minimal 
hard-sphere fluid solution $\eta^{\ssst m}_{\ssst\Lbd}(\bfr)$ of \refeq{fixptEQinLambda}, 
which exists also, satisfies $\eta^{\ssst m}_{\ssst\Lbd}(\bfr)<\ol\eta_{\wr}$, but
which is not a global  $\cP^{_\Lbd}_{\ag}$ maximizer in this $(\ag)$ region.
	Moreover, if we pick any $(\ag)_0$ satisfying \refeq{agTRIPLEsetLambda} and pick a large enough $\Lbd$ 
so that the globally $\cP$ stable solution of \refeq{fixptEQinLambda} is pointwise larger than  
$\eta^{\ssst m}_{\ssst\Lbd}(\bfr)$ for the chosen $(\ag)_0$ and $\Lbd$, then by the continuity of the map 
$(\ag)\mapsto P_{\ssst\Lbd}(\ag)$ and the continuity of the map 
$(\ag)\mapsto \cP^{_\Lbd}_{\ag}[\eta^{\ssst m}_{\ssst\Lbd}]$ restricted to\footnote{The
	map $(\ag)\mapsto \cP^{_\Lbd}_{\ag}[\eta^{\ssst m}_{\ssst\Lbd}]$ is generally not continuous 
	without the size restriction on $\eta^{\ssst m}_{\ssst\Lbd}$.
	For instance, think of the $ S$-shape sections of the solution diagram of the space-uniform 
	van der Waals problem \refeq{fixptEQvdW}.} 
$\eta^{\ssst m}_{\ssst\Lbd}<\ol\eta_{\wr}$, for $\Lbd$ as chosen and now fixed, there exists a whole 
finite-measure  $(\ag)$ neighborhood of $(\ag)_0$ in r.h.s.\refeq{agTRIPLEsetLambda} for which the globally 
$\cP$ stable hard-sphere fluid solution of \refeq{fixptEQinLambda} is pointwise larger than  
$\eta^{\ssst m}_{\ssst\Lbd}(\bfr)$, which in turn is not globally stable.
	Let this subset of r.h.s.\refeq{agTRIPLEsetLambda} be denoted by $^{\ssst \ell}\Theta_\bullf^{\ssst\Lbd}$.
	It is a forteriori contained in the multiple hard-sphere fluid region $\Theta_\bullf^{\ssst\Lbd}$.

	On the other hand, recall that according to Corollary 4.5 the hard-sphere fluid solution $\eta_{\ssst\Lbd}$
of \refeq{fixptEQinLambda} is unique if both of the following are true, 
$\wp^\ppr_{\CS}(\gamma_{\wr}) \alpha \Phi_{_\Lbd} > 1$ (with $\wp^\ppr_{\CS}(\gamma_{\wr})\ \approx 0.047$) 
and 
$\gamma < \gamma_{\!\ssst\Lbd}(\alpha)$ (with $\gamma_{\!\ssst\Lbd}(\alpha)$ given in Definition 4.4). 
	A unique hard-sphere fluid solution is necessarily the pointwise minimal solution,  
$\eta_{\ssst\Lbd}\equiv\eta^{\ssst m}_{\ssst\Lbd}$, and the conditions of Corollary 4.5 guarantee that 
$\eta_{\ssst\Lbd}\in C^0_{b,+}\cap\ol{\cal B}_{\ol\eta(\ag)}$, so the solution
is ``small'' in the sense that $\eta^{\ssst m}_{\ssst\Lbd}<\ol\eta_{\wr}$.
	If we supplement the conditions of Corollary 4.5 with the condition 
$\gamma +\alpha\norm{V}{1}\ol\eta_{\ssst fcc}^{\ssst\,cp} \leq \gamma_\fs$, then
no solution to the extended \refeq{fixptEQinLambda} exists which is somewhere in $\Lbd$ not fluid, and then
the unique hard-sphere fluid solution automatically is the unique maximizer of $\cP^{_\Lbd}_{\ag}$ for 
any compliant $(\ag)$ point. 

	Now note that the condition on $\alpha$ in Corollary 4.5 is fulfilled for all 
$(\ag)\in^{\ssst \ell}\Theta_\bullf^{\ssst\Lbd}$, and so is the no-non-fluid condition
$\gamma +\alpha\norm{V}{1}\ol\eta_{\ssst fcc}^{\ssst\,cp} \leq \gamma_\fs$. 
	Hence we conclude from the discussion in the previous two paragraphs that along any constant-$\alpha$ ray 
which begins in $^{\ssst \ell}\Theta_\bullf^{\ssst\Lbd}$ and continues to arbitrarily negative $\gamma$ values
there occurs a discontinuity in the map $\gamma\mapsto \{\eta^{\GC}_{\ssst\Lbd}(\bfr)\}$ from $\gamma$ to the set 
of global maximizers of $\cP^{_\Lbd}_{\ag}$ which are all fluid.
	Indeed, along any such ray the constant-$\alpha$ map $\gamma\mapsto \eta^{\ssst m}_{\ssst\Lbd}(\bfr)$ 
furnishes the unique global $\cP^{_\Lbd}_{\ag}$ maximizer when $\gamma$ is negative enough, i.e. 
$\eta^{\ssst m}_{\ssst\Lbd}(\bfr)\equiv \eta^{\GC}_{\ssst\Lbd}(\bfr)$ for $\gamma$ negative enough.
	Moreover, this map $\gamma\mapsto \eta^{\ssst m}_{\ssst\Lbd}(\bfr)$ 
extends continuously differentiably$^{\cite{amannB}}$ into the region 
$^{\ssst \ell}\Theta_\bullf^{\ssst\Lbd}$, 
for which a hard-sphere fluid solution $\eta^{\GC}_{\ssst\Lbd}(\bfr) > \eta^{\ssst m}_{\ssst\Lbd}(\bfr)$ of 
\refeq{fixptEQinLambda} is the global maximizer of $\cP^{_\Lbd}_{\ag}$, while 
$\eta^{\ssst m}_{\ssst\Lbd}(\bfr)$ is not.
	Furthermore, by the local $\cP$ stability of the pointwise minimal solutions$^{\cite{amannC}}$ a
branch of pointwise non-minimal global maximizers cannot bifurcate off of this continuously
differentiable branch of pointwise minimal small solution.
	Hence, some discontinuous change in the set of global maximizers must happen along each such ray,
as claimed. 
	We next clarify the nature of the discontinuity.

	For fixed suitable $\Lbd$ and $\alpha$ as just described, we now define 
$\gamma_\gl^{\ssst\Lbd}(\alpha)$ to be the supremum of $\gamma$ values for which 
$\eta^{\GC}_{\ssst\Lbd}(\bfr) \equiv \eta^{\ssst m}_{\ssst\Lbd}(\bfr)<\ol\eta_{\wr}$
{\it is the unique global maximizer of}
$\cP^{_\Lbd}_{\ag}$ {\it for all} $\gamma< \gamma_\gl^{\ssst\Lbd}(\alpha)$; clearly, 
$\gamma_{\!\ssst\Lbd}(\alpha)\leq\gamma_\gl^{\ssst\Lbd}(\alpha)
    <\hat\gamma^{\alg}_{\bullet\rm \ssst f}(\alpha\norm{V}{1})$.
	We also define
$_*\!\gamma_\gl^{\ssst\Lbd}(\alpha)$ to be the infimum of $\gamma$ values for which 
$\eta^{\ssst m}_{\ssst\Lbd}(\bfr)<\ol\eta_\wr$ {\it is not a global maximizer of} $\cP^{_\Lbd}_{\ag}$ 
{\it for all} $\gamma\in(_*\!\gamma_\gl^{\ssst\Lbd}(\alpha)\,,\,_*\!\gamma_\gl^{\ssst\Lbd}(\alpha)+\eps)$
for some $\eps>0$; clearly, 
$\gamma_\gl^{\ssst\Lbd}(\alpha)\leq_*\!\gamma_\gl^{\ssst\Lbd}(\alpha)
	<\hat\gamma^{\alg}_{\bullet\rm \ssst f}(\alpha\norm{V}{1})$.
	We next show that $_*\!\gamma_\gl^{\ssst\Lbd}(\alpha) = \gamma_\gl^{\ssst\Lbd}(\alpha)$.

	Indeed, suppose $_*\!\gamma_\gl^{\ssst\Lbd}(\alpha) \neq \gamma_\gl^{\ssst\Lbd}(\alpha)$.
	Then $\gamma_\gl^{\ssst\Lbd}(\alpha)<_*\!\gamma_\gl^{\ssst\Lbd}(\alpha)$, and now it follows from the
definitions of $\gamma_\gl^{\ssst\Lbd}(\alpha)$ and $_*\!\gamma_\gl^{\ssst\Lbd}(\alpha)$ that
$\eta^{\ssst m}_{\ssst\Lbd}(\bfr)<\ol\eta_\wr$ is a global maximizer of $\cP^{_\Lbd}_{\ag}$ for all 
$\gamma_\gl^{\ssst\Lbd}(\alpha)<\gamma<_*\!\gamma_\gl^{\ssst\Lbd}(\alpha)$, though not the unique one. 
	But then, not only are the values of $\cP^{_\Lbd}_{\ag}[\eta]$ the same for its pointwise minimal 
maximizer $\eta^{\ssst m}_{\ssst\Lbd}$ and for its other maximizer(s) $\eta^{\not{\ssst m}}_{_\Lbd}$, also 
the $\gamma$-derivatives of $\cP^{_\Lbd}_{\ag}[\eta]$ must be the same for $\eta^{\ssst m}_{\ssst\Lbd}$ and 
for $\eta^{\not{\ssst m}}_{_\Lbd}$.
	By the implicit function theorem, the derivative of $\gamma\mapsto \cP^{_\Lbd}_{\ag}[\eta_{_\Lbd}]$ 
exists along any constant-$\alpha$ section of a solution branch of \refeq{fixptEQinLambda},
except at the bifurcation points where it might or might not exist, but in any event either the left or 
right derivative w.r.t. $\gamma$ exists, then.
	Now, since any currently contemplated $(\ag)$ is not a bifurcation point for the pointwise 
minimal solution, the partial $\gamma$ derivative of $\cP^{_\Lbd}_{\ag}[\eta^{\ssst m}_{\ssst\Lbd}]$ 
exists at $(\ag)$. 
	Any other maximizer of $\cP^{_\Lbd}_{\ag}[\eta]$ belongs to a different solution branch, and we
may assume that in general it is not at a bifurcation point either, so the partial $\gamma$ derivative of 
$\cP^{_\Lbd}_{\ag}[\eta^{\not{\ssst m}}_{\ssst\Lbd}]$ generally exists at the contemplated $(\ag)$, too.
	Now, with the help of \refeq{fixptEQinLambda} one can easily show that, away from bifurcation points,
\beq
\partial_\gamma\cP^{_\Lbd}_{\ag}[\eta_{_\Lbd}]
=
\int_{\!\Lbd}\wp_\bullet^\pr
\bigl(\gamma -\big(\aV*\eta_{_\Lbd}\big)_{_{\!\Lbd}}\!\!(\bfr)\bigr)
d^3r ;
\label{PgammaDERIValongSOL}
\eeq
note that in terms of the functional for the total number of particles \refeq{Nfunc} we can re-express this derivative as
$
\partial_\gamma\cP^{\Lbd}_{\ag}[\eta_{_\Lbd}]
=
\cN^{\Lbd}[\eta_{_\Lbd}]
$.
	So we conclude that the two maximizers $\eta^{\not{\ssst m}}_{_\Lbd}$  and $\eta^{{\ssst m}}_{_\Lbd}$ of
$\cP^{_\Lbd}_{\ag}[\eta]$ also have the same $N$, i.e. 
$\cN^{\Lbd}[\eta^{\not{\ssst m}}_{_\Lbd}] = \cN^{\Lbd}[\eta^{{\ssst m}}_{_\Lbd}]$, which is impossible because
$\eta^{{\ssst m}}_{_\Lbd}$ is the pointwise minimal solution for the given $(\ag)$.
	This proves that $_*\!\gamma_\gl^{\ssst\Lbd}(\alpha) = \gamma_\gl^{\ssst\Lbd}(\alpha)$.
	Incidentally, the proof also shows that 
$\eta^{\ssst m}_{\ssst\Lbd}(\bfr)<\ol\eta_\wr$ {\it is not a global maximizer of} $\cP^{_\Lbd}_{\ag}$ 
{\it for all} $\gamma>\gamma_\gl^{\ssst\Lbd}(\alpha)$.

	Now, by the continuity of the maps $(\ag)\mapsto P_{\ssst\Lbd}(\ag)$ and 
$(\ag)\mapsto \cP^{_\Lbd}_{\ag}[\eta^{\ssst m}_{\ssst\Lbd}]$ at $(\alpha,\gamma_\gl^{\ssst\Lbd}(\alpha))$ it 
follows that $\alpha\mapsto\gamma_\gl^{\ssst\Lbd}(\alpha)$ is continuous in it's (restricted) domain of 
definition.\footnote{Once again, the restriction $\eta^{\ssst m}_{\ssst\Lbd}(\bfr)<\ol\eta_{\wr}$
	is vital; without it, the domain of definition of 
	$\alpha\mapsto\gamma_\gl^{\ssst\Lbd}(\alpha)$ can be extended to all $\alpha\in\RR_+$ but then this
	map is not continuous.}
	Moreover by the continuity of the maps $(\ag)\mapsto P_{\ssst\Lbd}(\ag)$ and 
$(\ag)\mapsto \cP^{_\Lbd}_{\ag}[\eta^{\ssst m}_{\ssst\Lbd}]$ at $(\alpha,\gamma_\gl^{\ssst\Lbd}(\alpha))$,
we also conclude that the pointwise minimal solution $\eta^{\ssst m}_{\ssst\Lbd}$ 
is certainly a global maximizer of $\cP^{_\Lbd}_{\ag}$ also at $(\alpha,\gamma_\gl^{\ssst\Lbd}(\alpha))$, and
then denoted $\eta^{\GC}_{\ssst\Lbd,g}$.
	However, the definition of $\gamma_\gl^{\ssst\Lbd}(\alpha)$ leaves it open
whether or not $\eta^{\ssst m}_{\ssst\Lbd}\,$ is the {\it unique} maximizer also at 
$(\alpha,\gamma_\gl^{\ssst\Lbd}(\alpha))$, in which case the ``sup'' in the definition of 
$\gamma_\gl^{\ssst\Lbd}(\alpha)$ could be replaced by ``max.''
	We next show that at $(\alpha,\gamma_\gl^{\ssst\Lbd}(\alpha))$ the global maximizers 
of $\cP^{_\Lbd}_{\ag}$ is not unique.

	Let $(\ag)$ be a point on a ray emanating from 
$^{\ssst \ell}\Theta_\bullf^{\ssst\Lbd}$ which is to the right of but near the curve 
$\gamma=\gamma_\gl^{\ssst\Lbd}(\alpha)$.
	Then, by the just proven fact that $\gamma_\gl^{\ssst\Lbd}(\alpha)=_*\!\gamma_\gl^{\ssst\Lbd}(\alpha)$,
and by their definitions, it follows that some hard-sphere fluid solution 
$\eta_{\ssst\Lbd}(\bfr) > \eta^{\ssst m}_{\ssst\Lbd}(\bfr)$ of 
\refeq{fixptEQinLambda} is a global $\cP^{_\Lbd}_{\ag}$ maximizer for each $\gamma>\gamma_\gl^{\ssst\Lbd}(\alpha)$
in a right $\eps$-neighborhood of $\gamma_\gl^{\ssst\Lbd}(\alpha)$. 
	For each such $\gamma>\gamma_\gl^{\ssst\Lbd}(\alpha)$ pick such a maximizer
$\eta_{\ssst\Lbd}(\bfr) (> \eta^{\ssst m}_{\ssst\Lbd}(\bfr))$ and consider the map 
$\gamma\mapsto \eta_{\ssst\Lbd}(\bfr)$. 
	Notice that the set $\{\gamma>\gamma_\gl^{\ssst\Lbd}(\alpha)\}$ is open.
	Since $\gamma\mapsto \wp_\bullet^\pr(\gamma)$ is monotonic increasing,  
each solution $\eta_{\ssst\Lbd}$ of \refeq{fixptEQinLambda} is a supersolution for 
\refeq{fixptEQinLambda} with	$\gamma$ replaced by $\gamma-\eps$.
	This implies that the branch  $\gamma\mapsto \eta_{\ssst\Lbd}(\bfr)$ of the 
globally stable hard-sphere fluid maximizers of $\cP^{_\Lbd}_{\ag}$ which are bigger than 
$\eta^{\ssst m}_{\ssst\Lbd}(\bfr)$, is pointwise monotonic increasing in $\gamma$.
	We conclude that the following limit exists pointwise and strongly in 
$C^0_b(\Lbd)\cap \ol{B}_{\ol\eta_\fs^{\ssst\,<}}$, 
\beq
\lim_{\gamma\downarrow \gamma_\gl^{\ssst\Lbd}(\alpha)} \eta_{\ssst\Lbd} =:
\eta^{\GC}_{\ssst\Lbd,\ell}.
\label{gammaDOWNlimGCeta}
\eeq
	The strong $C^0_b$ continuity of $\eta\mapsto\wp_\bullet^\pr\big(\gamma-(\aV*\eta)_\Lbd\big)$
implies that $\eta^{\GC}_{\ssst\Lbd,\ell}$ also solves \refeq{fixptEQinLambda}. 
	By the continuity of $\gamma\mapsto P_{\ssst \Lbd}(\ag)$ it 
follows that $\eta^{\GC}_{\ssst\Lbd,\ell}$ is also a global maximizer of $\cP^{_\Lbd}_{\ag}$.
	This proves that 
at $(\alpha,\gamma_\gl^{\ssst\Lbd}(\alpha))$ the global maximizers of $\cP^{_\Lbd}_{\ag}$ is not unique, 
interpreted as a first order phase transition in the sense of {\it mathematical co\"{e}xistence} of
two distinct global maximizers of $\cP^{_\Lbd}_{\ag}$, not to be confused with the {\it physical co\"{e}xistence} 
of two locally pure phases, separated by an interface, described by a single solution.

	It remains to prove that the phase transition is of first order also in the sense of Ehrenfest.
	In Ref.\cite{mkjkp} we showed that for fixed $\Lbd$, the map $(\ag)\mapsto P_{\ssst\Lbd}(\ag)$,
defined in \refeq{vpPinLambda}, is the limit of a family of functions which are convex in $\alpha$ and $\gamma$, and
so itself (bi-)convex in $(\ag)$, thus continuous and almost everywhere differentiable in both variables. 
	We now show that at $(\alpha,\gamma_\gl^{\ssst\Lbd}(\alpha))$ generally there is a kink in both, 
$\alpha\mapsto P_{_\Lbd}(\ag)$ and $\gamma\mapsto P_{_\Lbd}(\ag)$; we ignore the exceptional case 
$\gamma_\gl^{\ssst\Lbd}(\alpha)=c$.
	In that case by the bi-convexity of  $(\ag)\mapsto P_{_\Lbd}(\ag)$ both partial derivatives jump {\sl up}
when crossing the grand canonical phase transition curve from smaller to larger $\alpha$ or $\gamma$ values.

	By repeating almost verbatim the arguments used to prove that
$_*\!\gamma_\gl^{\ssst\Lbd}(\alpha) = \gamma_\gl^{\ssst\Lbd}(\alpha)$, 
one  proves that the $\gamma$-derivative of $P_{\ssst \Lbd}(\ag)$ jumps at $\gamma_\gl^{\ssst\Lbd}(\alpha)$. 
	Now, since we ignore when  $\gamma_\gl^{\ssst\Lbd}(\alpha)=$ constant, 
locally there exists $\gamma\mapsto \alpha_\gl^{\ssst\Lbd}(\gamma)$, the inverse function of 
$\gamma_\gl^{\ssst\Lbd}(\alpha)$, and arguing almost verbatim again, but now using
also that $V<0$, for $\gamma$ suitably fixed we find that
\beq
\lim_{\alpha\downarrow \alpha_\gl^{\ssst\Lbd}(\gamma)} \eta^{\GC}_{\ssst\Lbd} =
\eta^{\GC}_{\ssst\Lbd,\ell},
\label{alphaDOWNlimGCeta}
\eeq
too.
	Moreover, away from bifurcation points of a solution branch $(\ag)\mapsto\eta_{_\Lbd}$ of 
\refeq{fixptEQinLambda},
\beq
\partial_\alpha\cP^{_\Lbd}_{\ag}[\eta_{_\Lbd}]
=
- {\tst\frac{1}{2}}\,
\int_{\!\Lbd}\int_{\!\Lbd} V(|\bfr-\tilde\bfr|)\eta_{_\Lbd}(\bfr)\eta_{_\Lbd}(\tilde\bfr)\, 
d^3r\, d^3\tilde{r}.
\label{PalphaDERIValongSOL}
\eeq
	Incidentally, we can also re-express this derivative in terms of the functionals for the total 
number of particles \refeq{Nfunc} and total energy \refeq{Efunc} of a density function $\eta$, viz.
$
\alpha\partial_\alpha\cP^{\Lbd}_{\ag}[\eta_{_\Lbd}]
=
{\tst\frac{3}{2}}\cN^{\Lbd}[\eta_{_\Lbd}] - \cE^{\Lbd}_\alpha[\eta_{_\Lbd}]
$,
but we will not use this rewriting.
	Now, since $(\alpha,\gamma_\gl^{\ssst\Lbd}(\alpha))$ is not a bifurcation point for the pointwise 
minimal solution, also the partial $\alpha$-derivative of $\cP^{_\Lbd}_{\ag}[\eta^{\ssst m}_{\ssst\Lbd}]$ 
exists at $(\alpha,\gamma_\gl^{\ssst\Lbd}(\alpha))$.
	Since at $(\alpha,\gamma_\gl^{\ssst\Lbd}(\alpha))$ a pointwise larger global maximizer of
$\cP^{_\Lbd}_{\ag}[\eta]$ exists, too, which belongs to a solution branch which continues
to carry the global maximizers for some right neighborhood of $(\alpha,\gamma_\gl^{\ssst\Lbd}(\alpha)$,
it now follows from \refeq{PalphaDERIValongSOL} together with $V<0$ that also the $\alpha$-derivative
of $\cP_{\ssst\Lbd}(\ag)$ jumps.
	The transition is therefore of first order in the sense of Ehrenfest.~\QED

	The maximizer $\eta^{\GC}_{\ssst\Lbd,g}$ is of pointwise minimal type 
${\eta}^{\ssst\,m}_{\ssst\Lbd}(\bfr)$ and called the {\it gas} solution.
	The pointwise larger maximizer $\eta^{\GC}_{\ssst\Lbd,\ell}$ 
we call the {\it liquid} solution \refeq{fixptEQinLambda}, although our proof does not establish that
$\eta^{\GC}_{\ssst\Lbd,\ell}$ is quasi-uniform up to a boundary layer; numerically$^{\cite{mkjkp}}$ this
is the case, though.

	Theorem 6.1 and its proof do {\it not} establish that $\eta^{\GC}_{\ssst\Lbd,\ell}$
is the pointwise largest hard-sphere fluid solution ${\eta^{\ssst M}_{\ssst\Lbd}}$; more generally
it does not establish that ${\eta^{\ssst M}_{\ssst\Lbd}}$ is the global maximizer of 
$\cP^{_\Lbd}_{\ag}[\eta]$ for $(\ag)$ satisfying \refeq{agTRIPLEsetLambda} and 
$\Lbd$ sufficiently large.
	The proof only shows that under these conditions the global maximizer of 
$\cP^{_\Lbd}_{\ag}[\eta]$ is a hard-sphere fluid solution which is larger than 
the pointwise minimal solution $\eta^{\ssst m}_{\ssst\Lbd}$.
	While this necessarily implies that 
${\eta^{\ssst M}_{\ssst\Lbd}}(\bfr)>\eta^{\ssst m}_{\ssst\Lbd}(\bfr)$ $\forall\ \bfr\in\Lbd$
it does not even imply that 
$\cP^{_\Lbd}_{\ag}[\eta^{\ssst M}_{\ssst\Lbd}]\geq \cP^{_\Lbd}_{\ag}[\eta^{\ssst m}_{\ssst\Lbd}]$.
	This result in turn holds in strict form and in more generality, as we show next.

\smallskip
\noindent
{\bf Proposition 6.2:} {\it If $(\ag)$ lies in the $\varpi(\ag)>0$ subset of 
$\Theta^{\alg}_{\bullet\rm \ssst f}(\norm{V}{1})$ and $\Lambda$ is sufficiently large so that
\refeq{PfctnlDIFFestB} holds, then
$\cP^{_\Lbd}_{\ag}[\eta^{\ssst M}_{\ssst\Lbd}]> \cP^{_\Lbd}_{\ag}[\eta^{\ssst m}_{\ssst\Lbd}]$.
	Thus ${\eta^{\ssst M}_{\ssst\Lbd}}(\bfr)>\eta^{\ssst m}_{\ssst\Lbd}(\bfr)$ $\forall\ \bfr\in\Lbd$,
and an unstable third hard-sphere fluid solution is sandwiched inbetween.}

\smallskip
{\it Proof:}
	Pick any $(\ag)$ in the $\varpi(\ag)>0$ subset of 
$\Theta^{\alg}_{\bullet\rm \ssst f}(\norm{V}{1})$, and suppose $\Lbd$ is big enough so 
that \refeq{PfctnlDIFFestB} holds.
	Consider the iteration
$\eta^{(n+1)} = \wp^\pr_{\bullet}\big(\gamma - (\aV*\eta^{(n)})_{_{\!\Lbd}}\big)$,
starting from $\eta^{(0)}\equiv \ol\eta^{\ssst M}_{\vdW}<\ol\eta_\fs^{\ssst <}$.
	By Proposition 3.1 and 3.2 it iterates downward to the pointwise maximal solution
$\eta^{\ssst M}_{\ssst\Lbd}$ in the truncated cone 
$C^0_{b,+}(\ol\Lbd) \cap \ol{\cal B}_{\ol\eta^{\ssst M}_{\vdW}}$.
	As remarked after Proposition 3.1, {\it a priori} the minimal and maximal solutions may
coincide, but having proved \refeq{PfctnlDIFFestB}, this cannot happen because the functional
$\cP^{_\Lbd}_{\ag}[\eta]$ {\it increases} along any monotone converging iteration sequence.
	Indeed, setting $\dot\eta^{(n)}\equiv\eta^{(n+1)}-\eta^{(n)}$ for the difference of any two
subsequent iterates, and $\ave{\eta}^{(n)}\equiv(\eta^{(n+1)}+\eta^{(n)})/2$ for their arithmetical mean, 
if $\dot\eta^{(n)}(\bfr)\neq 0\ \forall\ \bfr\in\Lbd$, then by the mean value theorem (applied
to the $\wp$ integral) and a binomial identity (applied to the $V$ double integral) we have
\beq
\cP^{_\Lbd}_{\ag}[\eta^{(n+1)}] - \cP^{_\Lbd}_{\ag}[\eta^{(n)}] 
=  
\int_{\!\Lbd}\big(\wp_\bullet^\pr\bigl(\wtilde\gamma^{(n)}(\bfr)\bigr) - \ave{\eta}^{(n)}(\bfr)\big)
\big(-\aV* \dot\eta^{(n)}\big)_{_{\!\Lbd}}\!\!(\bfr) d^3r > 0,
\label{PfctnlDIFFitera}
\eeq
where $\wtilde\gamma^{(n)}(\bfr)= \gamma - \left(\aV* \Ave{\eta}^{(n)}\right)_{{\!\Lbd}}\!\!(\bfr)$
and $\Ave{\eta}^{(n)}(\bfr)$ is a (bounded) continuous function which takes values between the
smaller and the larger one of the two iterates $\eta^{(n)}(\bfr)$ and $\eta^{(n+1)}(\bfr)$; the 
inequality in \refeq{PfctnlDIFFitera} holds because $-V>0$ and, for monotonic iterations,
$\wp_\bullet^\pr\bigl(\wtilde\gamma^{(n)}(\bfr)\bigr) - \ave{\eta}^{(n)}(\bfr)$ 
has the same overall sign as $\dot\eta^{(n)}(\bfr)$.
	So, 
$\cP^{_\Lbd}_{\ag}[\eta^{\ssst M}_{\ssst\Lbd}]> \cP^{_\Lbd}_{\ag}[\eta^{\ssst m}_{\ssst\Lbd}]$,
which implies that
${\eta^{\ssst M}_{\ssst\Lbd}}\not\equiv\eta^{\ssst m}_{\ssst\Lbd}$, 
and therefore
${\eta^{\ssst M}_{\ssst\Lbd}}(\bfr)>\eta^{\ssst m}_{\ssst\Lbd}(\bfr)$
$\forall\ \bfr\in\Lbd$.
	Thus we have established the existence of at least two distinct hard-sphere fluid solutions
in the $\varpi(\ag)>0$ subset of $\Theta^{\alg}_{\bullet\rm \ssst f}(\norm{V}{1})$ when
$\Lbd$ is big enough so that \refeq{PfctnlDIFFestB} holds.

	Moreover, in this case, as limits of respective monotone iterations associated with 
the G\^{a}teaux derivative of $\cP^{_\Lbd}_{\ag}$ both the pointwise minimal solution 
${\eta^{\ssst m}_{\ssst\Lbd}}(\bfr)$ and the pointwise maximal solution 
$\eta^{\ssst M}_{\ssst\Lbd}(\bfr)$ are locally $\cP$ stable, or at most $\cP$ indifferent in exceptional cases, 
	Save such exceptional cases the existence of a third, unstable (under iterations and in $\cP$ sense)
solution sandwiched between $\eta^{\ssst\,m}_{\ssst\Lbd}(\bfr)$ and $\eta^{\ssst M}_{\ssst\Lbd}(\bfr)$ 
now follows, via the mountain pass lemma, from the local $\cP$ stability of these two solutions 
and the strong $C^0_b(\ol\Lbd)$ differentiability of the functional $\cP^{_\Lbd}_{\ag}[\eta]$.~\QED

\smallskip
\noindent
{\bf Remark:} Note that the assumptions in Proposition 6.2 do not imply that the 
global maximizer of $\cP^{_\Lbd}_{\ag}$ is a hard-sphere fluid solution; for this 
we need to assume more, e.g. \refeq{agTRIPLEsetLambda} as in the proof of Theorem 6.1.
	For the proof of Proposition 6.2 we therefore could {\it not} assume that
${\eta^{\ssst M}_{\ssst\Lbd}}\not\equiv\eta^{\ssst m}_{\ssst\Lbd}$, but instead had to (and did)
prove it anew.~\eQED

\smallskip
\noindent
{\bf Remark:} Variants of Theorem 6.1 and Proposition 6.2 and their proofs hold with
$\Theta^{\alg}_{\bullet\rm \ssst f}(\norm{V}{1})$ replaced by
$\Theta^{\alg}_{\bullet\rm \ssst f}(\Phi_{_\Lbd})$ and
$\ol\eta_{\vdW}^{\ssst\, m}$ by $\ol{\eta}^{\ssst\, m}_{\ssst\Lbd} <  \ol\eta_{\wr}$, as well as
$\ol\eta_{\vdW}^{\ssst M}$ by $\ol{\eta}^{\ssst M}_{\ssst\Lbd} > \ol\eta_{\wr}$.
      In that case the $L^1(\RR^3)$ integrability of $V$ is not required so that we can even
add $A_{\N}V_{\N}$ to $\aV$.
	If $V=V_{\N}$, then \refeq{fixptEQinLambda} can have {\it many more than three} hard-sphere
fluid solutions for the same $(\ag)$, cf. Ref.\cite{bsmkks,mkRMP}, even though the algebraic fixed point 
problem \refeq{FIXetaPHIeq} has at most three solutions, still.
	This shows that na\"{\i}ve inferences from the algebraic fixed point problem \refeq{FIXetaPHIeq}
onto the integral equation \refeq{fixptEQinLambda} are not to be drawn.~\eQED

\smallskip
\noindent
{\bf Remark:} Neither the proof of Theorem 6.1 nor the one of Proposition 6.2 imply that 
the pointwise maximal hard-sphere fluid solution of \refeq{fixptEQinLambda} is the global maximizer 
of $\cP^{_\Lbd}_{\ag}$ when the pointwise minimal solution is not.
	If one could show, possibly by the index theorems of Ref.\cite{amannA,nussbaum}, that in
the no non-fluid solutions regime at most two locally stable hard-sphere fluid solutions exist, 
but otherwise arbitrarily many unstable hard-sphere fluid solutions, then 
the pointwise maximal solution is the global maximizer whenever the pointwise minimal solution 
is not, and vice versa.~\eQED

	In the $(\ag)$ region where the pointwise minimal solutions of \refeq{fixptEQinLambda}
are {not} globally $\cP$ stable, by their local $\cP$ stability they are $\cP$ metastable. 
	Such metastability regions usually terminate at a {\it spinodal line},
the location of which  in $(\ag)$ space can be estimated.
	Namely, on the one hand we already know that a metastable ${\eta}^{\ssst\, m}_{\ssst\Lbd} < \ol\eta_{\wr}$
exists for each $(\ag)\in\Theta^{\alg}_{\bullet\rm \ssst f}(\Phi_{{\RR^3}})\cap\{\varpi(\ag)>0\}$
whenever $\Lambda$ is sufficiently large.
	On the other hand, by our nonexistence result of any solution which would take only 
hard-sphere fluid values, we also know that neither $\alpha$ nor $\gamma$ can be arbitrarily big.
	Yet, for bounded $\Lbd$ we can get a more subtle result, valid even if $A_{\N}V_{\N}$ is
added to $\aV$.

\smallskip
\noindent
{\bf Proposition 6.3:}
{\it Let $\aV =  A_{\W}V_{\W}+ A_{\Y}V_{\Y}+ A_{\N}V_{\N}$, with $A_{\W}$, $A_{\Y}$, and $A_{\N}$ 
non-negative, and let $v_{_{\!\Lbd}}>0$ be the spectral radius of $-(V*\,\cdot\,)_{_{\!\Lbd}}$ for
$\Lambda\subset\RR^3$ bounded.
	Assume that
\beq
\alpha v_{_{\!\Lbd}} 
\geq 
\min_{\ol\eta\in(0,1)}g_{_2}^\prime(\ol\eta)= g_{_2}^\pr(\ol\eta_{\wr})\approx 21.20.
\label{alphavCONDITIONforSMALLsol}
\eeq
	Let $\ol{\eta}_{\ssst<}=\ol{\eta}_{\ssst<}(\alpha v_{_{\!\Lbd}})$ denote the 
smallest solution to the equation
\beq
\alpha v_{_{\!\Lbd}}
 = g_{_2}^\prime(\ol\eta),
\label{etaGROSSkleinEQredo}
\eeq
and set
\beq
\hat\gamma(\alpha v_{_{\!\Lbd}}) =   g_{_2}(\ol{\eta}_{\ssst<}) - \alpha v_{_{\!\Lbd}}\ol{\eta}_{\ssst<}.
\label{gammaKLEINunten}
\eeq
	Let $\eta(\bfr)\leq \ol\eta_{\wr} \equiv g_{_2}^{-1}\left(\gamma_{\wr} \right) \approx 0.130$
be a small fluid solution of \refeq{fixptEQinLambda}.
	Then $\gamma < \hat\gamma(\alpha v_{_{\!\Lbd}})$.}

{\it Proof:} For any container $\ol\Lbd\subset\RR^3$ with finite Lebesgue measure $|\Lbd|$, each
kernel $\aV =  A_{\W}V_{\W}+ A_{\Y}V_{\Y}+ A_{\N}V_{\N}$, with $A_{\W}$, $A_{\Y}$, and $A_{\N}$ 
non-negative, is a Hilbert--Schmidt kernel (i.e. $V\in L^2(\Lbd\times\Lbd)$), 
and so the positive definite operator $- (V *\,\cdot\,)_{_{\!\Lbd}}$ is a compact operator on $L^2(\Lbd)$.
	By the Krein-Rutman theorem, the spectral radius $v_{_{\!\Lbd}}>0$ of $-(V*\,\cdot\,)_{_{\!\Lbd}}$ 
is the largest eigenvalue of $- (V *\,\cdot\,)_{_{\!\Lbd}}$, its eigenspace non-degenerate, and the 
corresponding eigenfunction nonvanishing everywhere.

	Now let $\eta^{\ssst\,m}_{\ssst\Lbd}(\bfr)\leq \ol\eta_{\wr}$ once again be the
pointwise smallest solution of \refeq{fixptEQinLambda}. 
	Since r.h.s.\refeq{fixptEQinLambda} is acting as a strictly convex function on the truncated cone
$C^0_{b,+}\cap \ol{\cal B}_{{\ol\eta}_{\wr}}$, we can apply Fujita's strategy$^{\cite{fujita}}$ as generalized
by Amann.$^{\cite{amannB}}$
	Let $\xi(\bfr)$ be the eigenfunction of $- (V *\,\cdot\,)_{_{\!\Lbd}}$
for $v_{_{\!\Lbd}}$, normalized as probability density function so that it integrates to~1.
	Let $\llangle\, \cdot\, \rrangle$ be the averaging functional w.r.t. $\xi$.
	Taking now the average of \refeq{fixptEQinLambda} with $\llangle\, \cdot\, \rrangle$ and
using Jensen's inequality, we find
\beq
\llangle \eta^{\ssst\,m}_{\ssst\Lbd}\rrangle
>
\wp_{\bullet}^\prime(\gamma +\alpha v_{_{\!\Lbd}}\llangle\eta^{\ssst\,m}_{\ssst\Lbd}\rrangle),
\label{fujitaINEQ}
\eeq
which cannot be satisfied if $\gamma\geq \hat\gamma(\alpha v_{_{\!\Lbd}})$.~\QED

	Proposition 6.3 implies the existence of a 
$\gamma_*^{_{\!\Lbd}}(\alpha)<\hat\gamma(\alpha v_{_{\!\Lbd}})$ such that 
\beq
\eta^{\ssst\,m}_{\ssst\Lbd,*}
:=
\lim_{\gamma\uparrow\gamma_*^{_{\!\Lbd}}(\alpha)}{\eta}^{\ssst\,m}_{\ssst\Lbd}|_\alpha,
\label{etaSTAR}
\eeq
which exists and solves \refeq{fixptEQinLambda}, satisfies the following alternative:
{either} $\|{\eta^{\ssst\, m}_{\ssst\Lbd,*}}\|_{^{C^0_b(\Lbd)}}\!\!=\ol\eta_{\wr}$, and then the
constant-$\alpha$ section of $(\ag)\mapsto{\eta}^{\ssst\,m}_{\ssst\Lbd}$ {\it may} continuously 
extend to $\gamma> \gamma_*^{_{\!\Lbd}}(\alpha)$, 
only then with ${\eta}^{\ssst\, m}_{\ssst\Lbd}(\bfr)\not\leq  \ol\eta_{\wr}$ for some $\bfr\in \Lbd$;
{or} $\|{\eta^{\ssst\, m}_{\ssst\Lbd,*}}\|_{^{C^0_b(\Lbd)}}\!\!<\ol\eta_{\wr}$, and then the 
constant-$\alpha$ section of $(\ag)\mapsto{\eta}^{\ssst\, m}_{\ssst\Lbd}$ is discontinuous, 
i.e. left and right limits of the map 
$\gamma\mapsto{\eta}^{\ssst\, m}_{\ssst\Lbd}|_{\alpha}$ at $\gamma_*^{_{\!\Lbd}}(\alpha)$ disagree.
	When the second alternative holds, the metastability region for the pointwise minimal gas-type
solutions terminates at the curve $\alpha\mapsto\gamma_*^{_{\!\Lbd}}(\alpha)$, which in this case
is the {\it spinodal curve for supersaturated vapor}.

	The computation of the function $\gamma_*^{_{\!\Lbd}}(\alpha)$ seems generally possible only
implicitly, through studying the family of pointwise minimal solutions ${\eta}^{\ssst\,m}_{\ssst\Lbd}(\bfr)$.
	However, its upper bound $\hat\gamma(\alpha v_{_{\!\Lbd}})$ can be easily computed when the 
spectral radius $v_{_{\!\Lbd}}$ is known. 
	The latter can be computed to any desired degree of precision by monotone iteration.

\noindent
{\bf Lemma 6.4:} {\it The spectral radius $v_{_{\!\Lbd}}$ of the positive operator 
$- (V *\,\cdot\,)_{_{\!\Lbd}}$ is given by
\beq
\ln v_{_{\!\Lbd}}
=
\lim_{n\to\infty}{\tst\frac{1}{n}}\ln\|{\xi^{(n)}}\|_{L^2(\Lbd)}
\label{specRAD}
\eeq
where 
$\xi^{(n+1)} = - (V*\xi^{(n)})_{_{\!\Lbd}}$ with $\xi^{(0)}\equiv 1$.
	Moreover, it is bounded by
\beq
- \llangle (V*1)_{_{\!\Lbd}}\rrangle
	\leq
v_{_{\!\Lbd}}
	< 
\norm{V(|\,\cdot\,|)}{L^1(\Lbd)}.
\label{specRADbounds}
\eeq}

{\it Proof of Lemma 6.4:}
	The identity \refeq{specRAD} is just the formula for the largest Lyapunov exponent 
($=\ln v_{_{\!\Lbd}}$) of our linear iteration, viewed as a dynamical system.
	The lower bound in \refeq{specRADbounds} is obtained from the Ritz type variational principle for 
$v_{_{\!\Lbd}}$ with the help of the trial function $\xi(\bfr)\equiv |\Lbd|^{-1/2}$, the upper 
bound by applying the sharp Young inequality$^{\cite{liebloss}}$ to this variational principle.~\QED

\noindent
{\bf Remark:} Since the evaluation of \refeq{specRAD} or \refeq{specRADbounds} in Lemma 6.4 may only be feasible 
numerically for general $\Lambda$, the following weaker estimates are of interest,
too:
\beq
\norm{V(|\,\cdot\,|)}{L^1(\Lbd)}
\leq 
\norm{V*1}{C^0_b(\Lbd)}
\leq 
\norm{V*1}{C^0_b(B_R)}
\leq 
\norm{V}{1},
\label{specRADestimatesRIGHT}
\eeq
with $|B_R| = |\Lbd|$.
	The first  upper bound ($=\Phi_{_\Lbd}$) is elementary.
	The second upper bound ($ = \norm{V(|\,\cdot\,|)}{L^1(B_R)}$)
follows from a simple radial rearrangement inequality; this bound
is explicitly evaluated in Appendix A.
	The third upper bound is again elementary but nontrivial {\it only if} $A_{\N}=0$.
	In that case, when $\Lbd\nearrow\RR^3$ in the sense of Fisher$^{\cite{fisher}}$, 
then both the lower bound and upper bound in \refeq{specRADbounds} converge to 
$\norm{V}{1} = (A_{\W}{\pi^2 / 4\varkappa^3} + A_{\Y}{4 \pi / \kappa^2})/\alpha$,
which therefore is the $\Lbd\nearrow\RR^3$ limit of the spectral radius.~\eQED

	We end this subsection with the observation that in the (at least) triplicity region of
hard-sphere fluid solutions of \refeq{fixptEQinLambda} where the pointwise smallest solutions
${\eta}^{\ssst\,m}_{\ssst\Lbd}(\bfr)$ are globally $\cP$ stable, the locally $\cP$ 
stable pointwise maximal solutions ${\eta}^{\ssst M}_{\ssst\Lbd}(\bfr)$ are $\cP$ metastable. 
	Clearly, the extent of this region has a lower $\gamma$ bound because of Corollary 4.5, and
a lower $\alpha$ bound because of Corollary 4.3.
	Moreover, since we imposed the sufficient condition \refeq{NOnonFLUIDcondition} for all solutions
to be fluid, we also have an upper bound on $\alpha$ given by $\alpha\norm{V}{1}<28.9$ (approximately);
indeed, if this bound is violated by $\alpha$, then for no choice of $\gamma$ satisfying 
\refeq{NOnonFLUIDcondition} (with $\norm{V}{1}$ in place of $\Phi_{_\Lbd}$)
is $(\ag)\in \Theta^{\alg}_{\bullet\rm \ssst f}(\norm{V}{1})$.
	Recall that this bound can be improved when better control over the solid branch of 
$\gamma\mapsto\wp_\bullet(\gamma)$ becomes available.
	The accurate determination of the boundary of this metastability region is generally feasible
only indirectly through numerical solution of the problem. 
	Numerical solution$^{\cite{mkjkp}}$ reveals that in this metastability region the fluid assumes the
shape of a giant liquid drop barely separated from the container walls by a thin layer of vapor. 
	This metastability boundary is a {\it spinodal curve} which represents the {\it smallest
giant liquid drop which can be contained in $\Lbd$ given $(\ag)$}.

\noindent
\section{$\cF$ STABILITY AND THE VAPOR $\lra$ DROP PHASE TRANSITION}
\smallskip
	In this section we discuss the thermodynamic stability of our non-uniform hard-sphere 
fluid solutions in bounded containers for the thermodynamic contact condition ``heat reservoir,''
i.e. what we called $\cF$ stability.

	Substituting the Carnahan--Starling approximation 
$p_{\bullf}^{\phantom{b}}(\ol\eta)=(g_{\ssst 1}^{-1}\circ g_{\ssst 2})(\ol\eta)$
into the entropy functional \refeq{Sfunc}, one can carry out the $\ol\eta$ integration to obtain
\beq
\cS^{_\Lbd}_{\bullf}[\eta] = {\tst\frac{11}{2}}\cN^{_\Lbd}[\eta]
 -
\int_{\!\Lbd} \eta(\bfr)
\left[\ln \eta(\bfr) +  \frac{3-2\eta(\bfr)}{ (1-\eta(\bfr))^2}\right] d^3r,
\label{SfuncCS}
\eeq
so for a hard-sphere fluid we have, within the Carnahan--Starling approximation,
\beq
\cF^{_\Lbd}_\alpha[\eta] 
= 
\cE^{_\Lbd}_{\alpha}[\eta] - \cS^{_\Lbd}_{\bullf}[\eta].
\label{FfuncBULL}
\eeq
	In the following, when we speak of a hard-sphere fluid density function $\eta(\bfr)$
as being globally or locally $\cF$ stable, we mean a global or local minimizer of \refeq{FfuncBULL}
under the constraint
\beq
\cN^{_\Lbd}[\eta]=N.
\label{NisNconstraint}
\eeq

\smallskip
\noindent
{\bf Proposition 7.1:}
{\it Any hard-sphere fluid density function $\eta(\bfr)$ which is locally or globally $\cF$ stable 
under the constraint \refeq{NisNconstraint} is a solution $\eta_{_\Lbd}(\bfr)$ of \refeq{fixptEQinLambda} for the 
same $\alpha$ but with $\gamma$ determined by the constraint \refeq{NisNconstraint}.
	Any globally (locally) $\cP$ stable solution of \refeq{fixptEQinLambda} is also 
globally (locally) $\cF$ stable.}

{\it Proof of Proposition 7.1:} Since the free energy functional \refeq{FfuncBULL} is strongly 
$C^0_b(\Lambda)$ differentiable and coercive, its local and global minimizers satisfy
the Euler--Lagrange equation for \refeq{FfuncBULL} under the constraint \refeq{NisNconstraint}.
	When this constraint is taken into account in the usual manner with the help of a
Lagrange multiplier $\gamma$, viz. the ``null functional'' $\cN^{_\Lbd}[\eta]-N$ is multiplied
by $\gamma$ and then subtracted from $\cF^{_\Lbd}_\alpha[\eta]$ and $\eta$ then varied unconditionally, 
a straightforward calculation gives the Euler--Lagrange equation
\beq
-\gamma+ (\aV*\eta)_{_\Lbd} + g_{\ssst 2}(\eta) = 0,
\label{ELeqF}
\eeq
which is precisely our \refeq{fixptEQinLambda} with the Carnahan--Starling approximation for the
hard-sphere fluid equation of state.
	So the local and global minimizers of \refeq{FfuncBULL} under the constraint \refeq{NisNconstraint}
are among the solutions of \refeq{fixptEQinLambda}, with $\gamma$ determined by \refeq{NisNconstraint}.

	As for the global $\cF$ stability, we note that 
the maximum $P_{_\Lbd}(\alpha,\gamma)$ of the pressure functional $\cP^{_\Lbd}_{\ag}[\eta]$
is also given by the Legendre--Fenchel transform$^{\cite{ellis}}$ (sending $N\to\gamma$)
\beq
P_{_\Lbd}(\alpha,\gamma) 
= 
\sup_{N}\,\Bigl\{\gamma  N - F_{_\Lbd}(\aN)\Bigr\},
\label{FtoPtransform}
\eeq
which, upon recalling the definition of $F_{_\Lbd}(\aN)$, can be rephrased as the variational principle
\beq
P_{_\Lbd}(\alpha,\gamma) 
= 
\sup_{\eta}\,\Bigl\{\gamma  \cN^{_\Lbd}[\eta]- \cF^{_\Lbd}_\alpha[\eta]\Bigr\}.
\label{FfuncToPtransform}
\eeq
	Since $P_{_\Lbd}(\alpha,\gamma)$ is given by the variational principle \refeq{vpPinLambda},
which {\it defines} the globally $\cP$ stable solutions $\eta_{_\Lbd}^{\GC}$ of \refeq{fixptEQinLambda}, 
it follows that each $\eta_{_\Lbd}^{\GC}$ also saturates the variational principle \refeq{FfuncToPtransform} 
--- for suppose to the contrary that 
$\gamma\cN^{_\Lbd}[\eta_{_\Lbd}^{\GC}]-\cF^{_\Lbd}_\alpha[\eta_{_\Lbd}^{\GC}]<P_{_\Lbd}(\alpha,\gamma)$,
then 
$\gamma\cN^{_\Lbd}[\eta_{_\Lbd}^{\GC}]-\cF^{_\Lbd}_\alpha[\eta_{_\Lbd}^{\GC}]<\cP^{_\Lbd}_{\ag}[\eta_{_\Lbd}^{\GC}]$, 
which we show to be impossible.
	Indeed, after some straightforward manipulations of \refeq{Ffunc}, given by the difference of \refeq{Efunc} and 
\refeq{Sfunc}, using {\it only} \refeq{fixptEQinLambda} in its reverse form \refeq{ELeqF}, and recalling \refeq{Pfctnl}, one finds that 
\beq
\gamma  \cN^{_\Lbd}[\eta_{_\Lbd}]- \cF^{_\Lbd}_\alpha[\eta_{_\Lbd}]
= 
\cP^{_\Lbd}_{\ag}[\eta_{_\Lbd}]
\label{gammaNminusFequalsP}
\eeq 
for {\it any} solution $\eta_{_\Lbd}$ of \refeq{fixptEQinLambda}, not just those which are
globally $\cP$ stable.
	So each globally $\cP$ stable $\eta_{_\Lbd}^{\GC}$ also saturates the variational principle \refeq{FtoPtransform}.
	But then each $\eta_{_\Lbd}^{\GC}$ also saturates the variational principle for global $\cF$ stability, 
with $N= \cN^{_\Lbd}[\eta_{_\Lbd}^{\GC}]$.

	A variation on the theme of this global stability proof gives the local $\cF$ stability
of locally $\cP$ stable solutions $\eta_{_\Lbd}$. 
	The proof again uses the identity \refeq{gammaNminusFequalsP},
valid for any solution $\eta_{_\Lbd}$ of \refeq{fixptEQinLambda}, but replaces 
$P_{_\Lbd}(\alpha,\gamma)$ in  \refeq{FtoPtransform} by $\cP^{_\Lbd}_{\ag}[\eta_{_\Lbd}]$
and the variation in the global Legendre--Fenchel transform  \refeq{FtoPtransform} by a restriction to 
a neighborhood of $\eta_{_\Lbd}$.~\QED

\smallskip\noindent
{\bf Remark:} 
Incidentally, \refeq{FtoPtransform} guarantees that $P_{_\Lbd}(\alpha,\gamma)$ is convex in $\gamma$.~\eQED

\smallskip\noindent
{\bf Remark:} The infimum of $\cF^{_\Lbd}_\alpha[\eta]$ 
under the constraint $N= \cN^{_\Lbd}[\eta]$ is {\it generally not} given by
the Legendre--Fenchel transform of $\cP^{_\Lbd}_{\ag}[\eta]$ (sending $\gamma\to N$).
	Put differently, $\cP^{_\Lbd}_{\ag}[\eta]$ and $\cF^{_\Lbd}_{\alpha}[\eta]$
are generally not convex duals of each other.
	As a spin-off of this, the reversal of the stability conclusion in Proposition 7.1 is
not allowed; i.e., not all globally (locally) $\cF$ stable solutions of \refeq{fixptEQinLambda} 
are also globally (locally) $\cP$ stable.~\eQED

\smallskip\noindent
{\bf Remark:}
	When $V\in L^1(\RR^3)$ we can take the infinite volume limit.
	The Legendre--Fenchel type variational principle \refeq{FfuncToPtransform} then becomes the 
thermodynamic variational principle$^{\cite{gatespenroseA,gatespenroseB}}$
\beq
\pi_\bullf(\alpha,\gamma) 
= 
\sup_{\eta\in C^0_b(\RR^3)}
\left\{\gamma \langle\eta\rangle_{_{\RR^3}} - f_\alpha[\eta]\right\},
\label{fTOpTRANSF}
\eeq
where, for each $\eta\in C^0_b(\RR^3)$, 
\beq
\langle\eta\rangle_{_{\RR^3}}
:=
\lim_{\Lbd\uparrow\RR^3} |\Lbd|^{-1}\cN^{_\Lbd}[\eta],
\label{nFUNCofNONCONSTeta}
\eeq
\beq
f_\alpha[\eta]
:=
\lim_{\Lbd\uparrow\RR^3} |\Lbd|^{-1}\cF^{_\Lbd}_\alpha[\eta].
\label{fFUNCofNONCONSTeta}
\eeq
	For spatially uniform density functions $\eta(\bfr)\equiv \ol\eta$, the 
functional \refeq{fFUNCofNONCONSTeta} for the free-energy density~:~temperature ratio of $\eta$
takes the simple van der Waals form
\beq
f_\alpha[\ol\eta]
=
\ol\eta g_{\ssst 2}(\ol\eta) - g_{\ssst 1}(\ol\eta) - {\tst\frac{1}{2}}\alpha\norm{V}{1}\ol\eta^2,
\label{fFUNCofCONSTeta}
\eeq
here with the local hard-sphere thermodynamics treated in the Carnahan--Starling approximation.
	Note that the infimum of \refeq{fFUNCofNONCONSTeta} under the constraint 
$\langle\eta\rangle_{_{\RR^3}}=\ol\eta$, denoted
\beq
f_\bullf(\alpha,\ol\eta)
:=
\inf_{\eta\in C^0_b(\RR^3)}\bigl\{f_\alpha[\eta]\,\big|\, \langle\eta\rangle_{_{\RR^3}}=\ol\eta \bigr\},
\label{fOFalphaUNDetaDEF}
\eeq
is {\it generally not} achieved by a constant function $\bfr\mapsto\ol\eta$,
but by a piecewise constant $\eta^{\PC}_{\vdW}(\bfr)\not\in C^0_b(\RR^3)$ 
(PC for ``petit canonical'' coming in handy), and satisfies the van der Waals--Maxwell formula
\beq
f_\bullf(\alpha,\ol\eta)
=
CH\{f_\alpha[\ol\eta]\},
\label{fOFalphaUNDeta}
\eeq
where $CH\{\,\cdot\,\}$ denotes the {\it convex hull}. 
	This formula for the thermodynamic free energy density~:~temperature ratio can be 
rigorously obtained by taking a 
van der Waals (Kac) limit with infinitely far ranging, infinitely weak pair interactions {\it after}
the thermodynamic (infinite volume) limit$^{\cite{fisher,ruelleA,ruelle}}$ has been taken, see
Ref.\cite{kacetal} for one-dimensional, Ref.\cite{JOELuOLI} for three-dimensional systems, both with Kac 
interations, and see  Refs.\cite{gatespenroseA,gatespenroseB} for larger classes of interactions.
	Formula \refeq{fOFalphaUNDeta} means that in the thermodynamic limit the van der Waals 
free energy density~:~temperature ratio is itself a Legendre transform, namely the Legendre 
transform w.r.t. $\gamma$ of the convex function $\gamma\mapsto \pi_\bullf(\alpha,\gamma)$.
	This example of {\it equivalence of ensembles} at the level of the 
thermodynamic functions free energy and pressure is {\it generally false}
for the finite volume functionals, as noted in the previous remark.
	For certain types of $V$ non-equivalence of ensembles in van der Waals-type theory occurs
even in a {\it coupled limit} of infinite volume  and infinitely far ranging, infinitely weak pair 
interactions.$^{\cite{gatespenroseA}}$~\eQED

	In our numerical investigations$^{\cite{mkjkp}}$ of the canonical non-uniform van der Waals theory 
for $V=V_{\W}$ and $\Lbd$ a ball of radius $50\varkappa^{-1}$ we found $\cF$ stable liquid drops surrounded
by a vapor atmosphere when $\alpha\norm{V}{1}\approx 31.2$.
	Note that for this $\alpha\norm{V}{1}$ value the sufficient no-non-fluid solutions condition 
$\gamma +\alpha\norm{V}{1}{\ol\eta}_{\ssst fcc}^{\ssst\,cp} \leq \gamma_\fs$ is violated for the 
relevant $\gamma$ values used to compute hard-sphere fluid solutions, yet 
solid solutions can nevertheless be ruled out with our more refined knowledge of the solid branch. 
	We remark that the droplet solutions that we found were all situated in the $(\aN)$-region where
$F_{_\Lbd}(\aN)$ displays the ``wrong'' convexity which is ``jumped over'' by the grand canonical phase transition.

	Interestingly enough, our numerical studies$^{\cite{mkjkp,mkRMP,bsmkks}}$  revealed that the change 
from quasi-uniform vapor state to droplet state in the canonical ensemble is not gradual but involves 
another first-order phase transition which is embedded in the $(\aN)$-region
 ``jumped'' by the grand canonical phase transition; see also Refs.\cite{mkRMP,bsmkks}.
	While a complete analytical proof of all the interesting details revealed by our numerical studies seems 
futile, our next theorem does assert the existence of a petit canonical first-order transition between a 
quasi-uniform vaporous  and a strongly non-uniform free energy minimizer with same $(\aN)$.
	It generalizes our proof$^{\cite{mkJSPa}}$ from regularized Newtonian interactions and simpler equation 
of state of the perfect gas to the hard-sphere equation of state and interactions which include the shorter
range van der Waals and Yukawa interactions.
	To state our theorem, we recall that for $(\ag)=(\alpha,\gamma_\gl^{\ssst\Lbd}(\alpha))$ the functional 
$\cP^{_\Lbd}_{\ag}[\eta]$ has two global maximizers in the hard-sphere fluid regime, the 
gas solution ${\eta}^{\GC}_{\ssst\Lbd,g}(\bfr)$ of the pointwise minimal type 
${\eta}^{\ssst\,m}_{\ssst\Lbd}(\bfr)$ and the liquid solution
${\eta}^{\GC}_{\ssst\Lbd,\ell}(\bfr)>{\eta}^{\GC}_{\ssst\Lbd,g}(\bfr)$.

\smallskip
\noindent
{\bf Theorem 7.2:} {\it	Let $\ol\Lbd$ be a convex container of macroscopic proportions, i.e.
$\diam(\Lbd)\gg 1$ and $\diam(\Lbd)/|\Lbd|^{1/3}= O(1)$, such that a ball domain $B$ of volume 
$|B| = |\Lbd|/8$ is a strict subset of $\Lambda$.
	Let $V\in L^1(\RR^3)$ and let $\alpha\norm{V}{1}\in (31-\eps,31+\eps)$.
	Then $\alpha$ is in the domain of the map $\alpha\mapsto \gamma=\gamma_\gl^{\ssst\Lbd}(\alpha)$, 
the grand canonical gas vs. liquid phase transition curve, and there exists an $N^{_\Lbd}_{\vd}(\alpha)\in 
	\big[\cN^{_\Lbd}[{\eta}^{\GC}_{\ssst\Lbd,g}],\cN^{_\Lbd}[{\eta}^{\GC}_{\ssst\Lbd,\ell}]\big)$
for which two distinct solutions of \refeq{fixptEQinLambda} minimize 
$\cF^{_\Lbd}_\alpha[\eta]$ globally under the constraint $\cN^{_\Lbd}[\eta]=N^{_\Lbd}_{\vd}(\alpha)$.
	The transition between the global $\cF$ minimizers is of first order in the sense of Ehrenfest, 
i.e. the partial derivatives of $(\alpha, N)\!\mapsto\! F_{_\Lbd} (\alpha, N)$ 
jump at the canonical phase transition curve $\alpha\!\mapsto\! N^{_\Lbd}_{\vd}(\alpha)$, 
provided the radial symmetric decreasing rearrangements of the two $\cF$ minimizers intersect at a single 
level value.}

\noindent
{\bf Remark:} 	One of the two global $\cF$ minimizers is of the pointwise minimal (given $\ag$) type 
${\eta}^{\ssst\,m}_{\ssst\Lbd}(\bfr)$ and represents the supersaturated vapor phase.
	Our proof will suggest that the other one is very likely of droplet type, having a high 
density (liquid) core surrounded by a low density (vapor) atmosphere, but our proof does not conclusively
establish the existence of such a solution type for \refeq{fixptEQinLambda}.
	Numerically$^{\cite{mkjkp}}$ such solutions do exist, and they do intersect the equal-$N$ vapor
solution at a single level value.~\eQED

\noindent
{\bf Remark:} The global $\cF$ minimizer representing a supersaturated vapor phase is $\cP$ metastable.
	The global $\cF$ minimizer representing a liquid drop surrounded by a vapor atmosphere is $\cP$ 
unstable.~\eQED

\noindent
{\bf Remark:} A compromise between taking the infinite volume limit $\Lbd\to\RR^3$ (in the sense of, e.g., Fisher)
and to work in a strictly finite domain is to work in $\RR^3$ but with the restriction that all densities 
are periodic w.r.t. to the 3-torus $\TT^3=\RR^3/\ZZ^3$.
	The canonical non-uniform van der Waals theory in $\TT^3$ was studied most recently in 
Ref.\cite{carlenETalB}.
	For strictly finite-range pair interactions and the equation of state of the lattice gas model, 
they proved the existence of minimizers of the so-called Gates-Lebowitz-Penrose free-energy functional
which, when restricted to a single fundamental cell, 
look like a liquid drop surrounded by a vapor atmosphere in a finite container $\ol\Lambda$.~\eQED

\smallskip
{\it Proof of Theorem 7.2:} 
	For our proof we apply the strategy of Ref.\cite{mkJSPa} where a canonical phase transition of the type as 
stated in Theorem 7.2 is proved for $V$ given by a class of regularizations of $V_{\N}$ and $\wp$ given by 
the perfect gas law --- except for the Ehrenfest part concerning the $\alpha$ derivative of $F_{_\Lbd}(\aN)$, 
for which we follow Ref.\cite{mkCMPa}.
	We note though that the more rapid decay of $V_{\W}(r)$ and $V_{\Y}(r)$ with $r$ and the more complicated 
local thermodynamics require much more delicate estimates in the current proof.
	In particular, the condition of Proposition 3.3 for the absence of not-all-fluid solutions is too
restrictive now, so that our full knowledge of the solid branch $\wp_{\bulls}(\gamma)$ will be brought in.
	With that we now begin our proof.

	First, by evaluating the algebraic van der Waals problem for our hard-sphere fluid one easily 
verifies that when $\alpha\norm{V}{1} \approx 31$ and $\Lambda$ is large, then $\alpha$ is in the domain of 
$\gamma_\gl^{\ssst\Lbd}(\alpha)$, the grand canonical gas vs. liquid phase transition curve described 
in Theorem 6.1.
	Just draw a family of straight parallel lines with slope $\approx 31$ into the second figure and note that the whole 
fluid triplicity $\gamma$-interval for this $\alpha$ value can be covered without intersecting the solid branch.
	Note that the condition of Proposition 3.3 for the absence of not-all-fluid solutions is violated, though.

	Moreover, since the two $\cP$ maximizers along the grand canonical phase transition curve are 
pointwise ordered,
${\eta}^{\GC}_{\ssst\Lbd,g}(\bfr)<{\eta}^{\GC}_{\ssst\Lbd,\ell}(\bfr)\ \forall\bfr\in\Lbd$, we conclude that
$\cN^{_\Lbd}[{\eta}^{\GC}_{\ssst\Lbd,g}]<\cN^{_\Lbd}[{\eta}^{\GC}_{\ssst\Lbd,\ell}]$ so that
the half-open $N$-interval stated in Theorem 7.2 is not empty.

	We next recall Proposition 7.1, according to which any globally $\cP$ stable solution of \refeq{fixptEQinLambda} 
is also globally $\cF$ stable.
	So in particular ${\eta}^{\GC}_{\ssst\Lbd,g}\equiv{\eta}^{\PC}_{\ssst\Lbd}$ is a global minimizer
of $\cF^{_\Lbd}_{\alpha}[\eta]$ under the constraint $N= \cN^{_\Lbd}[{\eta}^{\GC}_{\ssst\Lbd,g}]$.
	This global $\cF$ minimizer ${\eta}^{\PC}_{\ssst\Lbd}\equiv{\eta}^{\GC}_{\ssst\Lbd,g}$ 
is of the pointwise minimal (given $\ag$) type ${\eta}^{\ssst\,m}_{\ssst\Lbd}(\bfr)$ and 
situated on a fixed $\alpha$-section of the solution branch $(\aN)\mapsto {\eta}^{\ssst\,m}_{\ssst\Lbd}(\bfr)$
of quasi-uniform small solutions ($<\ol\eta_\wr$), given in terms of the invertible parameter representation 
$\gamma\mapsto {\eta}^{\ssst\,m}_{\ssst\Lbd}(\bfr)$ and $\gamma\mapsto N=\cN^{_\Lbd}[{\eta}^{\ssst\,m}_{\ssst\Lbd}]$
for each $\alpha$.
	This representation is well-defined because for fixed $\alpha$ the map 
$\gamma\mapsto {\eta}^{\ssst\,m}_{\ssst\Lbd}(\bfr)$ is pointwise increasing and 
(by the implicit function theorem) continuous (even continuously differentiable) 
in the half-open $\gamma$ interval $(-\infty,\hat\gamma_{\bullet\rm \ssst f}(\alpha \norm{V}{1})]$
containing $\gamma_\gl^{\ssst\Lbd}(\alpha)$, where $\hat\gamma_{\bullet\rm \ssst f}(\alpha \norm{V}{1})$
is the right limit \refeq{gammaRIGHTbull} of the van der Waals triplicity region 
$\Theta^{\alg}_{\bullet\rm \ssst f}(\norm{V}{1})$ of the hard-sphere fluid 
(recall our Propositions 3.1 and 3.4).
	This $\gamma$ interval maps into the $N$ interval 
$(0,\hat{N}(\alpha)]$, where $\hat{N}(\alpha):= \cN^{_\Lbd}[\hat{\eta}^{\ssst\,m}_{\ssst\Lbd}]$
and $\hat{\eta}^{\ssst\,m}_{\ssst\Lbd}$ is the pointwise minimal solution of \refeq{fixptEQinLambda}
for $\gamma = \hat\gamma_{\bullet\rm \ssst f}(\alpha \norm{V}{1})$.
	Note that $\cN^{_\Lbd}[{\eta}^{\GC}_{\ssst\Lbd,g}]<\hat{N}(\alpha)$, by Theorem 6.1.
	Moreover, since by Theorem 6.1 for each $\alpha$ in the domain of $\gamma_\gl^{\ssst\Lbd}$ the
map $\gamma\mapsto {\eta}^{\ssst\,m}_{\ssst\Lbd}(\bfr)$ furnishes the unique globally $\cP$ stable solution
${\eta}^{\GC}_{\ssst\Lbd}(\bfr)$ for each $\gamma<\gamma_\gl^{\ssst\Lbd}(\alpha)$, by Proposition 7.1 for 
each admissible $\alpha$ as stated in Theorem 7.2 the map $N\mapsto {\eta}^{\ssst\,m}_{\ssst\Lbd}(\bfr)$ 
then furnishes a globally $\cF$ stable solution ${\eta}^{\PC}_{\ssst\Lbd}(\bfr)$
for each $N\in\big(0, \cN^{_\Lbd}[{\eta}^{\GC}_{\ssst\Lbd,g}]\big]$,
with ${\eta}^{\PC}_{\ssst\Lbd}(\bfr)\equiv{\eta}^{\GC}_{\ssst\Lbd,g}(\bfr)$ at 
$N= \cN^{_\Lbd}[{\eta}^{\GC}_{\ssst\Lbd,g}]$.
	
	Furthermore, by the monotonicity of $\gamma\mapsto N= \cN^{_\Lbd}[{\eta}^{\ssst\,m}_{\ssst\Lbd}]$
and the pointwise minimality of ${\eta}^{\ssst\,m}_{\ssst\Lbd}$ (for given $\alpha,\gamma$), and by its 
uniqueness as solution of \refeq{fixptEQinLambda} for $\gamma< \gamma_{\!\ssst\Lbd}(\alpha)$ (see Corollary 4.5),
the fixed-$\alpha$ section of the branch of locally stable gas solutions 
$N\mapsto {\eta}^{\ssst\,m}_{\ssst\Lbd}(\bfr)$ furnishes 
the {\it unique} globally $\cF$ stable solution for each $N< \cN^{_\Lbd}[{\eta}^{\ssst \alpha}_{\ssst\Lbd}]$, 
where ${\eta}^{\ssst\alpha}_{\ssst\Lbd}$ is ${\eta}^{\ssst\,m}_{\ssst\Lbd}(\bfr)$ for 
$(\ag)=(\alpha,\gamma_{\!\ssst\Lbd}(\alpha))$.
	Now let $N^{_\Lbd}_{\vd}(\alpha)$ be the supremum over $N\in(0,\hat{N}(\alpha)]$ for which 
$N\mapsto {\eta}^{\ssst\,m}_{\ssst\Lbd}(\bfr)(<\ol\eta_\wr)$ furnishes the {\it unique globally $\cF$ stable 
solution for each} $N<N^{_\Lbd}_{\vd}(\alpha)$.
	Clearly, ${N}^{_\Lbd}_{\vd}(\alpha)\geq \cN^{_\Lbd}[{\eta}^{\ssst \alpha}_{\ssst\Lbd}]$.
	We also define $_*N^{_\Lbd}_{\vd}(\alpha)$ as the infimum over $N\in(0,\hat{N}(\alpha)]$ for which 
$N\mapsto {\eta}^{\ssst\,m}_{\ssst\Lbd}(\bfr)(<\ol\eta_\wr)$ {\it is not globally $\cF$ stable
for each $N\in(_*N^{_\Lbd}_{\vd}(\alpha)\,,\,_*N^{_\Lbd}_{\vd}(\alpha)+\eps)$ for some} $\eps>0$.
	Clearly, $_*N^{_\Lbd}_{\vd}(\alpha)\geq {N}^{_\Lbd}_{\vd}(\alpha)$.
	We show:	

\ \ \ (a) $_*N^{_\Lbd}_{\vd}(\alpha) = {N}^{_\Lbd}_{\vd}(\alpha)$;

\ \ \ (b) $\cN^{_\Lbd}[{\eta}^{\GC}_{\ssst\Lbd,g}]\leq N^{_\Lbd}_{\vd}(\alpha)<\hat{N}(\alpha)$;

\ \ \ (c) at $N^{_\Lbd}_{\vd}(\alpha)$ the global $\cF$ minimizer is not unique.

	To prove claim (a) suppose that $_*N^{_\Lbd}_{\vd}(\alpha)> {N}^{_\Lbd}_{\vd}(\alpha)$.
	Then from the definitions of $_*N^{_\Lbd}_{\vd}(\alpha)$ and ${N}^{_\Lbd}_{\vd}(\alpha)$ it
follows that $N\mapsto {\eta}^{\ssst\,m}_{\ssst\Lbd}(\bfr)(<\ol\eta_\wr)$ is globally $\cF$ stable for
all $N\in (N^{_\Lbd}_{\vd}(\alpha)\,,\, _*{N}^{_\Lbd}_{\vd}(\alpha))$, but at least one other global
$\cF$ minimizer exists for each such $N$ (given $\alpha$).
	It suffices to assume that exactly one other  global
$\cF$ minimizer $\eta^{\PC}_{_\Lbd}$ exists for each such $N$ (given $\alpha$).
	But then these two minimizers of $\cF^{_\Lbd}_{\alpha}[{\eta}]$ not only have the same 
$\cF^{_\Lbd}_{\alpha}$ value for each such $N$, also the derivative of 
$N\mapsto \cF^{_\Lbd}_{\alpha}[{\eta}_{\ssst\Lbd}]$ is the same for both minimizers.
	Now it follows right away from \refeq{gammaNminusFequalsP} that along a constant-$\alpha$ section of 
a solution branch of \refeq{fixptEQinLambda} we have
\beq
\partial_N\cF^{_\Lbd}_{\alpha}[\eta_{_\Lbd}]
=
\Gamma[\eta_{_\Lbd}],
\label{FrhoDERIValongSOL}
\eeq
where $\Gamma[\eta_{_\Lbd}]$ is the $\gamma$-value for which $\eta_{_\Lbd}$ solves \refeq{fixptEQinLambda}.
	So both hypothetical global $\cF$ minimizers solve \refeq{fixptEQinLambda} for the {\it same} 
$(\ag)=(\alpha,\Gamma[\eta^{\ssst m}_{_\Lbd}])$, but since $\eta^{\ssst m}_{_\Lbd}$ is the pointwise 
minimal solution at $(\ag)=(\alpha,\Gamma[\eta^{\ssst m}_{_\Lbd}])$, it follows that
$\cN^{_\Lbd}[{\eta}^{\PC}_{\ssst\Lbd}]>\cN^{_\Lbd}[{\eta}^{\ssst m}_{\ssst\Lbd}]$, which contradicts the
hypothesis that both density functions are global minimizers of $\cF^{_\Lbd}_{\alpha}[\eta]$ for the same
$(\aN)$.
	This proves that $_*N^{_\Lbd}_{\vd}(\alpha) = {N}^{_\Lbd}_{\vd}(\alpha)$;
incidentally, the same type of argument also proves that 
$N\mapsto {\eta}^{\ssst\,m}_{\ssst\Lbd}(\bfr)(<\ol\eta_\wr)$ {\it is not globally $\cF$ stable
for} $N>N^{_\Lbd}_{\vd}(\alpha)$.

	As for (b), to prove the first inequality we recall that 
by Proposition 7.1 and Theorem 6.1 we know that ${\eta}^{\ssst\,m}_{\ssst\Lbd}$ 
is a global minimizer of $\cF^{_\Lbd}_{\alpha}$ {\it for all} $N\in(0,\cN^{_\Lbd}[{\eta}^{\GC}_{\ssst\Lbd,g}])$.
	Now suppose that beside ${\eta}^{\ssst\,m}_{\ssst\Lbd}$
there exists a second global minimizer of $\cF^{_\Lbd}_{\alpha}$ for some $N_*$ satisfying
$\cN^{_\Lbd}[{\eta}^{\ssst \alpha}_{\ssst\Lbd}]\leq N_*<\cN^{_\Lbd}[{\eta}^{\GC}_{\ssst\Lbd,g}]$.
	But then, by the proof of point (a), it follows that 
$N\mapsto {\eta}^{\ssst\,m}_{\ssst\Lbd}(\bfr)(<\ol\eta_\wr)$ is not globally $\cF$ stable
for each $N>N_*$, which is a contradiction.
	This proves the first inequality in (b).

	To prove the second inequality in (b) we show that for $N= \hat{N}(\alpha)$ a droplet 
type density function has lower free energy than the vapor type solution $\hat{\eta}^{\ssst\,m}_{\ssst\Lbd}$ 
which defines $\hat{N}(\alpha)$, and by continuity this will be so also for some left neighborhood of
$\hat{N}(\alpha)$. 
	Since for each $N$ we will only compare densities which all integrate 
to the given $N$, we can ignore the $\cN^{_\Lbd}$ functionals in $\cF^{_\Lbd}_{\alpha}$ and compare  
\beq
\cA^{_\Lbd}_{\alpha}[\eta] = 
{\tst\frac{1}{2}}\int_{\!\Lbd}\int_{\!\Lbd}\aV(|\bfr-\tilde\bfr|) \eta(\bfr)\eta(\tilde\bfr)
\, d^3r\, d^3\tilde{r}
 +\!
\int_{\!\Lbd} \eta(\bfr)
\left[\ln \eta(\bfr) +  \frac{3-2\eta(\bfr)}{ (1-\eta(\bfr))^2}\right] d^3r
\label{AfuncCS}
\eeq
evaluated with ${\eta}^{\ssst\,m}_{\ssst\Lbd}$ versus its evaluation with some droplet like
density of the same $N$.

	First, let $N=\hat{N}(\alpha)$.
	Recalling the upper bound on the gas solutions 
${\eta}^{\ssst\,m}_{\ssst\Lbd}(\bfr)\leq \ol{\eta}_{\vdW}^{\ssst\,m}$ where the spatially 
uniform van der Waals solution is for the same $\gamma$ as ${\eta}^{\ssst\,m}_{\ssst\Lbd}$, 
we have in particular $\hat{\eta}^{\ssst\,m}_{\ssst\Lbd}(\bfr)\leq \hat{\ol{\eta}}_{\vdW}^{\ssst\,m}$.
	We apply this bound to the interaction integral, plus use the estimate 
$\llangle{(V*1)_{_\Lbd}}\rrangle_{\ssst\Lbd} > - \norm{V}{1}$.
	We also apply Jensen's inequality w.r.t. uniform 
spatial average to the (negative of the) entropy integral, noting the convexity of the map 
$x\mapsto x\ln x + x(3-2x)/(1-x)^2$, and use that 
$\llangle{\eta^{\ssst\,m}_{\ssst\Lbd}}\rrangle_{\ssst\Lbd}=N/|\Lbd|$ for all $N\in(0,\hat{N}(\alpha)]$.
	This yields the lower bound on $\cA^{_\Lbd}_{\alpha}[\hat\eta^{\ssst\,m}_{\ssst\Lbd}]$ given by
\beq
|\Lbd|^{-1}\cA^{_\Lbd}_{\alpha}[\hat\eta^{\ssst\,m}_{\ssst\Lbd}] \geq
- {\tst\frac{1}{2}}\alpha \norm{V}{1} \hat{\ol{\eta}}_{\vdW}^{\ssst\,m}{}^{\!\!2}
+ 
\tst\frac{\hat{N}}{|\Lbd|}
\biggl(\ln \tst\frac{\hat{N}}{|\Lbd|} 
+
\tst\frac{3-2\sst{\hat{N}/|\Lbd|}}{ \big(1-\sst{\hat{N}/|\Lbd|}\big)^2}
\biggr),
\label{AlowBOUNDgas}
\eeq
where we wrote $\hat{N}$ for $\hat{N}(\alpha)$. 
	Also $\hat{\ol{\eta}}_{\vdW}^{\ssst\,m}$ is a function of $\alpha$, and by \refeq{AVEolETAvdWMINUEolETAest} 
and $\llangle{\hat\eta^{\ssst\,m}_{\ssst\Lbd}}\rrangle_{\ssst\Lbd}=\hat{N}/|\Lbd|$ we have that
\beq
\hat{\ol\eta}^{\ssst\,m}_{\vdW}
=
\tst\frac{\hat{N}}{ |\Lbd|}\left(1+O[\oslash(\Lbd)^{-2/3}]\right),
\label{etavdWestimate}
\eeq
so that up to a correction of $O[\diam(\Lbd)^{-2/3}]$, 
we can substitute $\hat{N}/|\Lbd|$ for $\hat{\ol\eta}^{\ssst\,m}_{\vdW}$, or the other way round.
	On the other hand, by inserting into $\cA^{_\Lbd}_{\alpha}\left[{\eta}\right]$
a trial density of the type ``liquid drop with vapor atmosphere'' which integrates to
$\hat{N}$, we get an upper bound on the minimum of the reduced free energy functional 
for $\hat{N}$, given $\alpha$. 
	It  suffices to choose a spherically symmetric trial density without atmosphere,
\beq
\hat\eta_{_d}(\bfr) = \tst\frac{\hat{N}}{ |B|} \chi_{_{\ssst B}}(\bfr),
\label{dropCHI}
\eeq
where $B\subset \Lbd$ is a ball whose volume is determined by setting 
$\hat{N}/|B| = \hat{\ol\eta}_{\vdW}^{\ssst M}$, the pointwise largest van der Waals solution at  
$\hat\gamma(\alpha \norm{V}{1})$.
	This yields the upper bound
\bea
|\Lbd|^{-1} \inf_\eta \cA^{_\Lbd}_{\alpha}[\eta]  
\!\!\!&\leq& |\Lbd|^{-1}\cA^{_\Lbd}_{\alpha}[\hat\eta_{_d}]\nonumber\\
\!\!\!&=&
{\tst\frac{1}{2}}\alpha \llangle{(V*1)_{_B}}\rrangle_{\ssst B}\!\tst\frac{\hat{N}}{|B|}
\tst\frac{\hat{N}}{ |\Lbd|}
+ 
\tst\frac{\hat{N}}{ |\Lbd|}
\biggl(\ln \tst\frac{\hat{N}}{|B|} 
+
\tst\frac{3-2\sst{\hat{N}/|B|}}{\big(1-\sst{\hat{N}/|B|}\big)^2}
\biggr).
\label{AupBOUNDdrop}
\eea
	Subtracting \refeq{AlowBOUNDgas} from \refeq{AupBOUNDdrop} and using (see Appendix A.) that
\beq
\llangle{(V*1)_{_B}}\rrangle_{\ssst B} = -\norm{V}{1}(1-O[1/\diam(B)]),
\label{aveVest}
\eeq
and anticipating that $|\Lbd|/|B|= O[1]$ so that we can neglect the $O[1/\diam(B)]$ correction, 
we find that the upper bound \refeq{AupBOUNDdrop} on the infimum of $\cA$ is lower than the lower 
bound \refeq{AlowBOUNDgas} on the free energy of the gas solution at $N=\hat{N}$ when
\beq
\alpha \norm{V}{1} 
>
2\; \frac{\ln \tst\frac{|\Lbd|}{|B|} 
+ \tst\frac{3-2\sst{\hat{N}/|B|} }{ \big(1-\sst{\hat{N}/|B|}\big)^2}
- \tst\frac{3-2\sst{\hat{N}/|\Lbd|} }{ \big(1-\sst{\hat{N}/|\Lbd|}\big)^2}
}{
\tst\frac{\hat{N}}{ |B|}-\tst\frac{\hat{N}}{ |\Lbd|}},
\label{dropHATcriterion}
\eeq
up to a correction of $O[\diam(\Lbd)^{-2/3}]$.
	The criterion \refeq{dropHATcriterion} can be re-expressed as
\beq
\alpha \norm{V}{1} 
>
2\; \frac{\ln \frac{\hat{\ol\eta}_{\vdW}^{\ssst M}}{ \hat{\ol\eta}_{\vdW}^{\ssst m}} 
+ \tst\frac{3-2\sst{\hat{\ol\eta}_{\vdW}^{\ssst M}} }{ \big(1-\sst{\hat{\ol\eta}_{\vdW}^{\ssst M}}\big)^2}
- \tst\frac{3-2\sst{ \hat{\ol\eta}_{\vdW}^{\ssst m}} }{ \big(1-\sst{ \hat{\ol\eta}_{\vdW}^{\ssst m}}\big)^2}
}{
\tst{\hat{\ol\eta}_{\vdW}^{\ssst M}}-\tst{ \hat{\ol\eta}_{\vdW}^{\ssst m}}},
\label{dropHATcriterionVDW}
\eeq
up to a correction of $O[\diam(\Lbd)^{-2/3}]$.
	Now, for $\alpha\norm{V}{1}=31$ as stipulated in Theorem 7.2, the ratio 
$\hat{\ol\eta}_{\vdW}^{\ssst M}: \hat{\ol\eta}_{\vdW}^{\ssst m}\approx 9$, 
with $\hat{\ol\eta}_{\vdW}^{\ssst M}\approx 0.41$ and $\hat{\ol\eta}_{\vdW}^{\ssst m} \approx 0.045$.
	These values yield r.h.s.\refeq{dropHATcriterionVDW}$\approx 28.75< 31$, and also
$|\Lbd|/|B|\approx 9>8$ so that $B$ fits into $\Lbd$, satisfying the hypothesis of Theorem 7.2.
	This proves that the droplet-type density function has a lower free-energy~:~temperature ratio 
than the quasi-uniform vaporous solution of the same $N=\hat{N}(\alpha)$ for $\alpha\norm{V}{1}=31$.
	By continuity the regime where droplet type densities have lower free-energy~:~temperature ratio
than the quasi-uniform solutions extends to an open neighborhood 
of the chosen $\alpha$ for the corresponding $\hat{N}(\alpha)$.

	Second, by continuity again, the same conclusion also extends to some open left neighborhood of 
$\hat{N}(\alpha)$ for each such $\alpha$ in the neighborhood of the chosen $\alpha\norm{V}{1}=31$.
 	This completes the proof of claim (b). 

	Continuity and closedness arguments for the solution curves prove claim (c) in a similar fashion
of reasoning as used in the proof of Theorem 6.1.
	Here we also use Proposition 7.1, according to which 
the solution  ${\eta}^{\ssst\,m}_{\ssst\Lbd}$ is also locally $\cF$ stable
for each $\alpha$ and $N$ in the domain of the map $N\mapsto {\eta}^{\ssst\,m}_{\ssst\Lbd}$ because
the pointwise minimal (given $(\ag)$) solutions ${\eta}^{\ssst\,m}_{\ssst\Lbd}$ 
are locally $\cP$ stable (we ignore the exceptional cases when  ${\eta}^{\ssst\,m}_{\ssst\Lbd}$ is locally
$\cP$ indifferent).
	By its local $\cF$ stability, no bifurcation off of this gas branch occurs for $N < \hat{N}(\alpha)$, 
in particular not for $N=\cN^{_\Lbd}[{\eta}^{\GC}_{\ssst\Lbd,g}]\,\big(< \hat{N}(\alpha)\big)$.

	Henceforth we will write ${\eta}^{\ssst m}_{\ssst\Lbd}\equiv {\eta}^{\PC}_{\ssst\Lbd,v}$ for the 
quasi-uniform, vaporous $\cF$ minimizer, and  ${\eta}_{\ssst\Lbd}\equiv {\eta}^{\PC}_{\ssst\Lbd,d}$ for 
the non-quasi-uniform minimizer of (presumed) droplet type; we say ``presumed'' for,  strictly speaking 
we haven't shown that it is a droplet, although our above proof and the numerical evidence$^{\cite{mkjkp}}$ 
suggests it is.
	We remark that even though in our proof we worked with a trial droplet without atmosphere, 
any droplet type minimizer of $\cF^{_\Lbd}_{\alpha}[{\eta}]$ must solve \refeq{fixptEQinLambda}
and therefore must have a low-density atmosphere, as r.h.s.\refeq{fixptEQinLambda} is bounded away from $0$.

	Finally, we show that the canonical vapor versus droplet transition is of first order in the sense of
Ehrenfest, for which we need the hypothesized, yet empirically suggested, level intersection property.

	First, in our proof of point (a) above we showed that the 
constant-$\alpha$ derivatives of $N\mapsto\cF^{_\Lbd}_{\alpha}[\eta_{_\Lbd}]$ at $N=N_{\vd}(\alpha)$ cannot be 
the same for the quasi-uniform minimizer ${\eta}^{\PC}_{\ssst\Lbd,v}$ and for the non-quasi-uniform 
minimizer ${\eta}^{\PC}_{\ssst\Lbd,d}$.
	So our proof of point (a) above already proves that the constant-$\alpha$ derivative of 
$N\mapsto F_{_\Lbd}(\aN)$ is discontinuous at $N= N_{\vd}(\alpha)$.
	In fact, the constant-$\alpha$ derivative of $N\mapsto F_{_\Lbd}(\aN)$ jumps down when $N$ increases.
	This follows from \refeq{FrhoDERIValongSOL} and the monotonicity properties of the pointwise 
minimal solution branch of \refeq{fixptEQinLambda}.
	For suppose that $\Gamma[\eta^{\PC}_{_\Lbd,d}] \geq \Gamma[\eta^{\PC}_{_\Lbd,v}]$.
	Then, since $\eta^{\PC}_{_\Lbd,d}\not\equiv \eta^{\ssst m}_{_\Lbd,d}$, which denotes
the pointwise minimal solution at $(\alpha,\Gamma[\eta^{\PC}_{_\Lbd,d}])$, we have
$\cN^{_\Lbd}[\eta^{\PC}_{_\Lbd,d}]> \cN^{_\Lbd}[\eta^{\ssst m}_{_\Lbd,d}]\geq \cN^{_\Lbd}[\eta^{\PC}_{_\Lbd,v}]$,  
which contradicts the fact that $\cN^{_\Lbd}[\eta^{\PC}_{_\Lbd,d}] = \cN^{_\Lbd}[\eta^{\PC}_{_\Lbd,v}]$.
	Note that for this part of our proof of the Ehrenfest property we did not need to invoke that
the two global $\cF$ minimizers intersect only at a single level value.

	Next, to prove that the constant-$N$ map $\alpha\mapsto \cF^{_\Lbd}_{\alpha}[\eta_{_\Lbd}]$ has
a kink at $\alpha_{\vd}(N)$, where $N\mapsto \alpha = \alpha_{\vd}(N)$ is the local inverse function to 
the curve $\alpha\mapsto N= N_{\vd}(\alpha)$, and which exists locally unless the latter is constant, 
we adapt, and improve on, the strategy of Ref.\cite{mkCMPa}.
	First, being $\cF$ minimizers, the free-energy~:~temperature ratio is the same for the quasi-uniform 
minimizer ${\eta}^{\PC}_{\ssst\Lbd,v}$ and for the non-quasi-uniform minimizer ${\eta}^{\PC}_{\ssst\Lbd,d}$, 
i.e. $\cF^{_\Lbd}_{\alpha}[\eta^{\PC}_{_\Lbd,v}]=\cF^{_\Lbd}_{\alpha}[\eta^{\PC}_{_\Lbd,d}]$.
	Recalling the identity \refeq{gammaNminusFequalsP} we see that this implies the equality 
\beq
N(\Gamma[\eta^{\PC}_{_\Lbd,d}]-\Gamma[\eta^{\PC}_{_\Lbd,v}])
=
\cP^{_\Lbd}_{\alpha,\Gamma[\eta^{\PC}_{_\Lbd,d}]}[\eta^{\PC}_{_\Lbd,d}]
-
\cP^{_\Lbd}_{\alpha,\Gamma[\eta^{\PC}_{_\Lbd,v}]}[\eta^{\PC}_{_\Lbd,v}],
\label{NgammaDIFFisPdiff}
\eeq
where $\Gamma[\eta_{_\Lbd}]$ is the $\gamma$-value for which $\eta_{_\Lbd}$ is a solution of \refeq{fixptEQinLambda}.
	Now suppose that, for given $N$,  the derivative of 
$\alpha\mapsto \cF^{_\Lbd}_{\alpha}[\eta^{\PC}_{_\Lbd}]$ at $\alpha =  \alpha_{\vd}(N)$ 
is the same for both minimizers.
	By the implicit function theorem these derivatives exist along the constant-$N$ sections of the
solution branches of \refeq{fixptEQinLambda}; at  $\alpha =  \alpha_{\vd}(N)$ the derivative for $\eta^{\PC}_{_\Lbd,d}$
may have to be read as right-derivative, though generically it will be a derivative.
	Using the variational principle for $F_{_\Lbd}(\aN)$ and the definition of $\cF^{_\Lbd}_{\alpha}[\eta]$
we obtain\footnote{This formula may cause some temporary consternation, for a
		thermodynamic free-energy~:~temperature ratio should satisfy the ``thermodynamic relation'' 
		$\partial_\alpha\cF^{_\Lbd}_{\alpha}[\eta_{_\Lbd}] = \cE^{_\Lbd}_{\alpha}[\eta_{_\Lbd}]$
		along the globally $\cF$ stable solution branch of \refeq{fixptEQinLambda}, with
		$\cE^{_\Lbd}_{\alpha}[\eta_{_\Lbd}]$ given in \refeq{Efunc}.
		This puzzle is resolved by noticing that in our strictly classical setup we have 
		omitted even the minimal amount of quantum mechanics normally injected into classical
		statistical mechanics with the help of the de Broglie wavelength, as per 
		``normalization'' of the entropy and chemical potential; see our Appendix C.}
\beq
\partial_\alpha\cF^{_\Lbd}_{\alpha}[\eta_{_\Lbd}]
=
 {\tst\frac{1}{2}}\,
\int_{\!\Lbd}\int_{\!\Lbd} V(|\bfr-\tilde\bfr|)\eta_{_\Lbd}(\bfr)\eta_{_\Lbd}(\tilde\bfr)\, 
d^3r\, d^3\tilde{r}.
\label{FalphaDERIValongSOL}
\eeq
	Incidentally, for our $V<0$ \refeq{FalphaDERIValongSOL} shows that $\alpha\mapsto F_{_\Lbd}(\aN)$ is monotonic 
decreasing, but we won't need that.
	By \refeq{FalphaDERIValongSOL} we conclude that the hypothesized equality
$\partial_\alpha\cF^{_\Lbd}_{\alpha}[\eta^{\PC}_{_\Lbd,v}]
=
\partial_\alpha\cF^{_\Lbd}_{\alpha}[\eta^{\PC}_{_\Lbd,d}]$ for the two $\cF^{_\Lbd}_\alpha$ minimizers at the
same $(\aN)$ implies that their potential energy~:~temperature ratios  are the same, too.
	Inspection of the definition  \refeq{Pfctnl} of the pressure~:~temperature ratio functional 
of a density function $\eta$ now reveals
\beq
\cP^{_\Lbd}_{\ag}[\eta^{\PC}_{_\Lbd,d}] - \cP^{_\Lbd}_{\ag}[\eta^{\PC}_{_\Lbd,v}] 
=  
\int_{\!\Lbd} \big[p_\bullf^{\phantom{b}}\bigl(\eta^{\PC}_{_\Lbd,d}(\bfr)\bigr) 
-
		p_\bullf^{\phantom{b}}\bigl(\eta^{\PC}_{_\Lbd,v}(\bfr)\bigr)\bigr] d^3r, 
\label{PdiffEQUALpDIFF}
\eeq
where $p_\bullf^{\phantom{b}}\bigl(\eta\bigr)=g_{_1}(\eta)$.
	By inserting \refeq{PdiffEQUALpDIFF} into \refeq{NgammaDIFFisPdiff} we obtain the equality 
\beq
N(\Gamma[\eta^{\PC}_{_\Lbd,d}]-\Gamma[\eta^{\PC}_{_\Lbd,v}])
=
\int_{\!\Lbd} \big[p_\bullf^{\phantom{b}}\bigl(\eta^{\PC}_{_\Lbd,d}(\bfr)\bigr) 
-
		p_\bullf^{\phantom{b}}\bigl(\eta^{\PC}_{_\Lbd,v}(\bfr)\bigr)\bigr] d^3r.
\label{NgammaDIFFisDIFFp}
\eeq

	Recall that at the end of the first part of the Ehrenfest proof, i.e. of the discontinuity of 
the constant-$\alpha$ derivative of $N\mapsto \cF^{_\Lbd}_{\alpha}[\eta^{\PC}_{_\Lbd}]$, we showed that
the l.h.s.\refeq{NgammaDIFFisDIFFp}$<0$.
	We now complete our proof that the 
constant-$N$ derivative of $\alpha\mapsto \cF^{_\Lbd}_{\alpha}[\eta^{\PC}_{_\Lbd}]$ is discontinuous at
the canonical phase transition curve  {\it provided the radial symmetric decreasing rearrangements of the
two $\cF$ minimizers intersect at only a single density value} $x$ by showing that 
r.h.s.\refeq{NgammaDIFFisDIFFp}$\geq 0$ {\it under this provision.}

	For any density function $\eta(\bfr)\in C^0_b(\ol\Lbd)$, let
$^\star\eta(|\bfr|)$ denote its radially symmetric decreasing equimeasurable rearrangement supported in the
ball $B$ of volume $|B|=|\Lbd|$. 
	Then, using that
\beq
\int_{\!\Lbd} f\bigl(\eta(\bfr)\bigr) d^3r	
=
\int_B	f\bigl(^\star\eta(|\bfr|)\bigr)d^3r	
\label{equimeasureINT}
\eeq
for any continuous function $f:\RR\to\RR$, and using the mean-value theorem (with 
$\backslash\!\!\!\eta(|\bfr|)$ sandwiched between the two $\cF$ minimizers uniquely determined),
and invoking the hypothesized level intersection property of the two minimizers, we find 
\bea
\int_{\!\Lbd} \big[p_\bullf^{\phantom{b}}\bigl(\eta^{\PC}_{_\Lbd,d}(\bfr)\bigr) 
-
		p_\bullf^{\phantom{b}}\bigl(\eta^{\PC}_{_\Lbd,v}(\bfr)\bigr)\bigr] d^3r
&=&
\int_{\!B} \big[p_\bullf^{\phantom{b}}\bigl(^\star\eta^{\PC}_{_\Lbd,d}(|\bfr|)\bigr) 
-
		p_\bullf^{\phantom{b}}\bigl(^\star\eta^{\PC}_{_\Lbd,v}(|\bfr|)\bigr)\bigr] d^3r
\nonumber\\
&=&
\int_{\!B} p_\bullf^{\prime}\bigl(\backslash\!\!\!\eta(|\bfr|)\bigr)
 \big[^\star\eta^{\PC}_{_\Lbd,d}(|\bfr|) -^\star\eta^{\PC}_{_\Lbd,v}(|\bfr|)\bigr] d^3r
\nonumber\\
&\geq& 
 p_\bullf^{\prime}\bigl(x\bigr)
\int_{\!B}\big[^\star\eta^{\PC}_{_\Lbd,d}(|\bfr|) -^\star\eta^{\PC}_{_\Lbd,v}(|\bfr|)\bigr] d^3r
\nonumber\\
&=&0, 
\label{DIFFpESTIMATE}
\eea
where the inequality is readily proved by estimating the penultimate integral separately on the positive
and negative parts of its integrand.
	This already concludes the second part of the proof of the Ehrenfest property, but we supplement
the result by showing that the constant-$N$ derivative of $\alpha\mapsto F_{_\Lbd}(\aN)$ jumps down when 
$\alpha$ increases.

	Indeed, since for $\alpha< \alpha_{\vd}(N)$ the vaporous solution is the unique global $\cF$ minimizer,
suppose now that at $\alpha = \alpha_{\vd}(N)$ we have
$\partial_\alpha\cF^{_\Lbd}_{\alpha}[\eta^{\PC}_{_\Lbd,v}]
<
\partial_\alpha\cF^{_\Lbd}_{\alpha}[\eta^{\PC}_{_\Lbd,d}]$.
	But then by straightforward adaptation of our proof of the discontinuity of $\partial_\alpha F_{_\Lbd}(\aN)$
we now conclude that
\beq
0> N(\Gamma[\eta^{\PC}_{_\Lbd,d}]-\Gamma[\eta^{\PC}_{_\Lbd,v}])
=
\cP^{_\Lbd}_{\alpha,\Gamma[\eta^{\PC}_{_\Lbd,d}]}[\eta^{\PC}_{_\Lbd,d}]
-
\cP^{_\Lbd}_{\alpha,\Gamma[\eta^{\PC}_{_\Lbd,v}]}[\eta^{\PC}_{_\Lbd,v}]
>0,
\label{NgammaDIFFisPdiffREDO}
\eeq
and so the derivative $\partial_\alpha F_{_\Lbd}(\aN)$ must jump down at $\alpha = \alpha_{\vd}(N)$.

	The proof of Theorem 7.2 is complete.~\QED

	We end section VII with some comments regarding the proof of Theorem 7.2.

\smallskip
\noindent
{\bf Remark:} Since our proof of the Ehrenfest property relies on the provision of the 
level intersection property of the two global $\cF$ minimizers, it seems prudent to have a backup strategy 
just in case the provision turns out not to hold for non-spherical containers; also for spherical
containers it hasn't been proven yet, although in that case there is numerical evidence in its favor.$^{\cite{mkjkp}}$
	The following argument does not rely on the provision of Theorem 7.2, and could 
be completed with some sharper estimates.

	Namely, we use that $\eta^{\PC}_{_\Lbd,v}<\ol\eta_{\wr}$ is the pointwise minimal solution for 
$(\ag)=(\alpha,\Gamma[\eta^{\PC}_{_\Lbd,v}])$, so that for the same $(\ag)$ we have the bound
$\eta^{\PC}_{_\Lbd,v}\leq\ol\eta^{\ssst\,m}_{\vdW}(<\ol\eta_{\wr})$.
	Moreover, $\eta^{\PC}_{_\Lbd,v}$ is quasi-uniform in the sense that it is nearly constant except 
for a small boundary layer near $\partial\Lbd$, viz. (recalling \refeq{AVEolETAvdWMINUEolETAest})
\beq
\ol\eta^{\ssst\,m}_{\vdW}
-\llangle\eta^{\ssst m}_{\ssst\Lbd}\rrangle_{_{\!\Lbd}}
\leq 
O[\oslash(\Lbd)^{-2/3}],
\label{AVEolETAvdWMINUEolETAestREDO}
\eeq
where $\langle\eta^{\PC}_{_\Lbd,v}\rangle_{_{\!\!\Lbd}}= N/|\Lbd|$ is the uniform mean over $\Lbd$. 
	Furthermore, since both $\cF$ minimizers have equal ``mass'' $N$, we have the identity
\beq
\langle\eta^{\PC}_{_\Lbd,v}\rangle_{_{\!\!\Lbd}} 
=
\langle\eta^{\PC}_{_\Lbd,d}\rangle_{_{\!\!\Lbd}}.
\eeq
	Since $x\mapsto p_\bullf^{\phantom{b}}\bigl(x\bigr)$ is a positive, increasing, convex function,
dividing l.h.s.\refeq{DIFFpESTIMATE} by $|\Lbd|$ and applying
Jensen's inequality combined with these identities and estimates yields
\bea
\left\langle p_\bullf^{\phantom{b}}\bigl(\eta^{\PC}_{_\Lbd,d}(\bfr)\bigr) \right\rangle_{\!\Lbd} 
-
\left\langle p_\bullf^{\phantom{b}}\bigl(\eta^{\PC}_{_\Lbd,v}(\bfr)\bigr)\right\rangle_{\!\Lbd} 
\!&\geq&
p_\bullf^{\phantom{b}}\bigl(\left\langle \eta^{\PC}_{_\Lbd,d}\right\rangle_{_{\!\!\Lbd}}\bigr) 
-
p_\bullf^{\phantom{b}}\bigl(\ol\eta^{\ssst\,m}_{\vdW}\bigr) 
\nonumber\\
\!&=&
p_\bullf^{\phantom{b}}\bigl(\left\langle \eta^{\PC}_{_\Lbd,v}\right\rangle_{_{\!\!\Lbd}}\bigr) 
-
p_\bullf^{\phantom{b}}\bigl(\ol\eta^{\ssst\,m}_{\vdW}\bigr) 
\nonumber\\
\!&\geq& - O[\oslash(\Lbd)^{-2/3}],
\label{DIFFpAPPROX}
\eea
where the small error, due to the boundary layer effects, goes to zero as $\Lbd$ goes to $\RR^3$, but is
not identically zero.
	
	Thus, to complete this proof one would need to show that the difference 
$\Gamma[\eta^{\PC}_{_\Lbd,d}]-\Gamma[\eta^{\PC}_{_\Lbd,v}] < 0$ stays away from zero;
alternatively, the proof would be completed if one could control the error term in 
Jensen's inequality to the effect that
$\langle p_\bullf^{\phantom{b}}\bigl(\eta^{\PC}_{_\Lbd,d}(\bfr)\bigr) \rangle_{\!\Lbd} 
-
p_\bullf^{\phantom{b}}\bigl(\langle \eta^{\PC}_{_\Lbd,d}\rangle_{_{\!\!\Lbd}}\bigr) 
\geq C>0$ independently of sufficiently large $\Lbd$.~\eQED

\smallskip
\noindent
{\bf Remark:} In the limit of vanishing hard-sphere volume the local thermodynamics goes over into 
that of the perfect gas. 
	In this case the hard-sphere pressure~:~temperature ratio as function of $\ol\eta$ is simply the 
identity map, and then r.h.s.\refeq{NgammaDIFFisDIFFp} is identically zero, and the proof of the discontinuity 
of the constant-$N$ derivative of $\alpha\mapsto \cF^{_\Lbd}_{\alpha}[\eta^{\PC}_{_\Lbd}]$ is complete, then.
	This in fact is the proof of Ref.\cite{mkCMPa}.~\eQED

\smallskip
\noindent
{\bf Remark:} We note that the jumping down of the constant-$N$ derivative of 
$\alpha\mapsto F_{_\Lbd}(\aN)$ at $\alpha = \alpha_{\vd}(N)$ also implies (for our $V<0$) that
$\cE^{_\Lbd}_{\alpha}[\eta^{\PC}_{_\Lbd,v}]>\cE^{_\Lbd}_{\alpha}[\eta^{\PC}_{_\Lbd,d}]$, which
is seen by noting \refeq{FalphaDERIValongSOL} and recalling the definition \refeq{Efunc} of the energy, 
keeping in mind the constancy of $\cN^{_\Lbd}[\eta^{\PC}_{_\Lbd}]$ at $\alpha = \alpha_{\vd}(N)$.
	With the jumping down of the energy~:~temperature ratio at $\alpha = \alpha_{\vd}(N)$, the
constancy of $\cF^{_\Lbd}_{\alpha}[\eta{\PC}_{_\Lbd}]$ at $\alpha = \alpha_{\vd}(N)$ then in turn implies that 
$\cS^{_\Lbd}_{\bullf}[\eta^{\PC}_{_\Lbd,v}] > \cS^{_\Lbd}_{\bullf}[\eta^{\PC}_{_\Lbd,d}]$, i.e. the 
entropy jumps down also.~\eQED

\smallskip
\noindent
{\bf Remark:} Our proof of the canonical phase transition reveals two metastability regions
in its  $(\aN)$ neighborhood in which locally $\cF$ stable solutions of \refeq{fixptEQinLambda}
exist.
	Also these metastability regions should terminate at their {\it spinodal lines}.
	We have to leave the determination of their location in $(\aN)$ space for some future work.~\eQED

\noindent
{\bf ACKNOWLEDGEMENT:} This work was initially supported through GRANT NAG3-1414 to J.K.P. from NASA, 
and subsequently by NSF GRANT DMS 96-23220 to M.K.; it was finalized while M.K. was supported by 
the NSF through GRANT DMS 08-07705.
	M.K. thanks P. Mironescu for communicating his unpublished results, and him and E.A. Carlen
for interesting discussions.
	The authors thank the anonymous referee for pointing out Refs.\cite{albertibellettiniA,albertibellettiniB} 
and for comments which prompted us to improve our introduction.

\newpage
\section*{APPENDIX}
\medskip

\noindent
{A. THE INTERACTION INTEGRALS IN SPHERICAL GEOMETRY}
\medskip

	In spherical geometry we are in the position to obtain several explicit results.

\medskip\noindent
{\bf Lemma A.1:} {\it Let $\Lbd = B_R$ be a ball of radius $R$ centered at the origin. 
	Then
\beq
\begin{array}{rlr}
- (V_{\W}*1)_{_{\!B_R}}(\bfr)
&= 
{\displaystyle{\frac{\pi}{4\varkappa^3}}}
\biggl[\arctan \left(\varkappa (R + r)\right)+\arctan \left(\varkappa (R -r)\right)+ \biggr.
&\phantom{(A.1)}\\
&\hskip+3.5truecm 
\biggl.
{\displaystyle{\frac{ 2\varkappa R\left(\varkappa^2 (R^2-r^2)-1\right) }{ 
(\varkappa^2 (R^2+r^2)+ 1)^2- 4\varkappa^4R^2r^2}}}\biggr]\, ,
&({\rm A}.1) \\
- (V_{\Y}*1)_{_{\!B_R}}(\bfr)
& =
	{\displaystyle{\frac{4 \pi }{ \kappa^2}}} \left[ 1-(1+\kappa R)e^{-\kappa R}{\sinh(\kappa r)/ \kappa r}\right]\,,
&({\rm A}.2)\\
- (V_{\N}*1)_{_{\!B_R}}(\bfr)
& = 
	{2 \pi}\left(R^2 - {\tst\frac{1}{3}}r^2\right)\phantom{\Bigl(\Bigr)} \, .
&({\rm A}.3)
\end{array}
\nonumber
\eeq
	Setting $\bfr=0$ in (A.1), (A.2), and (A.3) produces
\beq
\begin{array}{rlr}
	\norm{(V_{\W}*1)_{_{\!B_R}}}{C^0_b}\!\!\!\!
=
	\norm{V_{\W}(|\,\cdot\,|)}{L^1(B_R)}\!\!\!\!\!\!\!& = 
{\displaystyle{ \frac{\pi}{2\varkappa^3}}}\left[\arctan \left({\varkappa R}\right)+\varkappa R{\displaystyle{\frac{\varkappa^2 R^2-1
								}{(\varkappa^2 R^2+ 1)^2}}}\right]\!\!,
&({\rm A}.4)\\
	\norm{(V_{\Y}*1)_{_{\!B_R}}}{C^0_b}\!\!\!\!
=	
	\norm{V_{\Y}(|\,\cdot\,|)}{L^1(B_R)} \!\!\!\!\!\!\!&=
	{\displaystyle{\frac{4\pi}{\kappa^2}}} \left[ 1-(1+\kappa R)e^{-\kappa R}\right],
&({\rm A}.5) \\
	\norm{(V_{\N}*1)_{_{\!B_R}}}{C^0_b}\!\!\!\!
=	
	\norm{V_{\N}(|\,\cdot\,|)}{L^1(B_R)} \!\!\!\!\!\!\!&= 
	{2 \pi R^2} \, .
&({\rm A}.6)
\end{array}\nonumber
\eeq
	Integrating (A.1), (A.2), and (A.3) over $B_R$ yields
\beq
\begin{array}{rlr}
\norm{(V_{\W}*1)_{_{\!B_R}}}{L^1(B_R)}
\!\!\!\!\!\!&
 = \!
{\displaystyle{ \frac{\pi^2}{ 6\varkappa^6}}}\!
\left[\!4\varkappa^3 R^3\arctan(2\varkappa R)-4\varkappa^2 R^2\!+\ln\Big(1+4\varkappa^2 R^2\Big)\!\right]\!\!,
&({\rm A}.7)\\
\norm{(V_{\Y}*1)_{_{\!B_R}}}{L^1(B_R)}
\!\!\!\!\!\!&
 =
	{\displaystyle{\frac{16\pi^2}{3\kappa^5}}}\kappa^3R^3 \left[1-(1+\kappa R)
{\displaystyle{\frac{3}{2}\!\left[\!\frac{1\!+\!e^{-2\kappa R}}{\kappa^2 R^2} - 
		\frac{1\!-\!e^{-2\kappa R}}{\kappa^3 R^3}\!\right]}}\!\right]\!\!,\;
&({\rm A}.8)\\
\norm{(V_{\N}*1)_{_{\!B_R}}}{L^1(B_R)}
\!\!\!\!\!\!&
= 
{\displaystyle{\frac{8 \pi^2}{ 45}}}R^5 \, .
&({\rm A}.9)
\end{array}
\nonumber
\eeq
}

\smallskip\noindent
{B. ASSOCIATED PARTIAL DIFFERENTIAL EQUATIONS}
\medskip

	For Yukawa and Newton kernels the integral equation \refeq{fixptEQinLambda} for the density 
$\eta(\bfr)$ in $\Lbd\subset\RR^3$ is equivalent to a semilinear elliptic$^{\cite{gilbargtrudinger}}$ 
PDE of second order together with consistent boundary condition for the corresponding 
chemical self potential per particle, $-V *\eta$.
	Thus, setting $ - V_{\Y}*\eta \equiv \psi$, we find from \refeq{fixptEQinLambda} that $\psi$ solves 
\beq
	- \Delta \psi (\bfr) 
= 
	4\pi \wp_{\bullet}^\pr\big(\gamma+\alpha\psi(\bfr)\big)-\kappa^2 \psi(\bfr)\, .
\hfill({\rm B.1})\nonumber
\eeq
	In the formal Newtonian limit $\kappa \to 0$ we have 
$V_{\Y}\to V_{\N}$, and (B.1) 
reduces to 
\beq
	- \Delta \psi (\bfr) 
= 
	4\pi \wp_{\bullet}^\prime \big(\gamma + \alpha \psi(\bfr)\big) \, .
\hfill({\rm B.2})\nonumber
\eeq
	In the low density limit, (B.1) reduces to 
\beq
- \Delta \psi (\bfr) = 4\pi \zeta e^{\alpha \psi(\bfr)} -\kappa^2 \psi(\bfr)\, ,
\hfill({\rm B.3})\nonumber
\eeq
and (B.2) to
\beq
- \Delta \psi (\bfr) = 4\pi \zeta e^{\alpha \psi(\bfr)} \, ,
\hfill({\rm B.4})\nonumber
\eeq
with $\zeta = e^\gamma$ the fugacity.
	In each case,  \refeq{VconvRHO} evaluated at $\partial\Lbd$ provides a nonlinear and nonlocal 
boundary condition for $\psi$, which makes it quite difficult to study these PDEs in general domains.

\smallskip
\noindent
{\bf Remark:} If in \refeq{fixptEQinLambda} one replaces $\wp_{\bullet}^\pr(\gamma)$ 
by the strictly convex function $\exp(\gamma)$, then the alternative stated after Proposition 6.3
ceases to exist and the map $\gamma\mapsto{\eta}^{\ssst\, m}_{\ssst\Lbd}|_{\alpha}\in{C^0_b(\Lbd)}$ 
actually terminates at $\gamma_*^{\ssst\Lbd}(\alpha)$; see Refs.\cite{fujita,amannB}
.~\eQED

	For spherically symmetric solutions, i.e. $\psi(\bfr) = \phi(r)$ in a ball of radius $R$, 
satisfying the regularity condition $\phi^\prime(0) =0$, the PDEs (B.1), (B.2), (B.3),  and (B.4)  
simplify to ODEs with nonlinear and nonlocal boundary conditions that read,
for (B.1):
\beq
\phi(R) = 4\pi \frac{e^{-\kappa R} }{ \kappa R}
\int_0^R r \sinh(\kappa r) \wp^\pr_\bullet \big(\gamma + \alpha \phi(r)\big)dr \, ,
\hfill({\rm B.5}) \nonumber
\eeq
for (B.2): 
\beq
	\phi(R) 
= 
	4\pi \frac{1}{ R}\int_0^R r^2 \wp^\pr_\bullet \big(\gamma +\alpha\phi(r)\big)dr 
\, 
\hfill({\rm B.6})\nonumber
\eeq
for (B.3): 
\beq
\phi(R) = 4\pi \zeta \frac{e^{-\kappa R} }{ \kappa R}\int_0^R r \sinh(\kappa r) e^{\alpha \phi(r)} dr \, ,
\hfill({\rm B.7})\nonumber
\eeq
for (B.4): 
\beq
\phi(R) = 4\pi \zeta \frac{1}{R}\int_0^R r^2 e^{\alpha \phi(r)} dr \, .
\hfill({\rm B.8})\nonumber
\eeq
	For spherical symmetry, (B.2) has nice scaling properties which facilitate its 
discussion and aid in its numerical integration on a machine; see Ref.\cite{bsmkks,mkRMP}.  
	Its low density limit (B.4) becomes the homologously invariant isothermal 
gaseous ball equation, which has been extensively studied by 
Emden,$^{\cite{emdenBOOK}}$ Chandrasekhar$^{\cite{chandrasekhar}}$ and others.$^{\cite{gelfand,josephlundgren}}$
	Such scaling properties are not shared by (B.1), or (B.3),  
for which numerical studies of radial solutions apparently have not yet been carried out. 

	The spherical version of \refeq{fixptEQinLambda}, with $V=V_{\W}$ in
\refeq{VconvRHO}, does not seem to reduce to an ODE, and numerical integration of the integral 
equation \refeq{fixptEQinLambda} with $V=V_{\W}$ are correspondingly more involved, see Ref.\cite{mkjkp}. 

	The special case $\Lbd=\RR^3$, with $\eta(\bfr)$ solving \refeq{fixptEQ},
is of interest in itself, as explained in the introduction. 
	Since equations (B.1), (B.2), (B.3), and (B.4) do not depend on $\Lbd$, the same PDEs cover
the case $\Lbd=\RR^3$.
	However, instead of taking the formal limits $\Lbd\to\RR^3$ for their self-consistent boundary conditions,
the situation is more subtle.
	We illustrate this with the spherically symmetric situation, with finite $R$ 
boundary conditions (B.5), (B.6), (B.7), and (B.8).
	In fact we need to drop (B.6) and (B.8), for their limits are infinite because the
respective equations (B.2) and (B.4) do not possess solutions with their right-hand side in $L^1(\RR^3)$.
	For (B.1) under spherical symmetry, we may or may not include the limit $R\to\infty$ of (B.5), 
which is easily shown to be zero because $\wp^\pr$ is bounded; similarly, for bounded radially symmetric
solutions of (B.3) the limit $R\to\infty$ of (B.7) vanishes.
	Yet if we do include the condition that $\psi(\bfr)\to 0$ as $|\bfr|\to\infty$, then
we throw out all the constant solutions 
$\bfr\mapsto\ol\psi_{\vdW}(\bfr)\equiv -\ol\eta_{\vdW}\norm{V_{\Y}}{1}$. 
	This shows that the spatially constant van der Waals densities $\bfr\mapsto \ol\eta_{\vdW}$ 
are more subtle limits of the finite volume non-uniform van der Waals densities, namely in the sense
of supnorm convergence on the members of any sequence of nested compact subsets of $\RR^3$, which 
sequence converges to $\RR^3$; of course, convergence is also weak, i.e. pointwise.

	In the {\it wide interface approximation},$^{\cite{vankampen,percusA,percusB}}$ 
for our short ranged $V_\W\in L^1(\RR^3)$ and $V_\Y\in L^1(\RR^3)$ the convolution
$V*\eta$ given by \refeq{VconvRHO} for $\Lbd=\RR^3$ can be expanded to second order, and 
the fixed point equation \refeq{fixptEQ} reduces to a PDE for $\eta$ (not $\psi$), viz.
\beq
-\alpha M_2(V)\Delta \eta(\bfr) + \alpha\norm{V}{1}\eta(\bfr) 
  = 
\wp_{\bullet}^{\prime}{}^{-1}\big(\eta(\bfr)\big) -\gamma , 
\hfill({\rm B.9})\nonumber
\eeq
where 
\beq
M_2(V) = \frac{1}{6} \int_{\RR^3} |x|^2 V(|x|)d^3r
\hfill({\rm B.10})\nonumber
\eeq
is the ``second moment'' of $V$.
	Notice that (B.9) is the Euler--Lagrange equation for a so-called Cahn--Hilliard functional,
studied recently in Ref.\cite{carlenETalA}.

	Our numerical studies of \refeq{fixptEQinLambda} for $\Lbd=B_R$ with $R=50$ and $\varkappa=1$ revealed that 
near the critical point one finds solutions with inhomogeneity scale $R$. 
	This leads us to the following (mildly vague) conjecture:

\noindent
{\bf Conjecture:} {\it For $(\ag)$ in some droplet neighborhood of the (weakly $\Lbd$-dependent) critical point, 
the wide interface approximation becomes asymptotically exact, in the sense that one finds droplet
solutions $\eta_{_\Lbd,d}$ of \refeq{fixptEQinLambda} which converge in a suitable but reasonable sense to solutions
of ({\rm B.9}) in some ``universal'' limit as $\Lbd\to\RR^3$.}

\smallskip\noindent
{C. REVERSAL TO THE DIMENSIONAL QUANTITIES OF PHYSICS}
\smallskip

	As conventional in chemical physics we have used dimensionless units in which the ``density'' 
$\ol\eta$ is actually the volume fraction occupied by all the microscopic balls.
	Thus, if $N$ balls, having volume $|b|$ each, are inside a container $\ol\Lbd$ 
of volume $|\Lbd|$, then $\ol\eta = N|b|/|\Lbd|$.
	Also, we have absorbed several ``constants of nature'' in our quantities, and moreover ignored
the usual heuristic injection of quantum mechanics as per the thermal de Broglie wavelength.
	To make contact with physics one needs to reconvert our dimensionless into dimensional variables. 
	It suffices to do the conversion for the model with the van der Waals interaction potential; the 
conversion for the model with Yukawa or Newton interactions is done entirely analogously.

	Thus, $|b|$ is dimensional (a volume), and  we have to make the following replacements
``dimensionless''$\to$ ``dimensional'' quantities:
$\bfr\to \bfr/|b|^{1/3}$ for the position vectors, and therefore all lengths --- in particular,
$\varkappa\to |b|^{1/3}\varkappa$; next, $\alpha \to \beta\alpha$ for the coupling constant~:~temperature ratio; 
$\gamma \to \beta\mu - \ln (\lambda_\dB^3/ |b|)$ for the chemical potential per particle~:~temperature ratio;
$p \to |b| \beta p$ for the pressure~:~temperature ratio; $\eta \to |b|\rho$ for the particle density.
	We now have $\beta = (\kB T)^{-1}$, with $T$ the temperature in degree Kelvin, and 
$\lambda_\dB$ is the thermal de Broglie wavelength.
	In the same vein, we need to replace 
$\ln \eta(\bfr)\to \ln\big( \rho(\bfr)/\ol\rho_{\dB}\big)$ in the entropy functional, where
$\ol\rho_{\dB} = (2\pi m \kB T)^{3/2}/h^3$ is the thermal ``de Broglie density.''

	For applications to, say, fluids made of the nobel elements, the physical ordering is 
$|b| < 4\pi \varkappa^{-3}/3$ and $\varkappa \diam(\Lbd)\gg 1$.
	Numerically, $|b|\approx 1$\AA$^3$, and $\varkappa^{-1}\approx 2$\AA\ seem reasonable, while
$\diam(\Lbd)\approx 10-10^2\mathrm{cm}$ seems a reasonable range of laboratory container sizes.
	Also, the dimensional van der Waals coupling constant $\alpha$ has physical dimension of energy,
numerically in the range of ``typical molecular binding energies'' of the natural gases, although of 
course there is no quantum mechanical formation of Ne$_2$, Ar$_2$, etc. molecules in nature.
	The attraction between Ne, Ar, etc. atoms is manifested most dramatically through the 
condensation / evaporation phase transition exhibited by these chemical elements of matter.

\noindent
{D. ERRATA FOR REFS.\cite{mkJSPa} and \cite{mkjkp}} 
\smallskip

	All corrections to our previous papers Refs.\cite{mkJSPa} and \cite{mkjkp} are easy to make.
	Those in the categories {\it typo} and {\it slip of pen} are just listed without comment, 
those which deserve a commentary are commented on in footnotes. 
	Expressions to be replaced are surrounded by quotation marks.

\medskip
\noindent
{D.a. \it Errata for Ref.\cite{mkJSPa}}
\smallskip

p.223: above (2.66), replace ``it has been shown$^{(1)}$'' 
by\footnote{In fact, there is a small mistake in ref. 1 of Ref.\cite{mkJSPa}
to the effect that the factor $(N-1)$ in (2.66) (quoted from ref. 1) is incorrect.
The correct factor is $ N$; see Ref.\cite{mkCPAM}.}
``it has been argued$^{(1)}$''. 

p.238: in (4.7a), replace ``$|U$'' by ``$|U|$''.

p.248: replace\footnote{The critical sentence is: ``That implies that there exists an 
	incompressible mapping $ \rho_1 \mapsto \rho_2$.'' While true for some types of phase transitions
	associated with symmetry breaking, it is not clear that such incompressible mappings exist in the 
	context of the theorem.
	M.K. is grateful to Elliott Lieb for kindly pointing this out.}

\begin{quote}
``Then also $s(\rho_1)= s(\rho_2)$. That implies that there exists an incompressible 
mapping $\rho_1 \mapsto \rho_2$. (Note that entropy is conserved for incompressible 
mappings.) As has been shown in ref. 10 (see also refs. 8 and 21), any 
given $\rho_0$ can be mapped incompressibly to a unique spherical minimizer $\rho_M$ of 
$e(\rho)$ with $s(\rho_0)= s(\rho_M)$. By construction both $\rho_1$ and $\rho_2$ minimize 
$e(\rho)$ under conservation of entropy; hence, $\rho_1 \equiv \rho_2$, in contradiction to 
the assumption that the densities are not identical.''
\end{quote}
\centerline{by\footnote{Notice that the correction given here not only avoids 
	the pitfall of the original proof, it also eliminates the requirement of the original proof
	that $\Lambda$ be spherical.
	This nonspherical argument, taken from Ref.\cite{mkCMPa}, is a special case of the argument in our
	proof of Theorem 7.2; see the penultimate remark  in section VII.}}
\begin{quote}
``Since also $\tilde{f}(\rho_1)= \tilde{f}(\rho_2)$ (where $\tilde{f}(\rho)$ is given in (3.28), 
here with $\psi\equiv 0$), we then conclude that  
$\int_\Lambda \exp(-\beta_{\tr} U*\rho_1)d^3r = \int_\Lambda \exp(-\beta_{\tr} U*\rho_2)d^3r$ as
well. Hence, $\,\eta(\beta_{\tr};\rho_1)\, =\, \eta(\beta_{\tr};\rho_2)\,\equiv\, \eta_{\tr}$ 
(see p. 238 for the definition of $\eta(\beta;\rho)$). But then, since 
$\rho_1 =\tilde{\rho}_{\beta_{\mathrm{tr}}}$, so that
$-\beta_{\tr}U*\rho_1 =-\beta_{\tr}U*\tilde{\rho}_{\beta_{\rm tr}}=\tilde\Psi_{\eta_{\tr}}$
is the unique pointwise minimal solution of (4.6) for this value of $\eta\,=\,\eta_{\tr}$, 
it follows that 
$\int_\Lambda \exp(-\beta_{\rm tr} U*\rho_1)d^3r < \int_\Lambda \exp(-\beta_{\rm tr} U*\rho_2)d^3r$, 
which is a contradiction.''
\end{quote}
\medskip

\noindent{D.b. \it Errata for Ref.\cite{mkjkp}}
\smallskip

p.1353: replace ``The Hilbert space'' by ``The space''

p.1364: in (6.19), replace ``$O[r_0/R]$'' by ``$O[(r_0/R)^3]$''

p.1376: in Ref.33, replace ``Phnys.'' by ``Phys.''

\bigskip
%
%
%
\newcommand{\amst}[1]{  {\sl   Am. Math. Soc. Transl.     }{\bf{#1}}, }
\newcommand{\arma}[1]{  {\sl   Arch. Rational Mech. Anal. }{\bf{#1}}, }
\newcommand{\apj}[1]{   {\sl   Astrophys. J.              }{\bf{#1}}, }
\newcommand{\asa}[1]{   {\sl   Astron. Astrophys.         }{\bf{#1}}, }
\newcommand{\ass}[1]{   {\sl   Astrophys. Space Sci.      }{\bf{#1}}, }
\newcommand{\ajp}[1]{   {\sl   Aust. J. Phys.             }{\bf{#1}}, }
\newcommand{\cmp}[1]{   {\sl   Commun. Math. Phys.        }{\bf{#1}}, }
\newcommand{\cpam}[1]{  {\sl   Commun. Pure Appl. Math.   }{\bf{#1}}, }
\newcommand{\grl}[1]{   {\sl   Geophys. Res. Lett.        }{\bf{#1}}, }
\newcommand{\ijqc}[1]{  {\sl   Int. J. Quant. Chem.       }{\bf{#1}}, }
\newcommand{\jcp}[1]{   {\sl   J. Chem. Phys.             }{\bf{#1}}, }
\newcommand{\jde}[1]{   {\sl   J. Diff. Eq.               }{\bf{#1}}, }
\newcommand{\jfa}[1]{   {\sl   J. Funct. Anal.            }{\bf{#1}}, }
\newcommand{\jgr}[1]{   {\sl   J. Geophys. Res.           }{\bf{#1}}, }
\newcommand{\jmp}[1]{   {\sl   J. Math. Phys.             }{\bf{#1}}, }
\newcommand{\jpp}[1]{   {\sl   J. Plasma Phys.            }{\bf{#1}}, }
\newcommand{\jsp}[1]{   {\sl   J. Stat. Phys.             }{\bf{#1}}, }
\newcommand{\hpha}[1]{  {\sl   Helv. Phys. Acta           }{\bf{#1}}, }
\newcommand{\mnras}[1]{ {\sl   Mon. Not. R. Astron. Soc.  }{\bf{#1}}, }
\newcommand{\phys}[1]{  {\sl   Physica                    }{\bf{#1}}, }
\newcommand{\phya}[1]{  {\sl   Physica A                  }{\bf{#1}}, }
\newcommand{\phyb}[1]{  {\sl   Physica B                  }{\bf{#1}}, }
\newcommand{\phyc}[1]{  {\sl   Physica C                  }{\bf{#1}}, }
\newcommand{\phyd}[1]{  {\sl   Physica D                  }{\bf{#1}}, }
\newcommand{\phye}[1]{  {\sl   Physica E                  }{\bf{#1}}, }
\newcommand{\prev}[1]{  {\sl   Phys. Rev.                 }{\bf{#1}}, }
\newcommand{\prea}[1]{  {\sl   Phys. Rev. A               }{\bf{#1}}, }
\newcommand{\preb}[1]{  {\sl   Phys. Rev. B               }{\bf{#1}}, }
\newcommand{\preC}[1]{  {\sl   Phys. Rev. C               }{\bf{#1}}, }
\newcommand{\pred}[1]{  {\sl   Phys. Rev. D               }{\bf{#1}}, }
\newcommand{\prel}[1]{  {\sl   Phys. Rev. Lett.           }{\bf{#1}}, }
\newcommand{\phfl}[1]{  {\sl   Phys. Fluids               }{\bf{#1}}, }
\newcommand{\phfa}[1]{  {\sl   Phys. Fluids A             }{\bf{#1}}, }
\newcommand{\phfb}[1]{  {\sl   Phys. Fluids B             }{\bf{#1}}, }
\newcommand{\pss}[1]{   {\sl   Planet. Space Sci.         }{\bf{#1}}, }
\newcommand{\rmp}[1]{   {\sl   Rev. Mod. Phys.            }{\bf{#1}}, }
\newcommand{\scr}[1]{   {\sl   Space Sci. Rev.            }{\bf{#1}}, }
\newcommand{\tams}[1]{  {\sl   Trans. Am. Math. Soc.      }{\bf{#1}}, }
\newcommand{\zph}[1]{   {\sl   Z. Phys.                   }{\bf{#1}}, }


%
\end{document}